\documentclass{aa}  
\usepackage[varg]{txfonts}

\usepackage{graphicx}
\usepackage{subfigure}
\usepackage{txfonts}

\usepackage{color}
\usepackage{textgreek} % \textalpha \textAlpha ...
\usepackage{multirow} 
\usepackage{natbib}
\bibpunct{(}{)}{;}{a}{}{,} % to follow the A&A style
\usepackage{lipsum}

\usepackage[colorlinks=true,
            linkcolor=blue, % initially red
            urlcolor=blue,
            citecolor=blue]{hyperref}
            
\usepackage{flushend} % To balanced the two column at the end

\begin{document} 
\title{\textit{Herschel} survey and modelling of externally-illuminated photoevaporating protoplanetary disks
\thanks{{\it Herschel} is an ESA space observatory with science instruments provided by European-led Principal Investigator consortia and with important participation from NASA.}}
%\titlerunning{Photoevaporation of protoplanetary disks} % Limited to about 60 characters
\author{
	J. Champion\inst{1,2} \and
	O. Bern\'e\inst{1,2} \and
	S. Vicente\inst{3,4} \and
	I. Kamp\inst{3} \and
	F. Le Petit\inst{5} \and
	A. Gusdorf\inst{6} \and
	C. Joblin\inst{1,2} \and
	J. R. Goicoechea\inst{7}
          }
\authorrunning{J. Champion et al.}

\institute{Universit\'e de Toulouse, UPS-OMP, IRAP, Toulouse, France \\
	\email{jason.champion@irap.omp.eu}
	\and CNRS, IRAP, 9 Av. colonel Roche, BP 44346, 31028 Toulouse Cedex 4, France
	\and Kapteyn Astronomical Institute, University of Groningen, Postbus 800, 9700 AV Groningen, The Netherlands
	\and Institute of Astrophysics and Space Sciences (IA), Tapada da Ajuda - Edificio Leste - 2$^\circ$ Piso, 1349-018 Lisboa, Portugal
	\and LERMA, Observatoire de Paris, PSL Research University, CNRS, UMR8112, F-92190 Meudon, France
	\and LERMA, Observatoire de Paris, \'Ecole normale sup\'erieure, PSL Research University, CNRS, Sorbonne Universit\'es, UPMC Univ. Paris 06, F-75231, Paris, France
	\and Grupo de Astrofisica Molecular, Instituto de Ciencia de Materiales de Madrid (CSIC), E-28049, Madrid, Spain}
	
%\date{Received 00 month 2016 / Accepted 00 month 2017}
\date{Received July 2016 / Accepted January 2017}

%======================= End of title part =======================%

%===========================================================%
%						      0. Abstract							 %
%===========================================================%

\abstract
{Protoplanetary disks undergo substantial mass-loss by photoevaporation, a mechanism which is crucial to their dynamical evolution. However, the processes regulating the gas energetics have not been well constrained by observations so far.}
{We aim at studying the processes involved in disk photoevaporation when it is driven by far-UV photons (i.e. 6 < E < 13.6 eV). }
{We present a unique \textit{Herschel} survey and new ALMA observations of four externally-illuminated photoevaporating disks (a.k.a. proplyds). For the analysis of these data, we developed a 1D model of the photodissociation region (PDR) of a proplyd, based on the Meudon PDR code and we computed the far infrared line emission.}
{With this model, we successfully reproduce most of the observations and derive key physical parameters, i.e. densities at the disk surface of about $10^6$ cm$^{-3}$ and local gas temperatures of about 1000 K. Our modelling suggests that all studied disks are found in a \textit{transitional} regime resulting from the interplay between several heating and cooling processes that we identify. These differ from those dominating in classical PDRs i.e. grain photo-electric effect and cooling by \big[OI\big] and \big[CII\big] FIR lines. This specific energetic regime is associated to an equilibrium dynamical point of the photoevaporation flow: the mass-loss rate is self-regulated to keep the envelope column density at a value that maintains the temperature at the disk surface around 1000 K. From the physical parameters derived from our best-fit models, we estimate mass-loss rates -- of the order of $10^{-7}$ M$_\sun$/yr -- that are in agreement with earlier spectroscopic observation of ionised gas tracers. This holds only if we assume photoevaporation in the \textit{supercritical} regime where the evaporation flow is launched from the disk surface at sound speed.}
{We have identified the energetic regime regulating FUV-photoevaporation in proplyds. This regime could be implemented into models of the dynamical evolution of protoplanetary disks.}

% Manque quelque chose sur les PAH

 % see http://www.aanda.org/index2.php?option=com_content&task=view&id=170&Itemid=184
	
\maketitle

%\textcolor{red}{(Useless text to avoid a fatal compilation error related to the box: pdfTeX error (ext4): "pdfendlink" ended up in different nesting level than "pdfstartlink".)}

%======================= End of abstract =======================%

%===========================================================%
%						  1. Introduction							 %
%===========================================================%

\section{Introduction}

Protoplanetary disks are found around young stars as a natural consequence of angular momentum conservation in a rotating collapsing core. During a few million years \citep{zuckerman_inhibition_1995, haisch_disk_2001}, they are the reservoirs of dust and gas for the formation and the growth of planets. After this relatively short time, only a small fraction of the disk initial mass remains in the form of planets. The main part is supposed to have been accreted onto the central star or physically removed from the disk and one proposed mechanism is photoevaporation \citep[see recent reviews in e.g.][]{alexander_dispersal_2014, gorti_disk_2015}. Indeed, disks are subject to highly energetic photons that can sufficiently heat the gas at the disk surface to make it overcomes the gravitational field leading to escape flows and mass-loss. 
%Ultraviolet (UV) radiation is generally split into two ranges: Far-UV (FUV) for photons with an energy between 6 and 13.6 eV; Extreme-UV (EUV) between 13.6 eV and 0.1 keV. %X-rays between 0.1 and 100 keV. 
%FUV photons have the capability to dissociate H$_\mathrm{2}$ molecules while EUV photons can ionise hydrogen atoms. 

Photoevaporation of disks caused by Extreme-UV photons (or EUV, $13.6$ eV $< h\nu < 0.1$ keV) arising from the central star was first studied by \citet{hollenbach_photoevaporation_1994, yorke_photoevaporation_1996, richling_photoevaporation_1997, clarke_dispersal_2001} who showed that it can create a gap in the inner part of the disk and erodes it. \citet{gorti_photoevaporation_2009} have investigated the case of Far-UV (or FUV, $6$ eV $< h\nu < 13.6$ keV) photons and found that they can cause significant erosion since they can remove the gas from the outer disk where most of the mass is located. 
%The role of X-rays (0.1 to 100 keV) was found minor \citep{alexander_effects_2004} but can enhance the effect of photoevaporation by FUV photons by a factor of 2 \citep{gorti_photoevaporation_2009}. 
Photoevaporation of disks caused by EUV and FUV photons from nearby massive stars has also been analytically and numerically modelled \citep{johnstone_photoevaporation_1998, richling_photoevaporation_1998, richling_photoevaporation_2000, storzer_photodissociation_1999}, in order to understand the morphology and evolution of externally illuminated disks, also known as proplyds \citep{odell_discovery_1993}. These latters generally appear as disks surrounded by a tear-drop-shaped ionised envelope with a bright head facing the main UV source and an elongated tail pointing away.
%These objects were first detected in the Orion nebula cluster with ground-based optical \citep{laques_detection_1979} and radio observations \cite[e.g.][]{churchwell_solar_1987}. In the 90's, high angular resolution observations with the \textit{Hubble} space telescope (HST) identified and resolved many of them \citep{odell_postrefurbishment_1994, mccaughrean_direct_1996, bally_externally_1998, smith_new_2005, ricci_hubble_2008}. They generally appear as disks surrounded by a tear-drop-shaped ionised envelope with a bright head facing \textTheta{}$^\mathrm{1}$ Ori C, the main UV source in the nebula, and an elongated tail pointing away. Such objects have also been observed and resolved in other HII regions such as in NGC 3603 \citep{brandner_hst/wfpc2_2000}, in the Carina nebula \citep{smith_numerous_2003} or in Cygnus OB2 \citep{wright_photoevaporating_2012}. 
After the work of \citet{johnstone_photoevaporation_1998} and \citet{storzer_[o_1998, storzer_photodissociation_1999}, supported by observations \citep[e.g.][]{bally_externally_1998, chen_2.12_1998}, it was suggested that FUV photons penetrate the disk, dissociate molecules and heat the gas to temperatures about 10$^\mathrm{2}$ to 10$^\mathrm{3}$ K, thus launching a thermal wind at the disk surface. EUV photons cannot penetrate this flow but ionise the gas on its surface and raise its temperature to 10$^\mathrm{4}$ K. The ionisation front, delimiting the flow that feeds the envelope of the proplyd, is generally located at several disk radii. 

Regions where a strong FUV radiation field drives the physics and chemistry are called PhotoDissociation Regions (PDRs). All of the previously cited models of photoevaporation that include FUV photons are based on PDR models. \citet{tielens_photodissociation_1985} have developed the first general model describing PDRs for a dense molecular cloud illuminated by a strong FUV radiation field with an intensity $G_\mathrm{0}$ between 10$^\mathrm{3}$ and 10$^\mathrm{6}$ times the Habing field \citep{habing_interstellar_1968}, which corresponds to the average FUV flux in the interstellar medium and is valued at $1.6 \times 10^{-6} ~\mathrm{W\,m^{-2}}$ when integrated between 91.2 and 240 nm. To compute the structure and emission of the cloud, the model includes the main heating mechanisms such as the photoelectric effect, carbon ionisation, H$_\mathrm{2}$ dissociation and the main cooling mechanisms by few fine-structure lines of atoms and low-lying rotational lines from molecules for example. Thanks to PDR modelling, \citet{storzer_photodissociation_1999} have been able to reproduce the observed intensity of H$_\mathrm{2}$ ($\nu = 1 - 0$) S(1) and make predictions for the major gas cooling lines in the far-infrared (\big[OI\big], \big[CII\big], etc.). However so far, these latter lines have never been observed towards proplyds. \citet{walsh_molecular_2013} modelled the molecular line emission from a protoplanetary disk illuminated by an external radiation field $G_\mathrm{0}=4\times10^5$. \citet{facchini_external_2016} recently published a study in which they use a 3D PDR model
to compute the temperature at the surface of an externally illuminated disk. Their study concerns mainly 
disks situated in regions of moderate UV fields (up to $G_\mathrm{0}=3000$). Those authors also do not compare  these models with observations as they are mostly interested in modelling dynamics.

We conducted the first and unique far-IR (FIR) survey of proplyds with the \textit{Herschel Space Observatory}. The \textit{Herschel} data were combined with multi-wavelength observations (optical to sub-mm) available in the literature and analysed in the frame of a PDR model specially developed for this case. In this paper the following questions are adressed: 
\begin{itemize}
\item Can the FIR emission of proplyds be modelled as a PDR? 
\item What are the physical properties of this PDR (density, temperature)? 
\item What are the key mechanisms driving the thermal balance (heating and cooling processes) and hence the photoevaporation? 
\end{itemize}
We start by introducing the observations (Sect. \ref{sect_obs}) and the model (Sect. \ref{sect_proplydModel}). We then confront them to extract the physical properties of the studied objects (Sect. \ref{sect_modelResults}) and use the model to investigate the main mechanisms driving the disk surface properties (Sect. \ref{sect_surfaceHeating}) and the consequences on photoevaporation and mass-loss rates (Sect. \ref{sect_photoevapation_and_massloss}). Sect. \ref{sect_conclusions} contains our concluding remarks.

%====================== End of introduction ======================%

%===========================================================%
%						  2. Observations						 %
%===========================================================%

\section{Observations} \label{sect_obs}
 
 \subsection{\textit{Herschel} survey}
 
Although it is well established that FUV photons play a key role in the gas heating in PDRs, there is no quantitative study of classical PDR tracers in proplyds so far, mostly due to their small sizes and strong background emission. Since these tracers are mostly found at far-IR wavelengths, and hence have to be observed from space, the ESA \textit{Herschel Space Observatory} \citep{pilbratt_Herschel_2010} was the first to provide a sufficient sensitivity and resolution to observe them. 
 
 \subsubsection{Targets}
 
To maximise the signal-to-noise ratio for the first survey of this kind, we have carefully selected three proplyds (see Table \ref{table_targets_properties} but note that 203-506 will be presented later) according to their relatively large size (several arcseconds in diameter) and their strong incoming FUV radiation field ($G_\mathrm{0} > 10^4$). These targets are briefly described below:

\begin{itemize}
\item One of the giant proplyd candidates of \citet{smith_numerous_2003} in the Carina nebula (Fig. \ref{fig_halpha_105-600}), namely 104632.9-600354 in their paper and hereafter called 105-600 following \citet{mesa-delgado_protoplanetary_2016}. It is located amongst the southern pillars at a projected distance of 25.4 arcminutes south-east from \texteta{} Carinae and at 1 arcminute from the Bochum 11 cluster to the west. It was first studied by \citet{vicente_physical_2009} and has been observed with HST/ACS in H\textalpha{} emission \citep{smith_HST/ACS_2010}, revealing a large envelope pointing towards \texteta{} Carinae and a collimated bipolar jet that confirms the pre-stellar nature of this object. %Near-IR adaptive optics images obtained with VLT/NACO (PI: S. Vicente, ID: 076.C-0874) in Br\textgamma{} (2.17 \textmu m), H$_\mathrm{2}$ (1-0) S(1) (2.12 \textmu m), [FeII] (1.64 \textmu m) and in the K$_\mathrm{S}$-band show a nearly elliptical bright structure (1900 $\times$ 920 AU or 0.83\arcsec $\times$ 0.4\arcsec) inside the envelope possibly tracing the location of the young stellar object. The H$_\mathrm{2}$ and K$_\mathrm{S}$-band images show emission within the envelope with an hour-glass structure carved by a strong stellar wind jet, also visible in the [FeII] image \citep{vicente_physical_2009}. 
The large size of this object as well as its high molecular content ($M \approx$ 0.35 M$_\sun$ , after \citealt{sahai_are_2012}) could suggest that the young stellar object is still embedded in the remnant molecular core while the morphology is similar to a proplyd. From this point forward, we will consider that the molecular core inside the object is a disk since the true nature is not important for our study.

\item HST10 (also known as 182-413, Fig. \ref{fig_halpha_HST10}) is the most studied proplyd and was one of the first bright objects in which an embedded disk was seen in silhouette \citep[e.g.][]{odell_discovery_1993, odell_hubble_1996} and in emission through the \big[OI\big] 6300 \AA{}  and H$_\mathrm{2}$ (1-0) S(1) lines \citep{bally_externally_1998,  bally_disks_2000, chen_2.12_1998}. It is located at a projected distance of 56\arcsec{} from the star \textTheta{}$^1$ Ori C to the SE, and at 32.6\arcsec{} from \textTheta{}$^2$ Ori A to the NW. It is a teardrop shaped proplyd containing a prominent nearly edge-on disk. % (i $80\deg$, PA 86°\deg; \citet{bally_disks_2000}).

\item 244-440 (Fig. \ref{fig_halpha_244440}) is the largest proplyd observed in the Orion nebula cluster. It exhibits a faint nearly edge-on silhouette disk \citep{vicente_size_2005} and a one-sided microjet drived by the central star \citep{bally_disks_2000}. It is located at a projected distance of 142\arcsec{} from the star \textTheta{}$^1$ Ori C and at 29\arcsec{} from \textTheta{}$^2$ Ori A.
\end{itemize}

\subsubsection{Observing strategy and data reduction}

\begin{figure*}
 \begin{center}
    \subfigure[105-600]{%
  \label{fig_halpha_105-600}
  \raisebox{0.00\height}{\includegraphics[width=.4\linewidth, height=.4\linewidth]{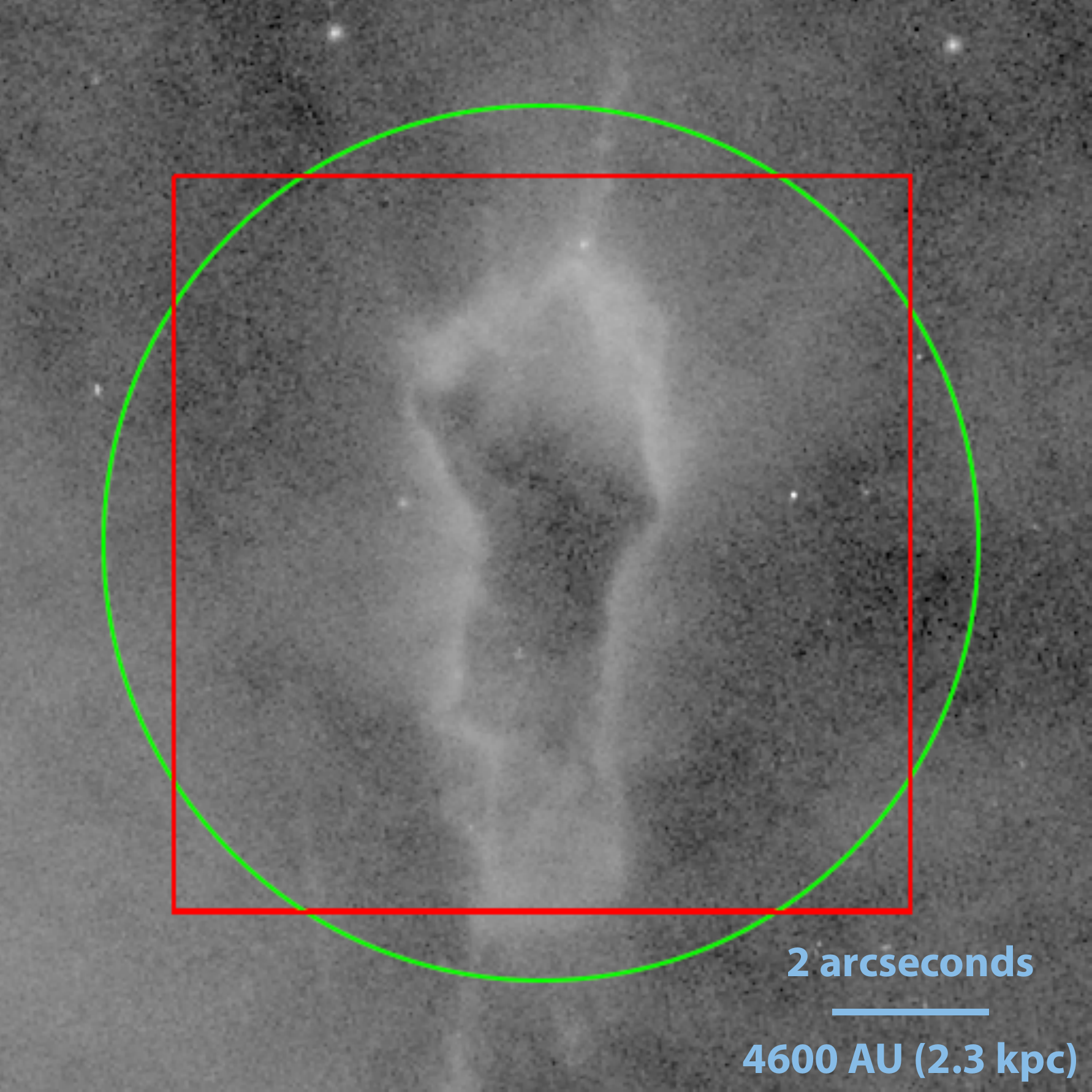}}
    }%
     \subfigure[HST10]{%
  \includegraphics[width=.4\linewidth, height=.4\linewidth]{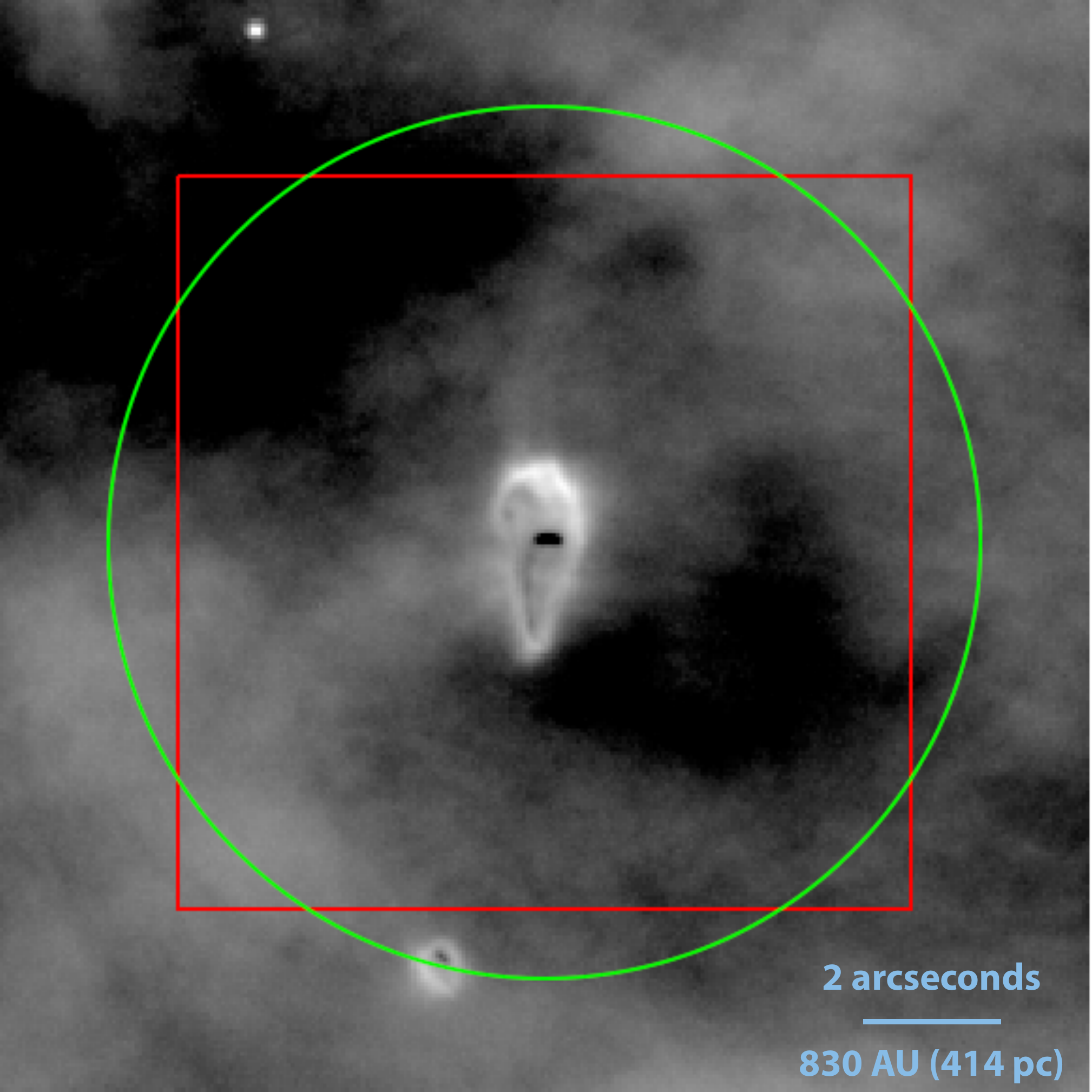}
  \label{fig_halpha_HST10}
  }%
   \\
    \subfigure[244-440]{% 
  \raisebox{0.00\height}{\includegraphics[width=.4\linewidth, height=.4\linewidth]{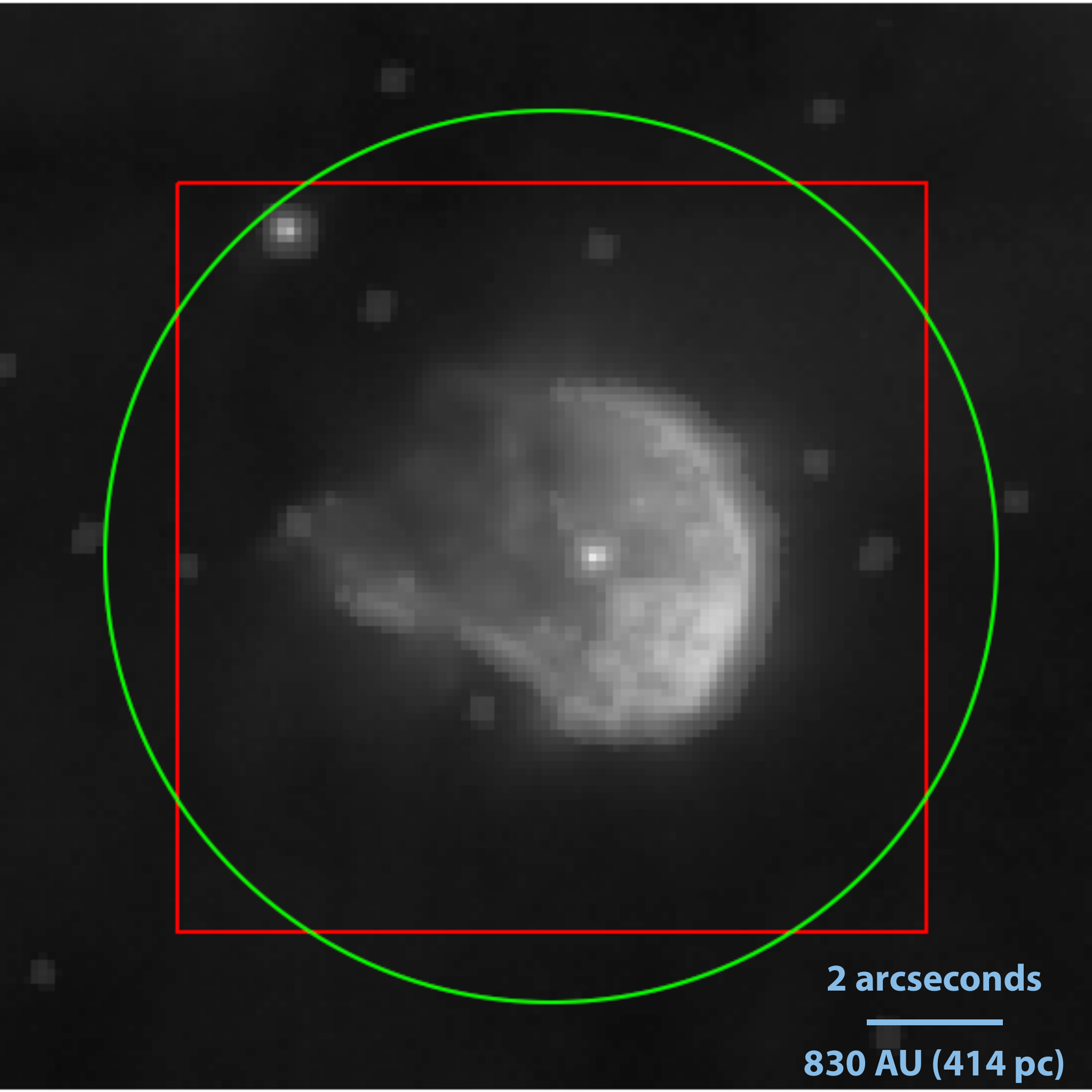}}
  \label{fig_halpha_244440}
  }% 
  \subfigure[203-506]{%
  \label{fig_halpha_203506}
  \raisebox{0.000\height}{\includegraphics[width=.4\linewidth, height=.4\linewidth]{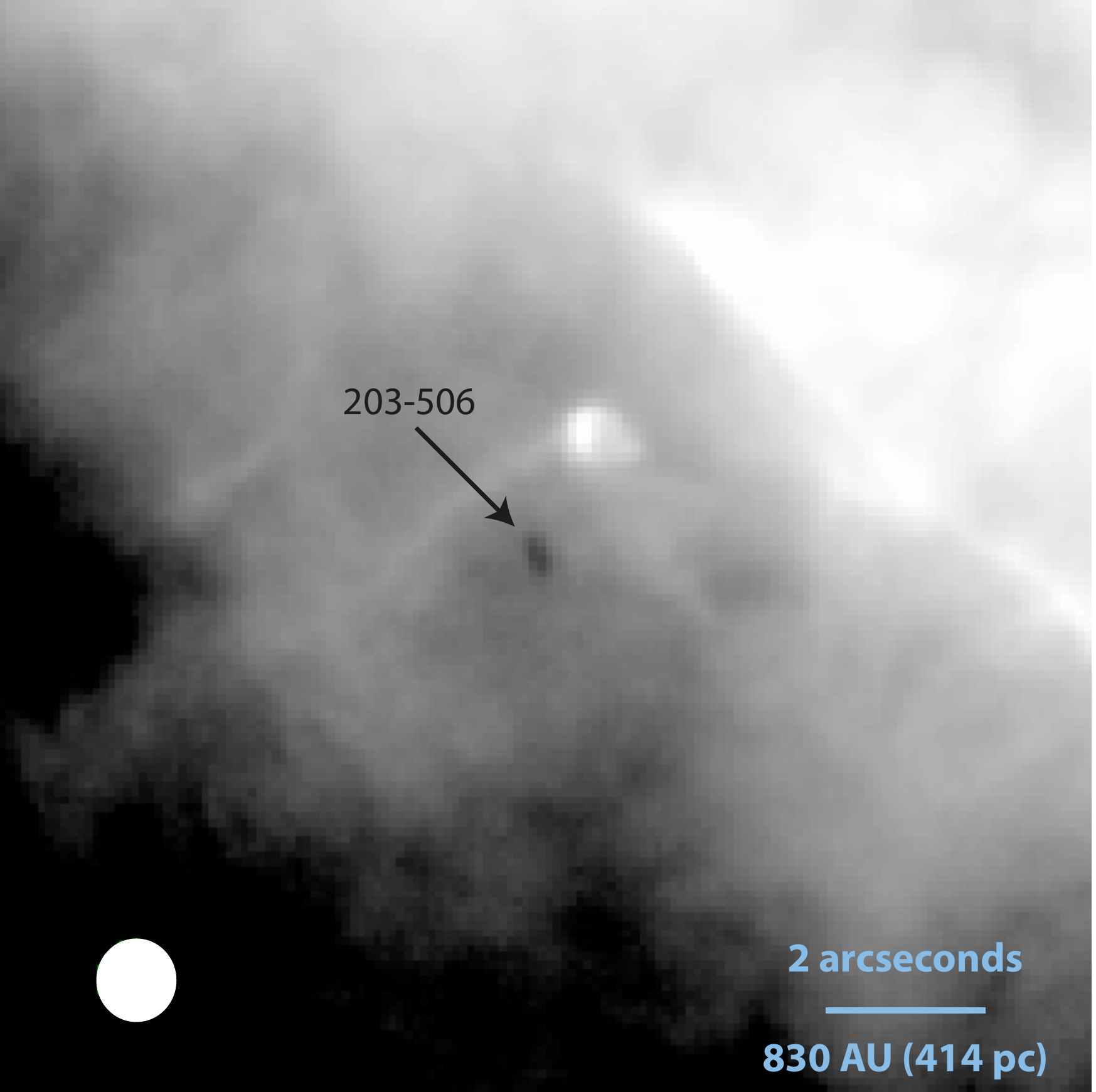}}
  }%
 \end{center}
 \caption{$14\arcsec \times 14\arcsec$ H\textalpha{} images of the observed proplyds from HST/WFPC2/F656N or WFC/F658N. Overlaid are the dimensions of one spaxel of PACS (red square of $9.4\arcsec \times 9.4\arcsec$) and the minimal FWHM of HIFI (green circle with a diameter of $11.3\arcsec$) for the ones observed with \textit{Herschel}. The ALMA beam of $0.99\arcsec\times{}0.97\arcsec$ is represented for 203-506.}
 \label{fig_halpha_images}
\end{figure*}

The main concern about observations of proplyds is that they are small objects so their emission is likely to be diluted in the \textit{Herschel} beam (Fig. \ref{fig_halpha_images}), and their emission can also be confused with emission from the nebula where they lie. Since the PDR density of proplyds is expected to be high at some locations, especially at the disk surface (about $10^6$ cm$^\mathrm{-3}$), some PDR tracers such as the \big[OI\big] 63 \textmu{}m and the high-$J$ CO lines ($J=$ 15-14, 17-16 and 19-18) are expected to be bright compared to background nebular emission (which has a lower density) and we thus proposed to observe them with PACS \citep{poglitsch_photodetector_2010}. With HIFI \citep{de_graauw_herschel-heterodyne_2010}, the \big[CII\big] 158 \textmu{}m line and fainter lower $J$ CO lines ($J=$ 7-6 and 10-9) were observed assuming that the proplyd emission can be spectrally separated from background/foreground nebular emission with a velocity resolution of $190$ m\,s$^\mathrm{-1}$ or less. Thanks to these two instruments, we were able to observe for the first time in proplyds selected excited CO lines spanning a wide range of densities (critical densities from 10$^\mathrm{5}$ to 10$^\mathrm{7}$ cm$^{\mathrm{-3}}$, see e.g. \citealt{yang_rotational_2010}), and the \big[OI\big] 63 \textmu{}m and \big[CII\big] 158 \textmu{}m fine-structure lines (critical densities from about 10$^\mathrm{3}$ to 10$^\mathrm{5}$ cm$^{\mathrm{-3}}$) which are generally the two major cooling lines in PDRs \citep{tielens_photodissociation_1985}. 

From one object or one line to another, different observing modes have been used to avoid contamination by the background emission (see Table \ref{table_targets_observations}). For HIFI, the mode "Dual Beam Switch" uses an OFF position (i.e. away from the source) on the sky if a nearby emission-free zone is expected, as for 105-600 mainly. If there is no such zone, the "Load Chop" mode allows to use internal loads as references and the mode "Frequency Switch" make its reference by switching between two frequencies in the local oscillator. In a similar way with PACS, we use the "Standard Chopping-nodding mode" for 105-600 assuming a vicinity free of emission while the "Unchopped grating scan" mode was used for the Orion proplyds that enables to go farther away from the nebula to find a clean OFF position.

The data reduction and parts of post-processing were done using the \textit{Herschel Interactive Processing Environment} \citep[HIPE version 13, see][]{ott_herschel_2010} in order to obtain level-2 calibrated spectra (Fig. \ref{fig_herschelLines}). Line profiles were then fitted using gaussian functions to extract the line integrated intensities and other characteristics such as the full width at half maximum (FWHM) and the velocity with respect to the local standard-of-rest, $v_\mathrm{LSR}$ (see Table \ref{table_lines_features}). Uncertainties have been calculated by quadratically summing the instrumental (see user manuals of the instruments for more details) and fitting uncertainties.

%%		 				2.1 selected targets						  %
%%--------------------------------------------------------------------------------------------------------%
%
%\subsection{Selected targets}  \label{subSubSect_obsTargets}
%
%
%Four sources have been considered in this study, namely the {\it Carina Candidate proplyd} (105-600 hereafter)
%situated in the Carina nebula; HST10 (or 182-413), 244-440, and 203-506 which are situated in the Orion Nebula. 
%The H$\alpha$ images for these four sources are presented in Fig.~\ref{fig_halpha_images}, and we
%briefly describe their general properties below.

\begin{table*}[]
  \begin{center} 
  \caption{General properties of the studied proplyds.}
  \label{table_targets_properties}
  \small
    \begin{tabular}{ccccccc}
 \hline\hline
 Proplyd name & RA & DEC & Apparent & Apparent & FUV field $G_\mathrm{0}$ & Heliocentric \\
  (parent nebula) & (J2000.0) & (J2000.0) & envelope size & disk size & (Habing field) & distance (parsec) \\
 \hline
 ~105-600 (Carina) & 10 46 32.97\tablefootmark{a} & $-$ 60 03 53.50\tablefootmark{a} & $9.5\arcsec\times{}3.7\arcsec$\tablefootmark{b} & ... & $2.2\times10^4$\tablefootmark{b} & $2300$\tablefootmark{c} \\
 HST10 ~~(Orion) & 05 35 18.22\tablefootmark{d} & $-$ 05 24 13.45\tablefootmark{d} & $2.6\arcsec\times{}1.0\arcsec$\tablefootmark{e} & $0.4\arcsec\times{}0.1\arcsec$\tablefootmark{e} & $2.4 \times 10^5$\tablefootmark{f} & $414$\tablefootmark{g} \\
 244-440 (Orion) & 05 35 24.38\tablefootmark{d} & $-$ 05 24 39.74\tablefootmark{d} & $5.6\arcsec$\tablefootmark{h} & $0.86\arcsec\times0.69\arcsec$\tablefootmark{h}  & $1\times10^5$\tablefootmark{b} & $414$\tablefootmark{g} \\
 203-506 (Orion) & 05 35 20.32\tablefootmark{d} & $-$ 05 25 05.55\tablefootmark{d} & ... & $0.75\arcsec\times{}0.61\arcsec$\tablefootmark{j}  & $2\times10^4$\tablefootmark{i} & $414$\tablefootmark{g} \\

 \hline
\end{tabular}
  \normalsize
  \end{center}
\tablefoot{Units of right ascension (RA) are hours, minutes, seconds and units in declination (DEC) are degrees, arcminutes and arcseconds (J2000.0). Sizes are given as the major axis diameter times the minor axis diameter of an elliptical object (only one value if circular). FUV fields are expressed in units of the Habing field which represents the estimated average FUV flux in the local interstellar medium \citep{habing_interstellar_1968}.} \\
\small
\tablefoottext{a}{\citet{sahai_are_2012}}, 
\tablefoottext{b}{\citet{vicente_physical_2009}}, 
\tablefoottext{c}{\citet{smith_numerous_2003}},
\tablefoottext{d}{\citet{ricci_hubble_2008}},
\tablefoottext{e}{\citet{chen_2.12_1998}},
\tablefoottext{f}{\citet{storzer_[o_1998}}, 
\tablefoottext{g}{\citet{menten_distance_2007}}, 
\tablefoottext{h}{\citet{vicente_size_2005}}, 
\tablefoottext{i}{mean value of \citet{marconi_near_1998} and \citet{walmsley_structure_2000}}, 
%\tablefoottext{*}{\textcolor{red}{references ?}}
\tablefoottext{j}{\citet{noel_dynamique_2003}}.
\end{table*}

\begin{table*}[]
  \begin{center} 
  \caption{Observing strategy for \textit{Herschel} spectroscopy.}
  \label{table_targets_observations}
  \small
   \begin{tabular}{cccccc}
 \hline\hline
 \multirow{2}{*}{Line} & \multirow{2}{*}{Target} & \multirow{2}{*}{Instrument} & \multirow{2}{*}{Observing mode} & \multirow{2}{*}{Band} & Beam size \\
  &  &  &  &  & (FWHM or spaxel witdth) \\
 \hline
 \multirow{3}{*}{\big[CII\big] 158 \textmu{}m}  & 105-600 & \multirow{3}{*}{HIFI - HRS} & Dual beam switch & \multirow{3}{*}{\shortstack{7b (1795.2 - 1902.8 GHz \\ or 158 - 177 \textmu{}m)}} & \multirow{3}{*}{11.3 \arcsec{}} \\
 & HST10 & & Load chop & & \\
 & 244-440 & & Load chop & & \\
 \hline
 \multirow{3}{*}{CO (7-6)}  & 105-600 & \multirow{3}{*}{HIFI - HRS} & Dual beam switch & \multirow{3}{*}{\shortstack{3a (807.1 - 851.9 GHz \\ or 352 - 372 \textmu{}m)}} & \multirow{3}{*}{28.4 \arcsec{}} \\
 & HST10 & & Frequency switch & & \\
 & 244-440 & & Frequency switch & & \\
 \hline
 \multirow{3}{*}{CO (10-9)}  & 105-600 & \multirow{3}{*}{HIFI - HRS} & \multirow{3}{*}{Frequency switch} & \multirow{3}{*}{\shortstack{5a (1116.2 - 1240.8 GHz \\ or 242 - 269 \textmu{}m)}} & \multirow{3}{*}{18.6 \arcsec{}} \\
 & HST10 & & & &\\
 & 244-440 & & & & \\
 \hline
 \multirow{3}{*}{CO (15-14), (17-16), (19-18)}  & 105-600 & \multirow{3}{*}{PACS} & Chopping nodding & \multirow{3}{*}{\shortstack{51-73 \textmu{}m and \\ 103-220 \textmu{}m}} & \multirow{3}{*}{9.4 \arcsec{}} \\
 & HST10 & & Unchopped grating scan & & \\
 & 244-440 & & Unchopped grating scan & & \\
 \hline
 \multirow{3}{*}{\big[OI\big] 63 \textmu{}m}  & 105-600 & \multirow{3}{*}{PACS} & Chopping nodding & \multirow{3}{*}{\shortstack{51-73 \textmu{}m and \\ 103-220 \textmu{}m}} & \multirow{3}{*}{9.4 \arcsec{}} \\
 & HST10 & & Unchopped grating scan & & \\
 & 244-440 & & Unchopped grating scan & & \\
 \hline
\end{tabular}

  \normalsize
  \end{center}
  ~
  \small 
  %\tablefoot{Beam sizes are given as FWHM for HIFI (circular beam) and as spaxel width for PACS (square beam).}
  %\tablefoot{More details about the observing strategy of this program (OT2\_oberne\_4) can be found in the HSPOT software. See the user manuals of the instruments for details about the observing modes.}
  %\tablefoottext{1}{HRS is the high resolution spectrometer of HIFI.} 
  \normalsize
\end{table*}

\subsubsection{Observing results}

 \begin{figure*}[]
     \begin{center}
        \subfigure[PACS spectroscopy - 105-600]{%
           \label{fig_herschelLines_PACS_105-600}
           \raisebox{+0.03\height}{\includegraphics[width=0.48\textwidth]{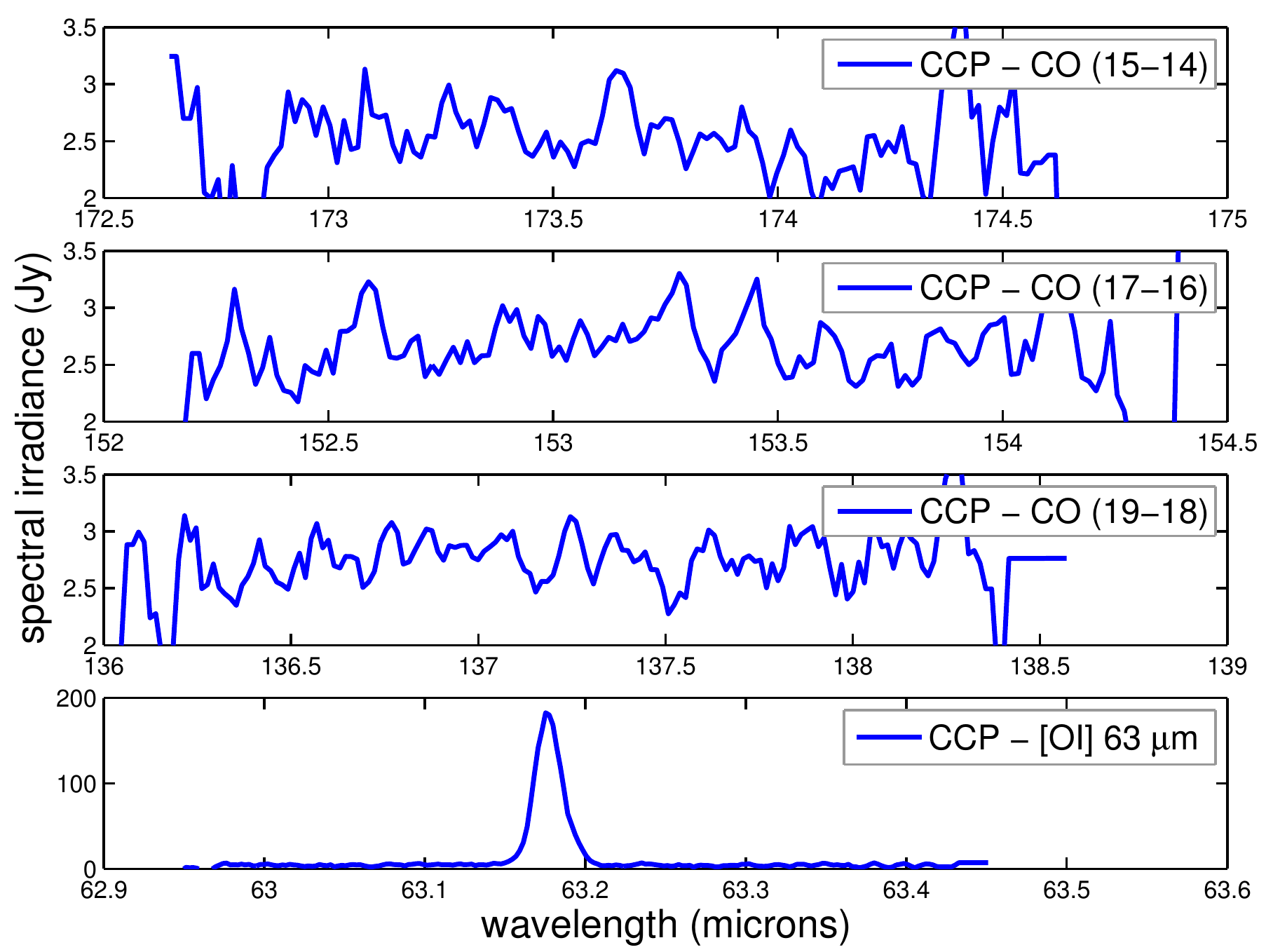}}
        } %
        \subfigure[HIFI spectroscopy - 105-600]{%
            \label{fig_herschelLines_HIFI_105-600}
            \includegraphics[width=0.48\textwidth]{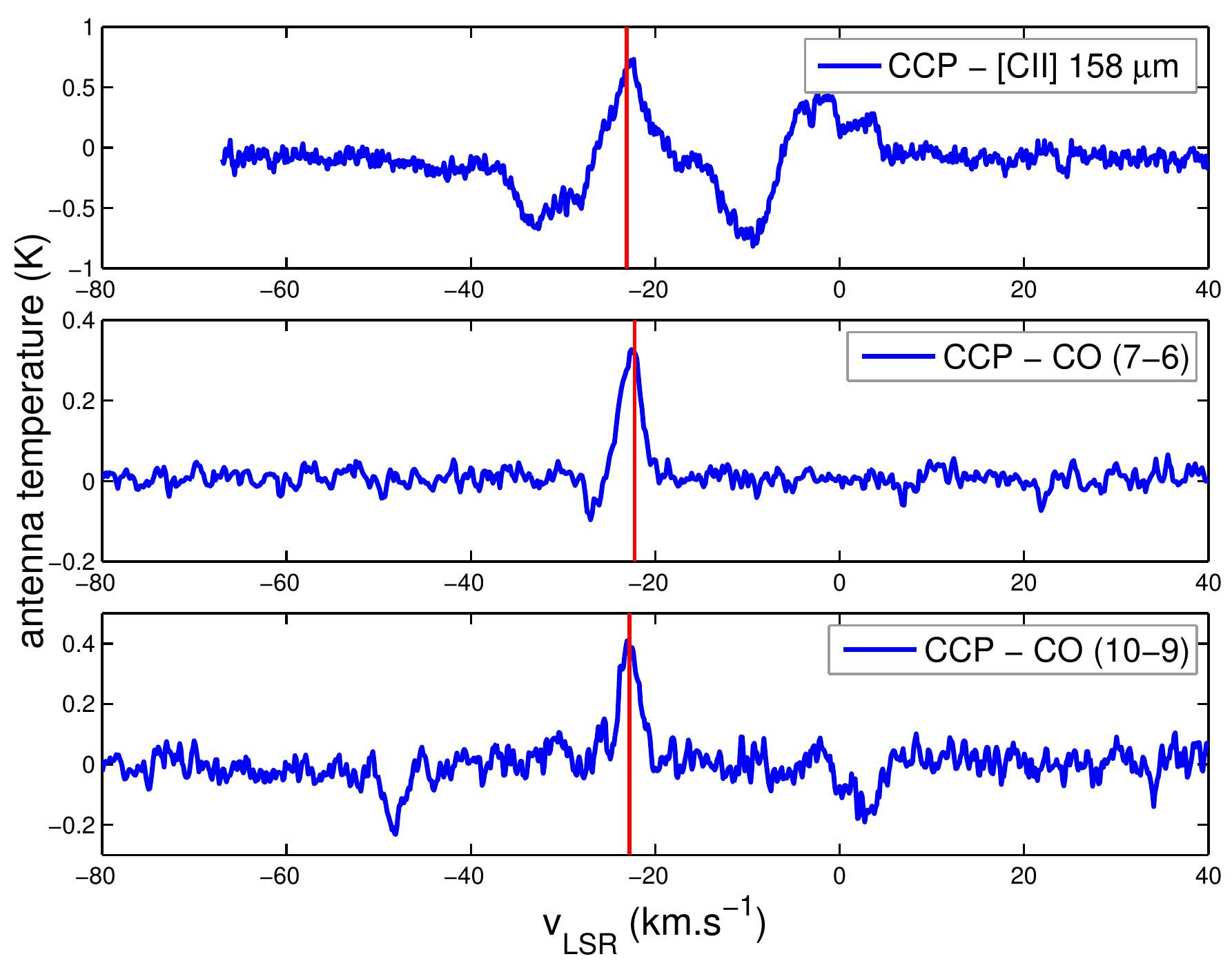}
        }%
        \\ %  ------- End of the first row ----------------------%
        \subfigure[PACS spectroscopy - HST10]{%
            \label{fig_herschelLines_PACS_HST10}
            \raisebox{+0.03\height}{\includegraphics[width=0.48\textwidth]{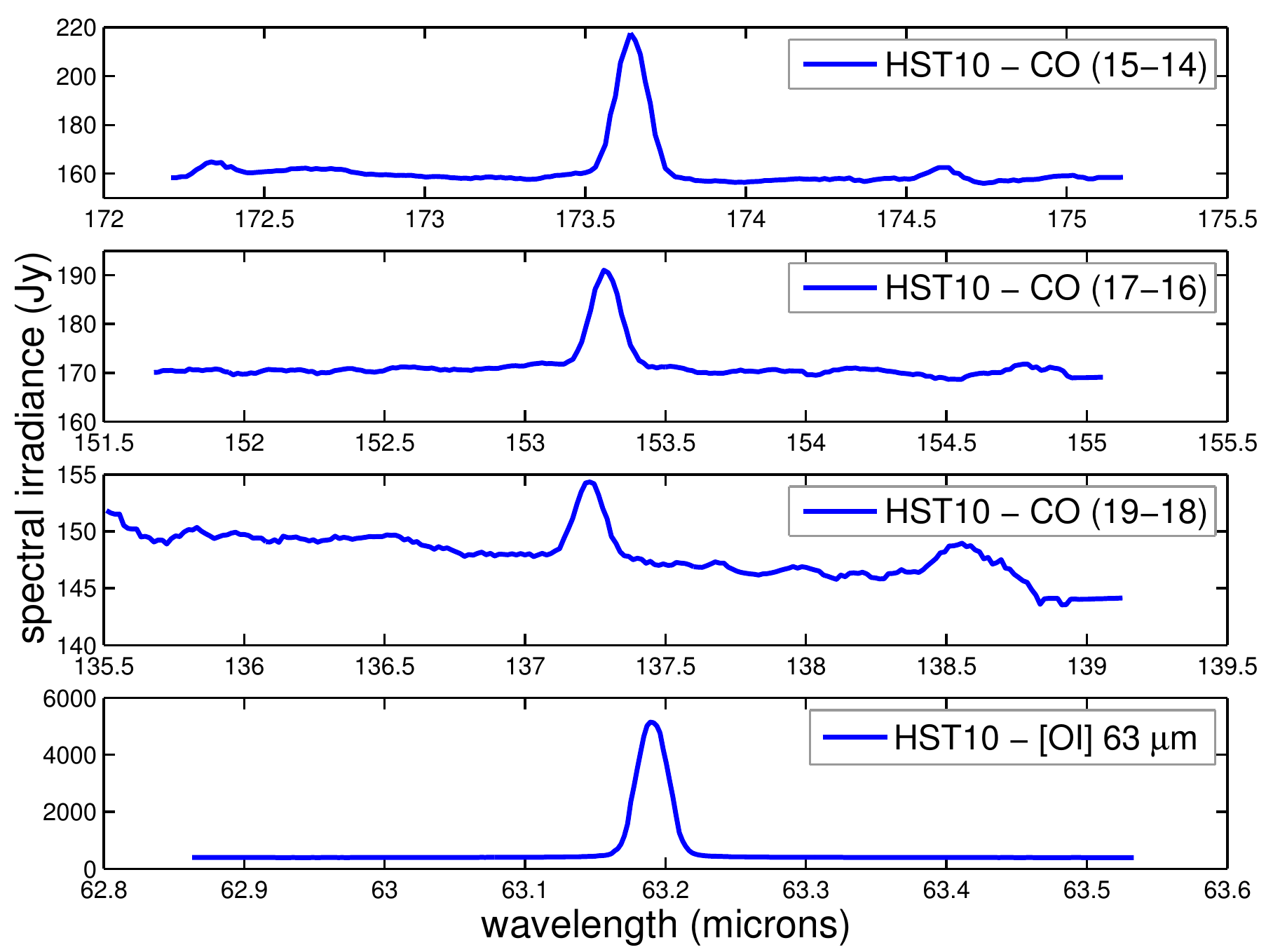}}
        }%
        \subfigure[HIFI spectroscopy - HST10]{%
            \label{fig_herschelLines_HIFI_HST10}
            \includegraphics[width=0.48\textwidth]{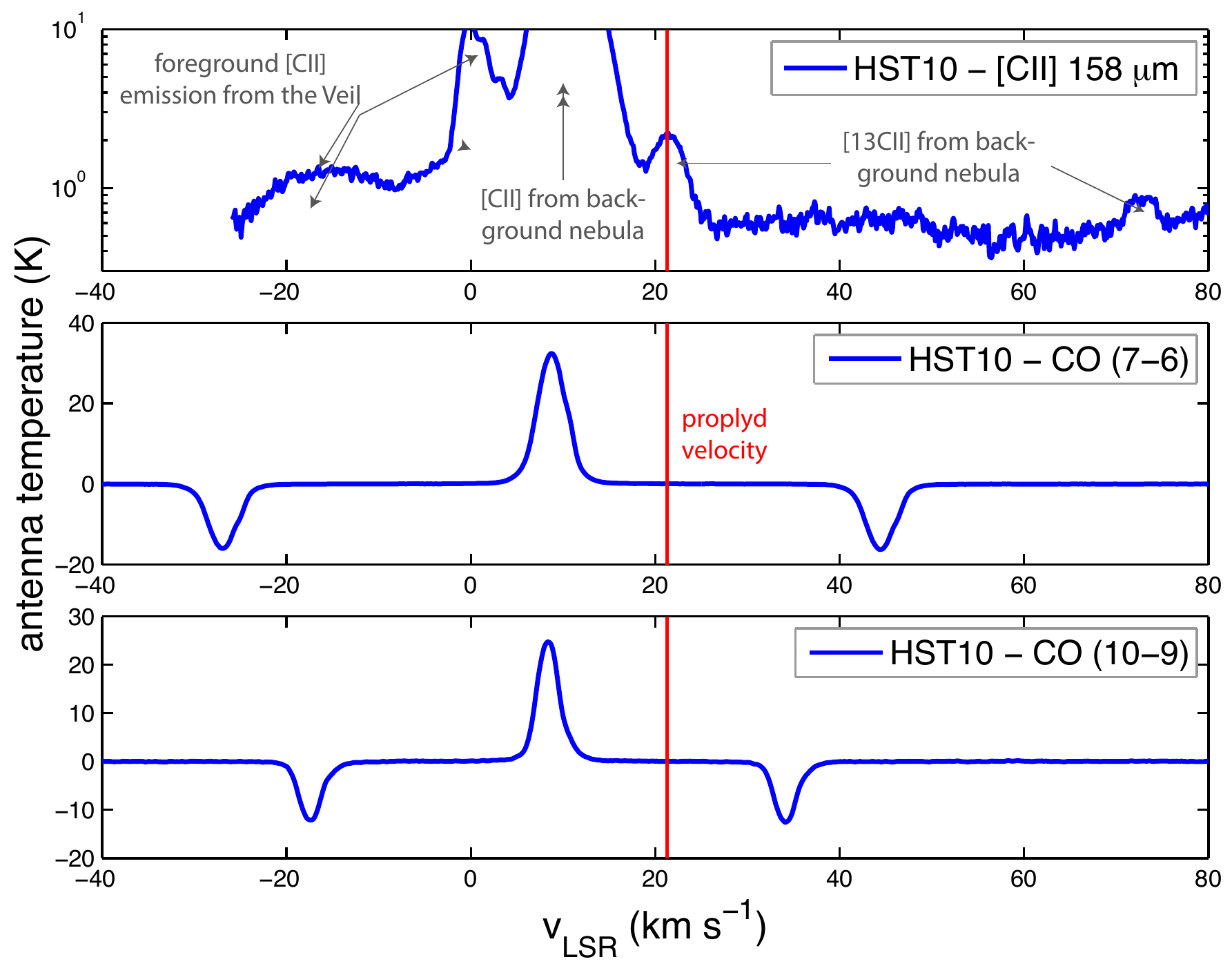}
        }%
        \\ %  ------- End of the second row ----------------------%
        \subfigure[PACS spectroscopy - 244-440]{%
            \label{fig_herschelLines_PACS_244-440}
            \raisebox{+0.03\height}{\includegraphics[width=0.48\textwidth]{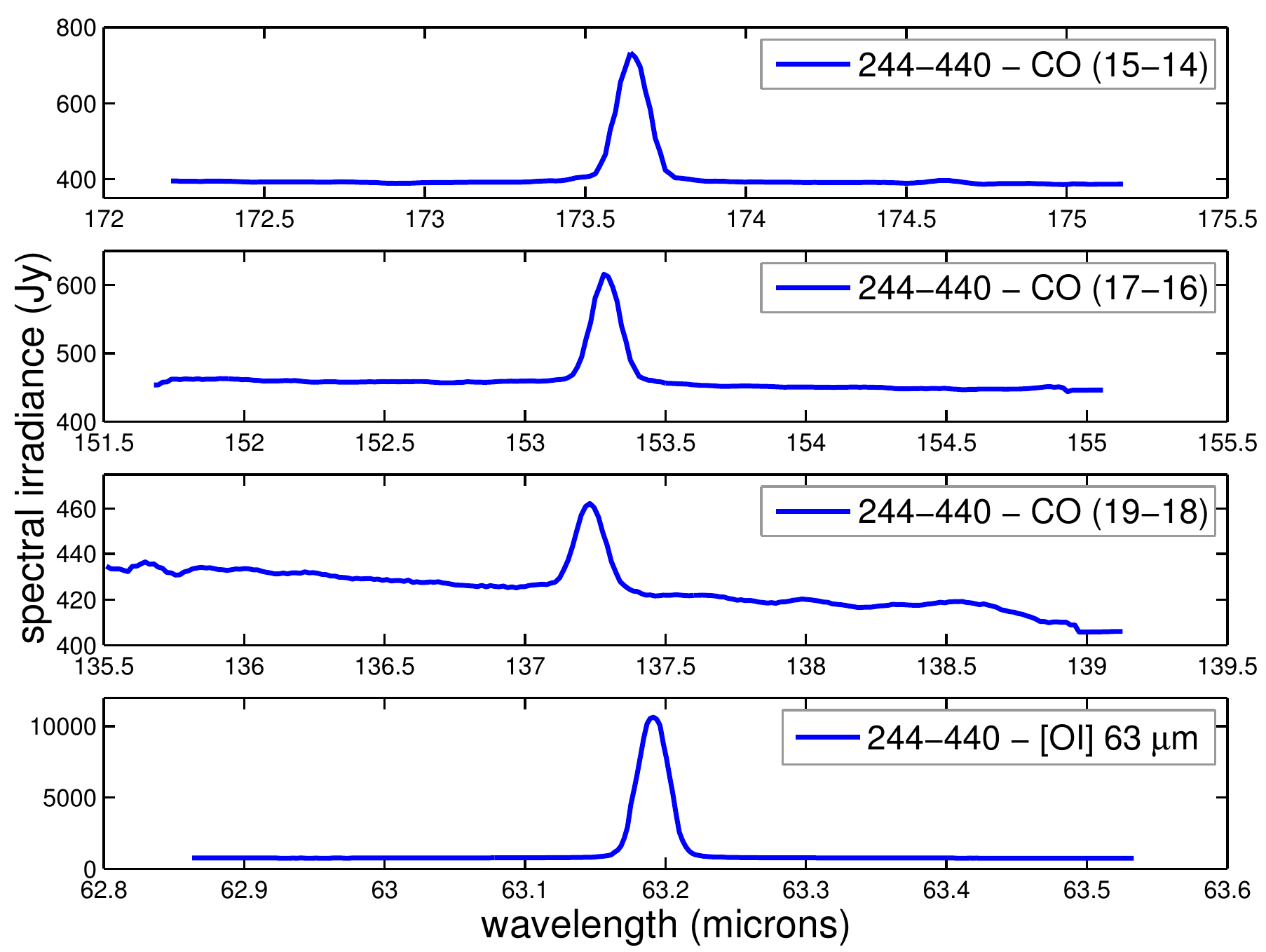}}
        }%
        \subfigure[HIFI spectroscopy - 244-440]{%
            \label{fig_herschelLines_HIFI_244-440}
            \includegraphics[width=0.48\textwidth]{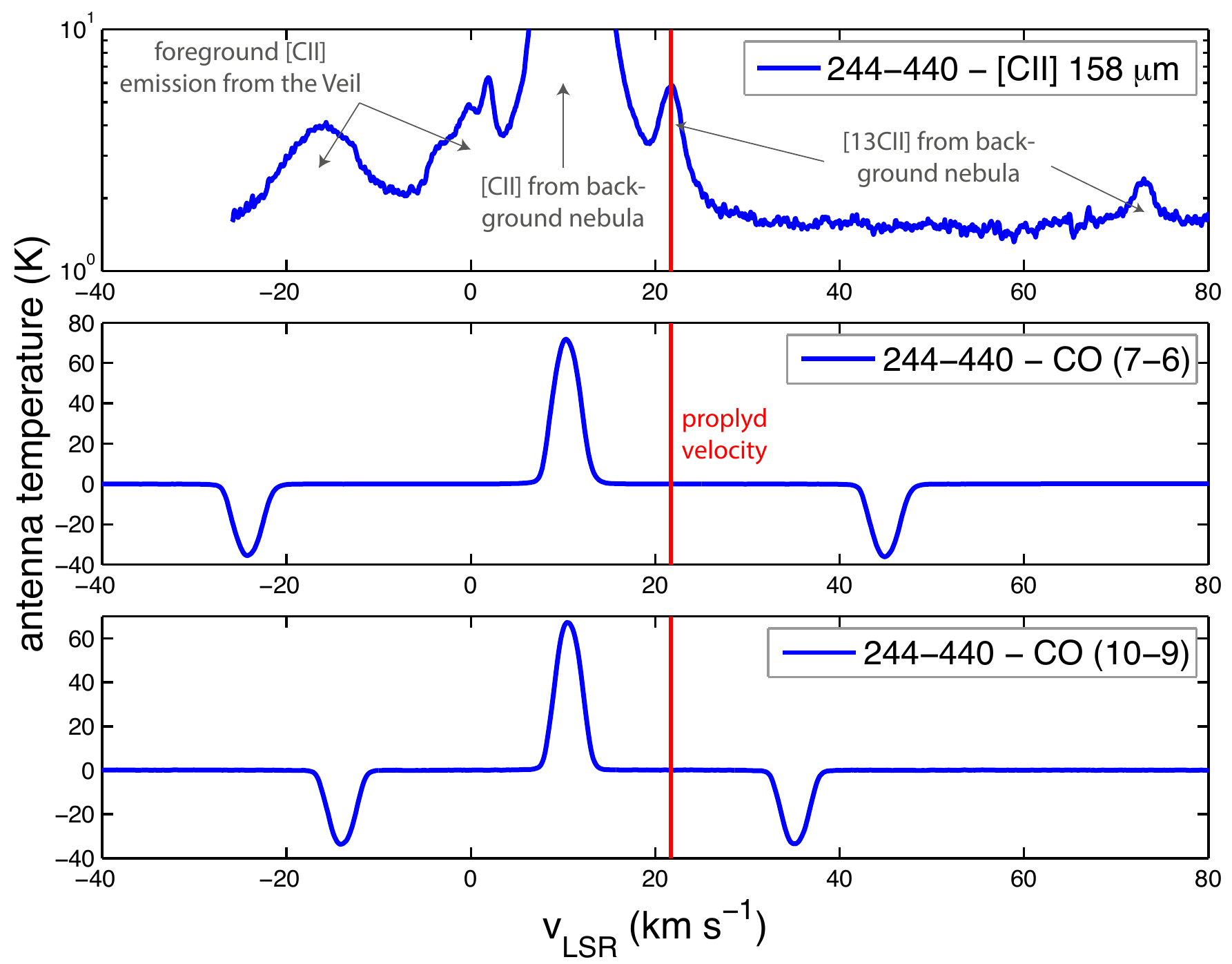}
        }%
    \end{center}
    \caption{Level-2 calibrated spectra for the three targets: 105-600, HST10 and 244-440 respectively from top to bottom. Spectrally unresolved lines observed with PACS are presented on the left and resolved lines observed with HIFI lied on the right. The red vertical bars on HIFI spectra indicate the v$_\mathrm{LSR}$ of the objects.}
   \label{fig_herschelLines}
\end{figure*}
 
\begin{table*}[]
  \begin{center} 
   \small
   \caption{Features of the observed lines (\textit{Herschel} plus published/archival data available).}
   \label{table_lines_features}
     \begin{tabular}{llcccccccr}
 \hline\hline
 \multirow{2}{*}{Target} & \multirow{2}{*}{Line} & Wavelength & \multirow{2}{*}{Instrument} & \multicolumn{3}{c}{Integrated intensity} & FWHM & $v_{\mathrm{LSR}}$ \\ & & (\textmu{}m) &  & (W m$^\mathrm{-2}$) & & (W m$^\mathrm{-2}$ sr$^\mathrm{-1}$) & (km s$^\mathrm{-1}$) & (km s$^\mathrm{-1}$)\\ 
\hline
 \hline
 \multirow{14}{*}{105-600} & \big[CII\big] 158 \textmu{}m & 157.74 & HIFI & ~~~2.39e-16~~~ &   (12\%) & ~~~3.68e-07\tablefootmark{c}~~~ & 6.4 & -23.1 \\ 
  & CO (7-6) & 371.65 & HIFI & ~~~1.18e-17~~~ &   (11\%) & ... & 2.7 & -22.7 \\ 
  & CO (10-9) & 260.24 & HIFI & ~~~2.26e-17~~~ &   (14\%) & ... & 2.5 & -22.8 \\ 
  & CO (15-14) & 173.63 & PACS & ~~~5.50e-18~~~ &   (24\%) & ... & ... & ... \\ 
  & CO (17-16) & 153.27 & PACS & ~~~7.59e-18~~~ &   (24\%) & ... & ... & ... \\ 
  & CO (19-18) & 137.20 & PACS & < 2.65e-17\tablefootmark{a}~ &   (20\%) & ... & ... & ... \\ 
  & \big[OI\big] 63 \textmu{}m & 63.18 & PACS & ~~~2.90e-15~~~ &   (12\%) & ~~~4.47e-06\tablefootmark{c}~~~ & ... & ... \\ 
  & CO (3-2) & 866.96 & APEX/FLASH+ & ~~~2.98e-18~~~ &  (20\%) & ... & 3.2 & -22.7 \\ 
  & CO (4-3) & 650.25 & APEX/FLASH+ & ~~~6.67e-18~~~ &  (20\%) & ... & 3.1 & -22.7 \\ 
  & CO (6-5) & 433.56 & APEX/CHAMP+ & ~~~1.26e-17~~~ &  (20\%) & ... & 3.5 & -22.6 \\ 
  & CO (7-6) & 371.65 & APEX/CHAMP+ & ~~~1.40e-17~~~ &  (20\%) & ... & 4.1 & -22.3 \\ 
  & \textsuperscript{13}CO (3-2) & 906.83 & APEX/FLASH+ & ~~~7.12e-19~~~ &  (20\%) & ... & 2.2 & -22.6 \\ 
  & HCO+ (4-3) & 840.41 & APEX/FLASH+ & ~~~2.97e-19~~~ &  (20\%) & ... & 2.0 & -22.6 \\ 
  & HCN (4-3) & 845.66 & APEX/FLASH+ & ~~~1.72e-19~~~ &  (20\%) & ... & 1.8 & -22.8 \\ 
 \hline
 \multirow{9}{*}{HST10} & \big[CII\big] 158 \textmu{}m & 157.74 & HIFI & < 2.27e-16\tablefootmark{b}~ &   (12\%) & ~~~4.72e-06\tablefootmark{b,c}~ & 3.9 & 21.3 \\ 
  & CO (7-6) & 371.65 & HIFI & < 4.32e-18\tablefootmark{a}~ &   (11\%) & ~~~5.84e-06\tablefootmark{a,c}~ & ... & ... \\ 
  & CO (10-9) & 260.24 & HIFI & < 2.63e-17\tablefootmark{a}~ &   (14\%) & ~~~3.56e-05\tablefootmark{a,c}~ & ... & ... \\ 
  & CO (15-14) & 173.63 & PACS & < 7.15e-16\tablefootmark{b}~ &   (13\%) & ~~~9.68e-04\tablefootmark{b,c}~ & ... & ... \\ 
  & CO (17-16) & 153.27 & PACS & < 3.47e-16\tablefootmark{b}~ &   (13\%) & ~~~4.70e-04\tablefootmark{b,c}~ & ... & ... \\ 
  & CO (19-18) & 137.20 & PACS & < 1.44e-16\tablefootmark{b}~ &   (13\%) & ~~~1.95e-04\tablefootmark{b,c}~ & ... & ... \\ 
  & \big[OI\big] 63 \textmu{}m & 63.18 & PACS & < 1.01e-13\tablefootmark{b}~ &   (12\%) & ~~~2.10e-03\tablefootmark{b,c}~ & ... & ... \\ 
  & H2 1-0 S(1) & 2.12 & HST/NICMOS & ~~~1.85e-18\tablefootmark{c}~ &  (20\%) & 2.50e-06~~ & ... & ... \\ 
  & [OI] 6300\AA{} & 0.63 & HST/WFPC2 & ~~~1.11e-17\tablefootmark{c}~ &  (34\%) & 1.50e-05~~ & ... & ... \\ 
 \hline
 \multirow{8}{*}{244-440} & \big[CII\big] 158 \textmu{}m & 157.74 & HIFI & < 4.68e-16\tablefootmark{b}~ &   (12\%) & ~~~8.09e-07\tablefootmark{b,c}~ & 2.2 & 21.7 \\ 
  & CO (7-6) & 371.65 & HIFI & < 2.69e-18\tablefootmark{a}~ &   (11\%) & ~~~2.46e-07\tablefootmark{a,c}~ & ... & ... \\ 
  & CO (10-9) & 260.24 & HIFI & < 1.53e-17\tablefootmark{a}~ &   (14\%) & ~~~1.40e-06\tablefootmark{a,c}~ & ... & ... \\ 
  & CO (15-14) & 173.63 & PACS & < 4.19e-15\tablefootmark{b}~ &   (13\%) & ~~~3.83e-04\tablefootmark{b,c}~ & ... & ... \\ 
  & CO (17-16) & 153.27 & PACS & < 2.72e-15\tablefootmark{b}~ &   (13\%) & ~~~2.49e-04\tablefootmark{b,c}~ & ... & ... \\ 
  & CO (19-18) & 137.20 & PACS & < 8.16e-16\tablefootmark{b}~ &   (13\%) & ~~~7.45e-05\tablefootmark{b,c}~ & ... & ... \\ 
  & \big[OI\big] 63 \textmu{}m & 63.18 & PACS & < 2.06e-13\tablefootmark{b}~ &   (12\%) & ~~~3.55e-04\tablefootmark{b,c}~ & ... & ... \\ 
  & OH 84 \textmu{}m & 84.45 & PACS & ~~~6.83e-18~~~ &  (13\%) & ~6.24e-07\tablefootmark{c}~ & ... & ... \\ 
 \hline
 \multirow{3}{*}{203-506} & HCO+ (4-3) & 840.41 & ALMA & ~~~3.04e-20\tablefootmark{c}~ &  (10\%) & 3.60e-09~~ & 4.0 & 9.5 \\ 
  & H2 1-0 S(1) & 2.12 & CFHT & ~~~7.01e-18\tablefootmark{c}~ &  (11\%) & 8.30e-07~~ & ... & ... \\ 
  & [OI] 6300\AA{} & 0.63 & HST/WFPC2 - MUSE & ~~~1.01e-16\tablefootmark{c}~ & ... & 1.20e-05~~ & ... & ... \\ 
 \hline
\end{tabular}
   \normalsize
  \end{center}
\tablefoot{1\textsigma{}-relative uncertainties of integrated intensities are given in parentheses after each value and take into account instrumental accuracy and fitting uncertainties. Other uncertainties are not indicated, they are below 5\% for the full width at half maximum (FWHM) and are estimated to be smaller than a percent for the velocity with respect to the local standard of rest ($v_\mathrm{LSR}$). \\
  \tablefoottext{a}{Upper limit determined by the noise (peak of the line supposed to be smaller than three times the noise level).}  \\
  \tablefoottext{b}{Upper limit determined by the observed background emission.} \\
  \tablefoottext{c}{Values converted using the apparent size given in Table \ref{table_targets_properties} and assuming that all the lines are coming from the disk except for \big[CII\big] 158 \textmu{}m and \big[OI\big] 63 \textmu{}m lines that are supposed to originate in the envelope.} }
\end{table*} 

Fig. \ref{fig_herschelLines} shows the spectra extracted from the \textit{Herschel} data. A summary of all detected lines and upper limits is given in Table \ref{table_lines_features}. 

For the Carina candidate proplyd, 105-600, PACS spectra are not contaminated by the surrounding nebula so that the target emission is directly observed (Fig. \ref{fig_herschelLines_PACS_105-600}). The \big[OI\big] 63 \textmu{}m line is clearly visible while the sensitivity of the instrument allows the detection, at the limit, of the CO (15-14) and CO (17-16) lines (with peaks about 3 - 4\textsigma) but is insufficient to detect the CO (19-18) line. Estimation of the integrated intensities, or their upper limits, can be obtained with reasonable uncertainties (about 24\%, see Table \ref{table_lines_features}) since the theoretical FWHM of the lines, related to instrumental characteristics, are known. Emission lines observed by HIFI are well spectrally resolved and with high signal-to-noise ratios (Fig. \ref{fig_herschelLines_HIFI_105-600}). However, the observing mode chosen for the \big[CII\big] 158 \textmu{}m line has caused a contamination from the background emission of the HII region. The strategy to remove this contamination and extract the \big[CII\big] 158 \textmu{}m line flux is detailed in Appendix \ref{app_CII_CPP}.
 
In the case of the two Orion proplyds observed with \textit{Herschel} (HST10 and 244-440), the low-resolution spectroscopy obtained by PACS (Figs. \ref{fig_herschelLines_PACS_HST10} and \ref{fig_herschelLines_PACS_244-440}) is highly contaminated by the nebula so that the observed emissions mainly come from the background and not from the targets. Only upper limits of the target emission can thus be extracted for the CO (15-14), CO (17-16), CO (19-18) and \big[OI\big] 63 \textmu{}m lines. The high-resolution spectroscopy provided by HIFI enables to spectrally resolve the emission of the Orion proplyds and the emission from the nebula. This is clearly visible with the \big[CII\big] 158 \textmu{}m fine-structure line, on Figs. \ref{fig_herschelLines_HIFI_HST10} and \ref{fig_herschelLines_HIFI_244-440}, where the targets emission lines are detected at a $v_\mathrm{LSR}$ about 21 - 22 km\,s$^\mathrm{-1}$ and located on the wing of the nebula emission line that peaks close to a $v_\mathrm{LSR}$ of 10 km\,s$^\mathrm{-1}$. Unfortunately, the velocities of the Orion proplyds shift the \big[CII\big] lines at the same frequency of the strongest [$^\mathrm{13}$CII] line from the UV-illuminated surface of Orion molecular could \citep{goicoechea_velocity-resolved_2015}. Here again, only upper limits can thus be extracted. For HST10, the maximum brightness of the \big[CII\big] 158 \textmu{}m line, after correcting the beam dilution, is 4.72 $\times$ 10$^\mathrm{-6}$ W\,m$^\mathrm{-2}$\,sr$^\mathrm{-1}$ which is the same order of magnitude as the one predicted by \citet{storzer_photodissociation_1999} who probably overestimated it.
At low velocities, several emission lines are observed: three components close to a $v_\mathrm{LSR}$ of 0 km\,s$^\mathrm{-1}$ and one located about -15 km\,s$^\mathrm{-1}$. Those lines likely correspond to the emission of the $^\mathrm{12}$C$^+$ from the Veil in front of the nebula seen in HI emission and absorption at those velocities \citep{van_der_werf_tearing_2013}. The line at a $v_\mathrm{LSR}$ about 70-75 km\,s$^\mathrm{-1}$ is another $^\mathrm{13}$C$^+$ line from the nebula \citep{ossenkopf_herschel/hifi_2013}. Even with the high spectral resolution of HIFI, no emission from HST10 and 244-440 is visible for the CO (7-6) and CO (10-9) transitions. The peak in the corresponding spectra is the emission from the nebula and the negative signatures are ghosts related to the observing mode (frequency switch, see Table \ref{table_targets_observations}). Constraining upper limits of the emission from the targets can still be derived (see Table \ref{table_lines_features}) assuming that the width of the lines should be smaller than the one detected from the $^\mathrm{12}$C$^+$ emission and that the peak of the lines is below the detection threshold, i.e. three times the noise level.
 
Level-2 calibrated spectra from the PACS line spectroscopy also allow to extract the continuum emission which is underlying the line (see Fig. \ref{fig_herschelLines}). The continuum extraction was done for the Carina candidate proplyd (105-600) only because the far-infrared signal from the two Orion proplyds is drowned in the emission of the nebula. In a five-by-five spaxels image of PACS, the emission from the target is estimated by extracting the flux in the central spaxel, correcting it for losses to the neighbouring spaxels because of the point spread function, and subtracting the mean observed flux in the 16 external spaxels. The obtained values are presented in Table \ref{table_CCP_continuum}. 

 \begin{table}[]
  \begin{center} 
    \small
    \caption{Continuum emission of the Carina candidate proplyd 105-600.}
    \label{table_CCP_continuum}
      \begin{tabular}{ccr@{~}lcc}
 \hline\hline
 Wavelength & \multirow{2}{*}{Instruments} & \multicolumn{2}{c}{Continuum} & AOR\tablefootmark{a} or \\
 (\textmu{}m) & & \multicolumn{2}{c}{emission (mJy)}& reference \\
 \hline
 3.55 & IRAC & 3.5~ & (11 \%) & 12914432 \\ 
 4.49 & IRAC & 3.5~ & (11 \%) & 12914432 \\ 
 5.73 & IRAC & 13.9~ & (11 \%) & 12914432 \\ 
 7.87 & IRAC & 28.3~ & (11 \%) & 12914432 \\ 
 24.00 & MIPS & \multicolumn{2}{c}{ < 1144.2\tablefootmark{b}} & 23788800 \\ 
 57.95 & PACS & 1153.5~ & (36 \%) & OT2\_oberne\_4 \\ 
 63.22 & PACS & 839.4~ & (45 \%) & OT2\_oberne\_4 \\ 
 68.70 & PACS & 1013.1~ & (25 \%) & OT2\_oberne\_4 \\ 
 70.00 & PACS & 1483.0~ & (11 \%) & 1342255062 \\ 
 137.25 & PACS & 1086.4~ & (15 \%) & OT2\_oberne\_4 \\ 
 153.19 & PACS & 1144.2~ & (16 \%) & OT2\_oberne\_4 \\ 
 160.00 & PACS & 1035.0~ & (11 \%) & 1342255062 \\ 
 173.75 & PACS & 1122.7~ & (21 \%) & OT2\_oberne\_4 \\ 
 189.50 & PACS & 1007.7~ & (43 \%) & OT2\_oberne\_4 \\ 
 250.00 & SPIRE & 50.3\tablefootmark{b} & (14 \%) & 1342255061 \\ 
 350.00 & SPIRE & 10.9\tablefootmark{b} & (36 \%) & 1342255061 \\ 
 350.00 & LABOCA & 196.0~ & (34 \%) & \cite{sahai_are_2012} \\ 
 870.00 & SABOCA & \multicolumn{2}{c}{ < 40.0~} & \cite{sahai_are_2012} \\ 
 \hline
\end{tabular}
    \normalsize
  \end{center}
\tablefoot{1\textsigma{}-relative uncertainties of the continuum emission are given inside parentheses after each value, in case of detection, and take into account instrumental accuracy and post-processing uncertainties. For the MIPS data, upper limit corresponds to the maximal signal detected plus three times the noise level. \\
 \tablefoottext{a}{Astronomical observation request.} \\
 \tablefoottext{b}{Values in MJy/sr.} } 
\end{table}

%		 			2.3 Complementary data				 	  	  %
%--------------------------------------------------------------------------------------------------------%

%\subsection{Complementary data} \label{subSect_obsComplData}

%\subsubsection{Proplyds of the \textit{Herschel} survey}
\subsubsection{Complementary data} \label{subSect_obsComplData}

For the Carina candidate proplyd 105-600, the dataset is complemented with optical to submillimeter data from other instruments. From the online archives, we retrieved the available data of the \textit{Spitzer}/IRAC and MIPS (PI: N. Smith, program-ID: 3420, 30848, \citealt{smith_spitzer_2010}) instruments, plus the photometry obtained by \textit{Herschel}/PACS and SPIRE (PI: S. Molinari, program-ID: OT2\_smolinar\_7). From this data, we have extracted the continuum emission by direct measurement of the spectral radiance or by aperture photometry (Table \ref{table_CCP_continuum}).  We also retrieved the data of the Atacama pathfinder experiment (APEX) 12-m telescope from \citet{sahai_are_2012} who used the LABOCA and SABOCA bolometers (beam size of respectively 7.8\arcsec{} and 19\arcsec{}) to measure continuum emission and CHAMP+ and FLASH+ spectrometers (beam sizes from 7.7\arcsec{} to 18.6\arcsec{}) to observe molecular lines. Finally, we used the optical HST/ACS H\textalpha{} image (PI: N. Smith, program-ID: 10475, \citealt{smith_HST/ACS_2010}). Performing photometry at the ionisation front, we obtained a brightness (including H\textalpha{} and \big[NII\big] 6583 \AA{} lines) of $4 \times 10^{-7}$ W\,m$^\mathrm{-2}$\,sr$^\mathrm{-1}$ which has been used to estimate the electron density at the ionisation front of $n_\mathrm{e} \leq 680$ cm$^\mathrm{-3}$. 

For the  proplyd 244-440, we also included the OH 84 \textmu{}m line detected in a PACS survey of the Orion Bar and described in \citet{parikka_spatial_2016}. %\citet{parikka_properties_2015}.
%HST10 has been observed in the H$_\mathrm{2}$ 1-0 S(1) line by \citet{chen_2.12_1998}, and we use their value for the intensity of this line. The [OI] $6300\AA{}$ line
%was observed with HST and MUSE. The HST data was published in \citet{bally_externally_1998} and we use the value given in their table 4. The optical [OI] line was also observed with MUSE on the VLT \citep{weilbacher_muse_2015}, with a lower angular resolution than HST. We extract the intensity from the publicly released data. The intensity derived from the MUSE data is a factor of $\sim$ 3 lower than the one derived from HST. We attribute this discrepancy to dilution of unresolved emission structures in the MUSE beam. The [OI] line is subject to extinction on the line of sight. A possibility to correct this effect (however it is not
%obvious that this is applicable to proplyds in Orion) is to use the extinction map of the Orion nebula derived from the MUSE data by \citep{weilbacher_muse_2015}.  
%Overall, due to the different instruments used and possible effects of extinction, we derive a range of possible values of the [OI] $6300\AA{}$ (Table \ref{table_lines_features}).
HST10 has been observed in the H$_\mathrm{2}$ 1-0 S(1) line by \citet{chen_2.12_1998}, and we use their value for the intensity of this line. The \big[OI\big] $6300$ \AA{} line
was observed with HST and we use the value given in the table 4 of \citet{bally_externally_1998}. 

\subsection{ALMA detection of the 203-506 proplyd}

%Three proplyds were selected for the \textit{Herschel} survey, one candidate to be a giant proplyd in the Carina nebula and two confirmed proplyds in the Orion nebula cluster. 
Recently, coauthors of this paper have serendipitously detected intense HCO$^+$ (4-3) emission towards the 203-506 proplyd with ALMA. This detection was quite surprising since only a few, and generally very massive, proplyd silhouette disks have been detected with ALMA in HCO$^+$ and reported in the literature \citep{williams_ALMA_2014}. %since other proplyds are not detected with ALMA in HCO$^+$ \citep{bally_ALMA_2015}. 
Since HCO$^+$ is a PDR tracer, and given that other PDR tracers have been detected towards 203-506 (see below), we have decided to include it in our study. It also complements our sample quite well since it is a silhouette proplyd \citep{bally_disks_2000}, i.e. seen against the bright background in visible. The disk is edge-on and situated just south of the Orion Bar. %It is a nearly edge-on disk located just south of the bright Orion bar and seen in silhouette against the background nebular light.
Its surface is bright in \big[OI\big] 6300 \AA{} and a faint \big[OI\big] jet is also visible \citep{bally_disks_2000}. It does not exhibit any visible envelope as the others probably because it is lying in the neutral PDR resulting from the photodissociation of the Orion bar. The fact that the ionisation front of the M42 HII region has not yet reached 203-506 suggests that this disk has emerged out of the molecular cloud only very recently and hence that it has been exposed to UV irradiation for a relatively short time compared to classical proplyds. The 203-506 proplyd was detected as part of the ALMA cycle 1 project dedicated to the observation of the Orion bar (PI: J. Goicoechea). The observations were conducted in Band 7 with an angular resolution of 1\arcsec{} insufficient to spatially resolve the disk. The detailed analysis and data reduction are presented in \citet{goicoechea_compression_2016}. 203-506 was clearly detected in the HCO$^+$ (4-3) line. For other tracers with lower critical densities such as low-$J$ CO lines in Band 7 of ALMA, the emission is dominated by the Orion molecular cloud and hence the disk cannot be detected. 
The derived integrated intensity for the HCO$^+$ (4-3) line is presented in Table \ref{table_lines_features}. Emission from the H$_\mathrm{2}$ 1-0 S(1) line was also observed  with the BEAR instrument \citep{noel_new_2005} at the Canada France Hawaii Telescope. This data is presented in \citet{noel_dynamique_2003} and we use their intensity. The  \big[OI\big] $6300$ \AA{} line was also observed by HST. The line integrated intensity was extracted from the HST image and corrected from extinction ($A_\mathrm{V} = 1.26$) using the extinction map of the Orion nebula derived from the MUSE data by \citet{weilbacher_muse_2015}. The intensity is reported in Table \ref{table_lines_features}.

\section{1D-model of the PDR in an externally-illuminated protoplanetary disk} \label{sect_proplydModel}

%In this section we present a model to predict the gas and dust emission from the PDR of an externally 
%illuminated protoplanetary disk, i.e. the cocoon and the surface layer of the disk.
Since their discovery, proplyds have been extensively studied with the help of PDR models in order to determine their physical properties \citep[e.g.][]{johnstone_photoevaporation_1998, storzer_photodissociation_1999, richling_photoevaporation_2000, adams_photoevaporation_2004, clarke_photoevaporation_2007, walsh_molecular_2013}.
This approach is well-suited when trying to explain the FIR lines tracing the warm gas present at the surface layer of the disk and in the envelope. In this section we describe our modelling methodology.
%In this section we analyse the gas and dust emission using state of the art models. %The proplyds is described as a PDR consisting of an envelope and the surface layer of the disk.

%					 4.1 Parameters of the model					  %
%--------------------------------------------------------------------------------------------------------%

\subsection{Proplyds as PDRs}\label{subSect_modelParam}

\begin{figure}[]
  \begin{center}
    \includegraphics[width=0.45\textwidth]{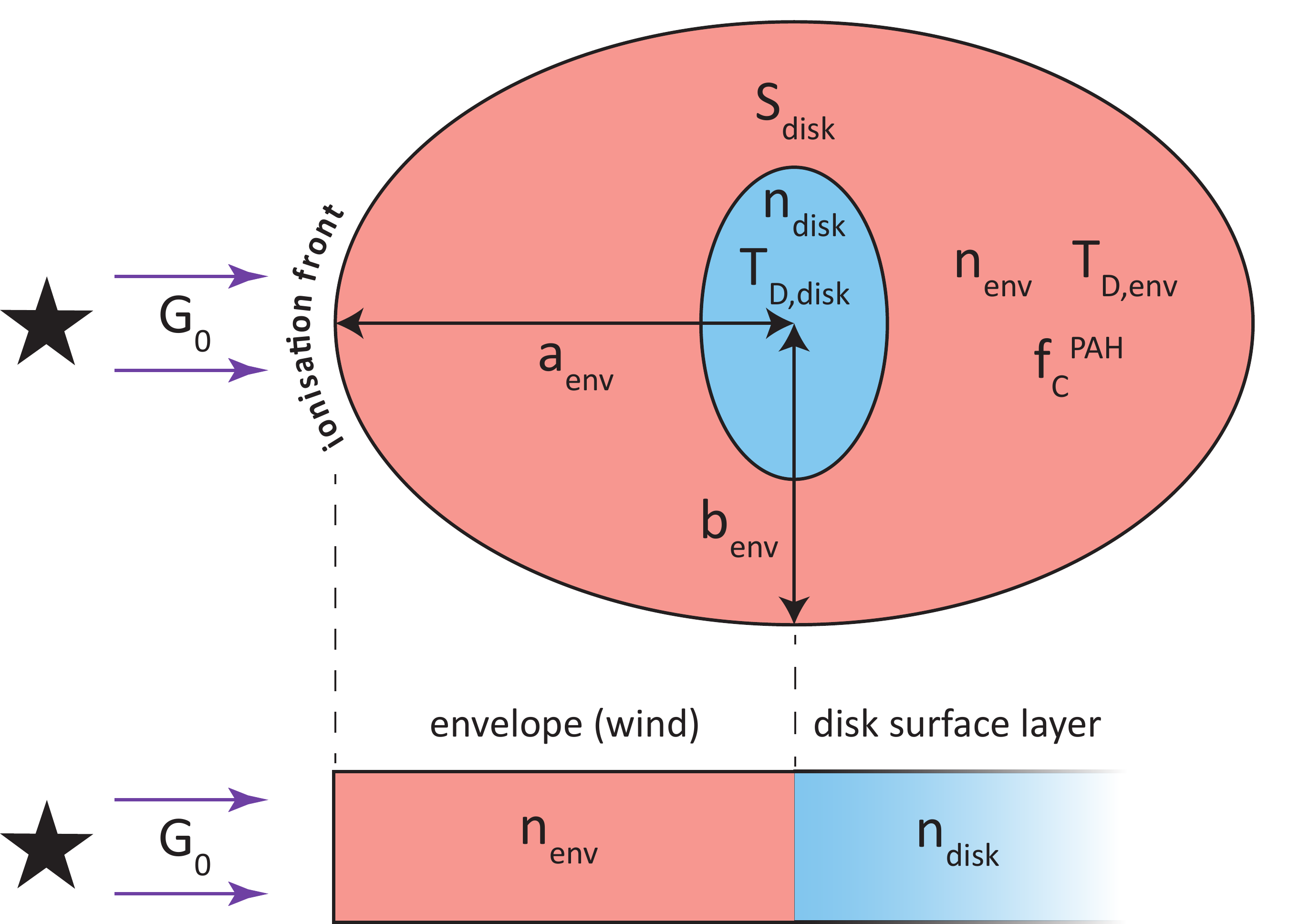}
   \end{center}
   \caption{Top: schematics of the apparent 2D geometry with parameters used in our adapted model of an externally-illuminated protoplanetary disk composed of a disk (blue) surrounded by an envelope (red). Bottom: 1D corresponding slab used to model the PDR of the proplyd.}
    \label{fig_model_geometry}
\end{figure}

Observations (Sect. \ref{sect_obs}) and preliminary modelling we performed showed that two different components, with different densities, are necessary to describe a proplyd and correspond very likely to the two distinct regions of the PDR, respectively the disk surface layer and the envelope. Given that, the geometry and parameters that we finally used are summarised in Fig. \ref{fig_model_geometry} and Table \ref{table_model_param}. We consider that a proplyd consists of an envelope (fed by the photoevaporation of the disk) with the apparent shape of an ellipse with semiaxes $a_\mathrm{env}$, $b_\mathrm{env}$, and defined by a mean density $n_\mathrm{env}$, 
a mean dust temperature $T_\mathrm{D,env}$ and $f\mathrm{_C^{PAH}}$ the fraction of elemental carbon in the gas phase locked in Polycyclic Aromatic Hydrocarbons (PAHs).
 At the center of the envelope lies the disk with an apparent surface $S_\mathrm{disk}$,
a mean dust temperature $T_\mathrm{D,disk} < T_\mathrm{D,env}$ and a mean density $n_\mathrm{disk}$ in the surface layer. Note that we are not modelling the properties of the bulk of the disk since we limit ourselves to the PDR which corresponds to the external layers of the proplyd (envelope and disk surface). Finally, the object is irradiated by FUV photons along the major-axis. The corresponding FUV radiation field $G_\mathrm{0}$ is expressed in units of the Habing field. %(estimated average FUV flux in the local interstellar medium). 
These parameters are used to predict both the dust spectral energy distribution (Sect. \ref{subSect_modelDust}), and the gas emission (Sect. \ref{subSect_modelPDR}). 

\begin{table*}[!ht]
  \begin{center} 
    \small
    \caption{Parameters of the models for the four studied proplyds.}
    \label{table_model_param}
      \begin{tabular}{lccccc}
 \hline\hline
 Symbol & Description & 105-600 & HST10 & 244-440 & 203-506 \\
 \hline
$a_\mathrm{env} $  						& Envelope  semi-major axis 					& $4.75\arcsec$ 								& $1.3\arcsec$ 									& $2.8\arcsec$ 					  				& ... \\
$b_\mathrm{env}$						& Envelope  semi-minor axis 					& $1.85\arcsec$ 								& $0.5\arcsec$ 									& $2.8\arcsec$ 									& ... \\
$n_\mathrm{env}$ 						& Envelope density 						& $8.5\times10^3$ cm$^\mathrm{-3}$\tablefootmark{a} 	& $2.5 \times10^5$ cm$^\mathrm{-3}$\tablefootmark{a}  	& $1\times10^5$ cm$^\mathrm{-3}$\tablefootmark{a}	& $4.5 \times10^4$ cm$^\mathrm{-3}$\tablefootmark{a} \\
$T_\mathrm{D,env}$ 					& Mean dust temperature in the envelope 		& $39.8$ K\tablefootmark{a} 						& ... 											& ... 											& ... \\
$f\mathrm{_C^{PAH}}$ 					& Fraction of gas phase C locked in PAHs 	& $0.24$ \%\tablefootmark{a} 						& $0.08$ \%\tablefootmark{b} 						& ... 											& ... \\
$S_\mathrm{disk}$ 						& Apparent surface of the disk 					& $3.82$ arcsec$^2$\tablefootmark{a} 				& $0.0314$ arcsec$^2$ 							& $0.47$ arcsec$^2$ 							& $0.36$ arcsec$^2$ \\
$n_\mathrm{disk}$ 						& Mean disk surface layer density 			& $7\times10^5$ cm$^\mathrm{-3}$\tablefootmark{a} 	& $4\times10^6$ cm$^\mathrm{-3}$\tablefootmark{a} 	& $2\times10^6$ cm$^\mathrm{-3}$\tablefootmark{a} 	& $1\times10^6$ cm$^\mathrm{-3}$\tablefootmark{a} \\
$T_\mathrm{D,disk}  (< T_\mathrm{D,env})$ 	& Mean dust temperature in the disk 			& $19.5$ K\tablefootmark{a} 						& ... 											& ... 											& ... \\
$G_\mathrm{0}$						& Radiation field 						& $2.2\times10^4$ 								& $2.4 \times 10^5$								& $1.0 \times 10^5$								& $2.0 \times 10^4$ \\
 \hline
\end{tabular}

    \normalsize
  \end{center} 
  ~ \\
  \noindent \tablefoottext{a}{Values obtained from the best-fit model.} \\
  \tablefoottext{b}{Value derived by \citet{vicente_polycyclic_2013}.}
 \end{table*}
 
 %Some of them are known (e.g. size or FUV field, see Table \ref{table_targets_properties}) while others will be determined by comparison between models and observations.

%			      4.2 Spectral energy distribution of the dust			  	  %
%--------------------------------------------------------------------------------------------------------%

\subsection{Dust spectral energy distribution} \label{subSect_modelDust}

Interstellar dust is mainly composed of small grains \citep[10 nm to 0.1 \textmu{}m, see e.g.][]{compiegne_global_2011} of silicates and carbonaceous compounds. 
Disks form from this medium and evolve with time in a way that their grains get larger and their dust-to-gas mass ratio increases \citep[see e.g. the review of][]{williams_protoplanetary_2011}. The dust properties in the photoevaporation flow (envelope) are expected to differ from that of the disk because of a physical selection in size \citep[e.g.][]{owen_imprint_2011}. However, %without constraining observational data regarding dust properties for our studied objects and since proplyds are very likely young objects (< $10^5$ years), 
we will initially assume that their dust populations are quite primitive and close to the ones of the ISM. We will then discuss the effects of grain evolution (Sect. \ref{subSect_modelPDR} and App. \ref{app_grainParam}).

The far-infrared spectrum of proplyds is dominated by the emission of the largest grains from that range, which are at thermal equilibrium with the radiation field. 
The thermal component of their spectral energy distribution can be modelled knowing the opacity $\tau_{\nu}$ and the mean dust temperature ($T_\mathrm{D,env}$, $T_\mathrm{D,disk}$\footnote{Contrary to the PDR approach which is limited to the surface layer of the disk, the mean dust temperature in the disk $T_\mathrm{D,disk}$ is valid for the bulk of the disk.}). Indeed, for a given dust temperature $T_\mathrm{D}$, the emitted spectral radiance follows the relation 
\begin{equation}
L_\nu{}\left(T_\mathrm{D}\right) = \tau{}_\nu{} \ B_\nu{}(T_\mathrm{D}),
\end{equation}
where $B_\nu{}$ is Planck's law. The dust opacity $\tau{}_\nu{}$ at the frequency $\nu{}$, is defined by
\begin{equation}
\tau{}_\nu{} = \tau{}_{\nu{}_0} \left(\frac{\nu{}}{\nu{}_0}\right)^\beta{},
\end{equation}
where $ \tau{}_{\nu{}_0}$ is the dust opacity at the reference frequency $\nu{}_0$ and $\beta$ is the spectral index. Following \citet{planck_collaboration_planck_2011}, we used a spectral index of 1.8 and a reference wavelength of 250 \textmu{}m (1200 GHz) for which they found that, in the molecular phase of the ISM, the total column density of hydrogen, $N_\mathrm{H}$, is related to the dust opacity by the empirical relation
\begin{equation} \label{eq_effSec}
\sigma = \frac{\tau{}_{250}}{N_\mathrm{H}} = 2.32 \pm{} 0.3 \times{} 10^{-25} ~\mathrm{cm}^\mathrm{2}.
\end{equation}
The mean column density of hydrogen is related to the total mass of gas and dust by
\begin{equation} \label{eq_massFromNh}
M = N_\mathrm{H} ~\Omega{} ~\mu{} ~m_\mathrm{H},
\end{equation}
where $\Omega{}$ is the observed surface in the instrumental beam knowing the distance to the source, $\mu{} ~m_\mathrm{H}$ the mean particle mass per hydrogen atom with $m_\mathrm{H}$ the mass of one hydrogen atom and $\mu{} = 1.4$.  Finally, for each component (disk and envelope), the emission can be written as a function of the total mass $M$ and the dust temperature $T_\mathrm{D}$,
\begin{equation}
L_\nu{}\left(T_\mathrm{D}\right) = \sigma \left(\frac{M}{\Omega{} ~\mu{} ~m_\mathrm{H}}\right) \left(\frac{\nu{}}{\nu{}_\mathrm{250}}\right)^\beta{} B_\nu{}(T_\mathrm{D}).
\end{equation}

The mid-infrared dust emission is dominated by PAHs which are found in the envelope and disk surface of proplyds \citep{vicente_polycyclic_2013}. This component is included in the model using the PAH emission modelled with the DustEM code
\citep{compiegne_global_2011}.We use the standard PAH population given by them, i.e. a log-normal size 
distribution of neutral molecules and one more for singly-charged molecules. The intensity of PAH emission then only depends on the incident FUV radiation 
field $G_\mathrm{0}$, the abundance of PAHs given as $f\mathrm{_C^{PAH}}$ and the total mass of the envelope $M_\mathrm{env}$. 
%~\%$\footnote{The PAH-to-gas mass ratio, parameter of the model, is estimated from the fit and then converted into the fraction of elemental carbon in the gas phase locked in PAHs assuming an elemental abundance $\left.\left(\left[\mathrm{C}\right]/\left[\mathrm{H}\right]\right)\right|_\mathrm{gas} = \left(1.61 \pm 0.17 \right)\times{} 10^{-4}$ from \citet{sofia_interstellar_2004}.}

%			      4.3 Line emission using the Meudon PDR code	  	  	  %
%--------------------------------------------------------------------------------------------------------%

\subsection{Line emission using the Meudon PDR code} \label{subSect_modelPDR}

To predict emission lines from a PDR, one needs a model that correctly describes geometrical and micro-physical elements. Doing both with accuracy is not currently possible for computational time reasons. In this study, we choose to favor a proper description of the physics and chemistry rather than geometry, also taking into account that we do not have enough angular resolution to perform a careful geometrical analysis. From this perspective, the 1D Meudon code is well suited since it is constantly updated to include latest improvements on the microphysics (e.g. H$_\mathrm{2}$ formation and excitation, photoelectric effect, grain temperature fluctuations etc. \citep{le_petit_model_2006, le_bourlot_surface_2012, bron_surface_2014, bron_efficient_2016}.
%In order to \textbf{correctly predict the atomic fine structure line and molecular line emission, we decided to emphasise the use of a very detailed PDR code rather than a complex geometry. We thus included in our model the Meudon PDR code \citep{le_petit_model_2006, Goicoechea_penetration_2007, gonzalez_garcia_radiative_2008, le_bourlot_surface_2012} which 
The Meudon PDR code is a 1D-model that considers a cloud made of a stationary plane-parallel slab of dust and gas illuminated by a given radiation field. The model iteratively solves the UV radiative transfer, thermal balance and chemistry at each point of the cloud. Once the computation of atoms and molecules abundances and physical conditions is done for each point in the cloud, the model can calculate the local emissivities of lines. In our case, the integration of this quantity gives us the brightness of the disk surface layer and the brightness of the envelope. Weighted those quantities by the size of the corresponding emitting region enables us to derive the line integrated intensity that can be compared with the spatially unresolved observations that we have.

Here, we use the 1.4.4\footnote{\url{http://pdr.obspm.fr/PDRcode.html}} public version of the code. We use the default values for the dust-to-gas mass ratio, dust properties, cosmic ray ionisation rate and turbulent velocity (Table \ref{table_pdrModel_fixedParam}), as well as elemental abundances appropriate for the ISM \citep[see][]{le_petit_model_2006}. However, grain properties in proplyds are likely to differ from those found in the ISM. Indeed, the smallest grains or PAHs seem to be underabundant in proplyds compared to the ISM: this is observed for HST10 by \citet{vicente_polycyclic_2013} and for 105-600 in this study (see Sect. \ref{subSubSect_CCP_dust}). This  observational fact is in agreement with the low detection rate of PAHs in isolated disks around young low-mass stars \citep{geers_C2D_2006, oliveira_spitzer_2010}. %while they are routinely detected in Herbig Ae/Be disks \citep[e.g.][]{acke_iso_2004}. 
We set the minimum grain radius to 3 nm in the model, i.e. excluding the smallest particles such as PAHs, for consistency with that observational fact and assuming that this is a general trend for proplyds. In the PDR model, the main impact of this modification is to remove the contribution of these small grains to the photo-electric heating and to the H$_\mathrm{2}$ formation rate. This parameter has not a significant impact on the modelled line fluxes because proplyds are dense PDRs where the grain photoelectric effect may not always dominate the heating (this will be discussed in App. \ref{app_grainParam} according to our results). 
Grain growth inside an evolving object tends to modify the grain distribution by increasing the maximum grain radius $a_\mathrm{max}$, lowering the absolute value of the power-law index and finally it reduces the FUV extinction curve. Since these parameters can not be constrained accurately based on current observations of proplyds, we have explored their effects by changing independently: 
\begin{itemize}
\item the maximal radius $a_\mathrm{max}$ up to 3 \textmu{}m, consistently with observations of Orion proplyds \citep{shuping_evidence_2003, vicente__2009};
\item the power-law index down to 3.0 \citep[e.g.][]{testi_dust_2014};
\item the FUV extinction using various curves of \citet{fitzpatrick_analysis_1988} including a case with a dust FUV extinction cross section per H nucleus $\sigma_\mathrm{ext} \simeq 8 \times 10^{-22}$ cm$^\mathrm{2}$ as used for a sample of proplyds in the Orion nebula by \citet{storzer_photodissociation_1999}. 
\end{itemize}
% with a dust FUV extinction cross section per H nucleus starting from $\sigma_\mathrm{ext} = 2.16 \times 10^{-21}$ cm$^\mathrm{2}$ (default value for the galaxy) down to $7.8 \times 10^{-22}$ cm$^\mathrm{2}$ close to the value of $8.2 \times 10^{-22}$ cm$^\mathrm{2}$ observed for a sample of proplyds in Orion nebula \textcolor{red}{SH1998}. 
The effects on the model results are discussed in App. \ref{app_grainParam} and the general conclusion is that the changes do not affect the model results significantly. Hence, we keep the default ISM properties throughout this paper. 

The input parameters of the Meudon code are the density profile as a function of visual extinction $A_\mathrm{V}$,
and the incoming FUV radiation field. These parameters are related to those of the proplyd model presented in Sect. {\ref{subSect_modelParam}}
in the following way. The impinging radiation field is set by the value $G_0$ arriving at the surface of the envelope. The density profile between the envelope surface, or ionisation front, at $A_\mathrm{V}=0$ and the disk surface
at $A_\mathrm{V}^\mathrm{jump}$ is constant with a value $n=n_\mathrm{env}$. At the envelope - disk interface (i.e. at at $A_\mathrm{V}^\mathrm{jump}$), the density jumps to $n=n_\mathrm{disk}$ and is kept constant up to $A_\mathrm{V}^\mathrm{max}$.
Since the parameter $A_\mathrm{V}^\mathrm{jump}$ gives the position of the envelope - disk interface, it is also the visual extinction of the envelope and is set by
\begin{equation}
A_\mathrm{V}^\mathrm{jump}= \frac{N_\mathrm{env}}{\kappa} =\frac{a_\mathrm{env}\times n_\mathrm{env}}{\kappa},\label{Eq_AvJump}
\end{equation}
where $\kappa$ is the parameter which relates the envelope column density $N_\mathrm{env}$  to the extinction at the density jump, 
and can be calculated from Table~\ref{table_pdrModel_fixedParam}, $\kappa=N_\mathrm{H}/A_\mathrm{V}=1.87\times10^{21}$ cm$^{-2}$ (assuming ISM properties). 
%The column density of the envelope is therefore defined by $N_{env}=\kappa~A_\mathrm{V}^{jump}$. 
The envelope column density is defined by $N_\mathrm{env} = n_\mathrm{env} \times a_\mathrm{env}$.
Finally, the value of $A_\mathrm{V}^\mathrm{max}$ is set to 10. The exact value is not critical but we have carefully checked that it is high enough. Indeed, most of the UV radiation is rapidly absorbed and most of the emission arises from regions of low extinction so increasing this parameter does not impact the results.
\begin{table}[]
  \begin{center} 
    \small
    \caption{Default values of some parameters used in the Meudon PDR code.}
    \label{table_pdrModel_fixedParam}
      \begin{tabular}{lr}
 \hline\hline
 Parameter & Value \\
 \hline
 Cosmic ray ionisation rate & 5 $\times$ 10$^\mathrm{-17}$ s$^\mathrm{-1}$ \\
 Turbulent velocity (Doppler broadening) & 2 km s$^\mathrm{-1}$ \\
 R$_\mathrm{V} = \mathrm{A}_\mathrm{V} / \mathrm{E(B-V})$ & 3.10 \\
 N$_\mathrm{H} / \mathrm{E(B-V)}$ & 5.8 $\times$ 10$^\mathrm{21}$ cm$^\mathrm{-2}$ \\
 Dust-to-gas mass ratio $\delta$ & 0.01 \\
 Power-law index of the grain size distribution & 3.50 \\
 Minimum grain radius $a_\mathrm{min}$ & 3 nm \\
 Maximum grain radius $a_\mathrm{max}$ & 300 nm \\
 \hline
\end{tabular}

    \normalsize
  \end{center}
 \end{table}
 
%			      			3.4 Fitting strategy	  	  	  			  %
%--------------------------------------------------------------------------------------------------------% 

\subsection{Fitting strategy} \label{subSect_modelFit}

In order to estimate the unknown parameters of the model (see Table \ref{table_model_param}), we are looking for the set of values that best fit the observations based on a chi-square minimisation. For the spectral energy distribution of the dust, the model numerically converges to the best solution in the space of parameters using a Nelder-Meadwith non-linear minimisation script. For the line emission, we run the model hereabove on a grid for $n_\mathrm{env}$ and $n_\mathrm{disk}$ in a range of realistic values. The apparent size of the disk $S_\mathrm{disk}$ is adjusted in the case of 105-600 while is known for the other proplyds. The best model is the one which minimizes the chi-square between observed and predicted line integrated intensities.

%======================== End of Model ========================%

%===========================================================%
% 			    		 5. Results of proplyd models		 
%===========================================================%

\section{Results of proplyd models} \label{sect_modelResults}

In this section, we present the results of the comparison between the models and observations. We start with 105-600, for which the largest number of detections is available
and follow with the Orion proplyds.

%			   		 5.1 Carina candidate proplyd		  	  	  	  %
%--------------------------------------------------------------------------------------------------------%

\subsection{Proplyd 105-600} \label{subsect_resultsFor105-600}

\subsubsection{Dust spectral energy distribution} \label{subSubSect_CCP_dust}

\begin{figure}[]
  \begin{center}
    \includegraphics[width=\linewidth]{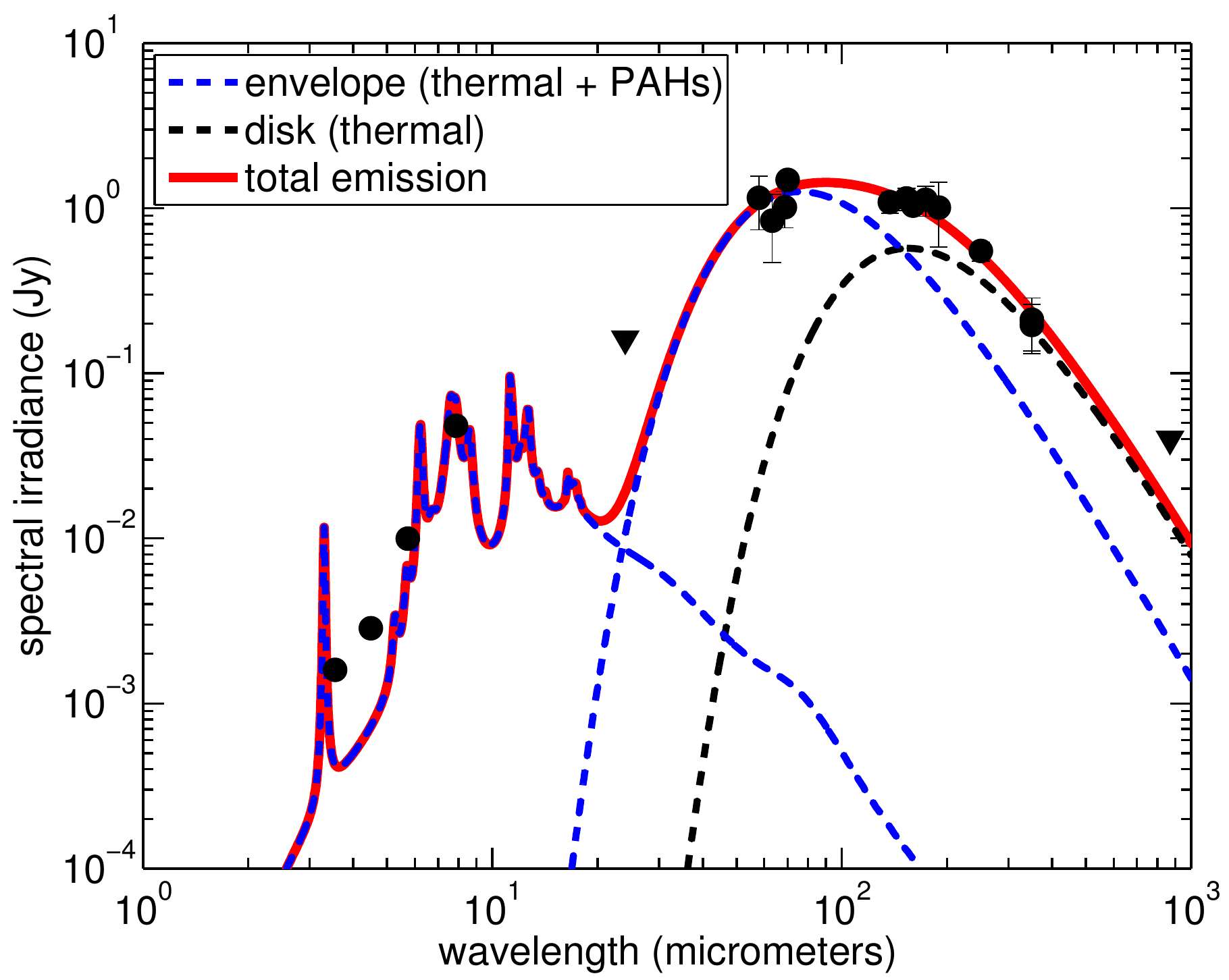}
  \end{center}
  \caption{Best fit to the observed spectral energy distribution of the dust in 105-600 with a model including two components for the thermal emission and one for the emission of PAHs. Dots are detections while triangle represents upper limits.}
    \label{fig_sedFinalFit}
\end{figure}

The spectral energy distribution of 105-600 is shown in Fig. \ref{fig_sedFinalFit}. The results of the fit with the model to the observed SED
 are given in Table \ref{table_sedParam}. Under the assumptions described in Sect. \ref{subSect_modelDust}, the first component, the disk, is composed of a cold dust population ($\sim$ 20 K) and a total mass (dust and gas) of 
 0.67 M$_\sun$. The envelope corresponds to a hotter dust population ($\sim$ 40 K) and contains a total mass of 0.048 M$_\sun$. 
This mass estimate is compared to other methods in Sect.~\ref{subSect_massEst} which yield similar 
values both for the disk and envelope.
The estimated  PAH abundance in the envelope is $f\mathrm{^{PAH}_C} = 0.0024$, about 30 times less than in the PDR of the 
NGC 7023 reflection nebula \citep{berne_formation_2012}. The depletion of PAHs within a proplyd-like object has already been observed in HST10 \citep{vicente_polycyclic_2013} with a relative abundance three times lower than in 105-600, i.e. about 90 times less than in NGC7023. 
These low abundances in proplyd-like object could result from PAH destruction by the strong UV radiation field  or sedimentation of PAHs 
in the disk but this is currently still unclear.

\begin{table}[]
  \begin{center} 
    \small
    \caption{Parameters derived from the fit of the spectral energy distribution for 105-600.}
      \label{table_sedParam}
      \begin{tabular}{lcc}
 \hline\hline
 Parameters & Best value & Interval at 1\textsigma \\ 
\hline
 $M_\mathrm{disk}$ (M$_\sun$) & 0.67 & 0.46 - 0.97 \\ 
 $T_\mathrm{D,disk}$ (K) & 19.5 & 17.9 - 21.4 \\ 
 $M_\mathrm{env}$ (M$_\sun$) & 0.048 & 0.036 - 0.062 \\ 
 $T_\mathrm{D,env}$ (K) & 39.8 & 37.7 - 42.4 \\ 
 $f\mathrm{^{PAH}_C}$ (\%)  & 0.24 & 0.22 - 0.25 \\ 
 \hline
\end{tabular}
      \
    \normalsize
  \end{center}
  \tablefoot{Confidence intervals at 1\textsigma{} are calculated based on a method with constant \textchi$^\mathrm{2}$ boundaries as confidence limits \citep[see][chap. 15.6]{press_numerical_2007}.}
\end{table} 

\subsubsection{Line emission} \label{subSubSect_CCP_gas}

\begin{figure*}[]
     \begin{center}
        \subfigure[Dispersion of the modelled line fluxes from the grid of densities.]{%
            \label{fig_obsVsMod_fit_gridInDensity}
            \includegraphics[width=0.42\textwidth]{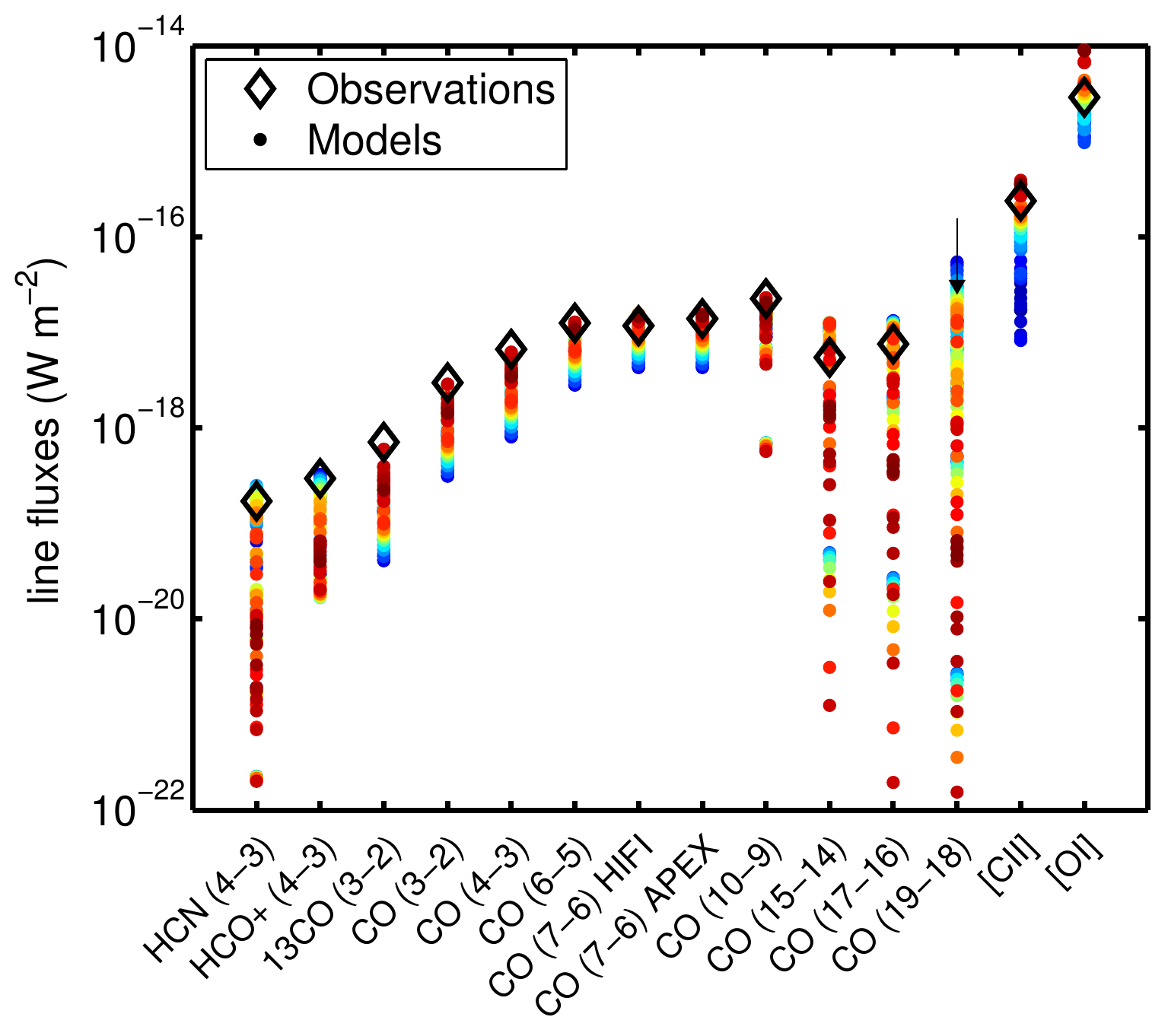}
        }%
        \subfigure[Normalized \textchi{}$^2$.]{%
           \label{fig_obsVsMod_contour_chi2}
           \raisebox{+0.1\height}{\includegraphics[width=0.42\textwidth]{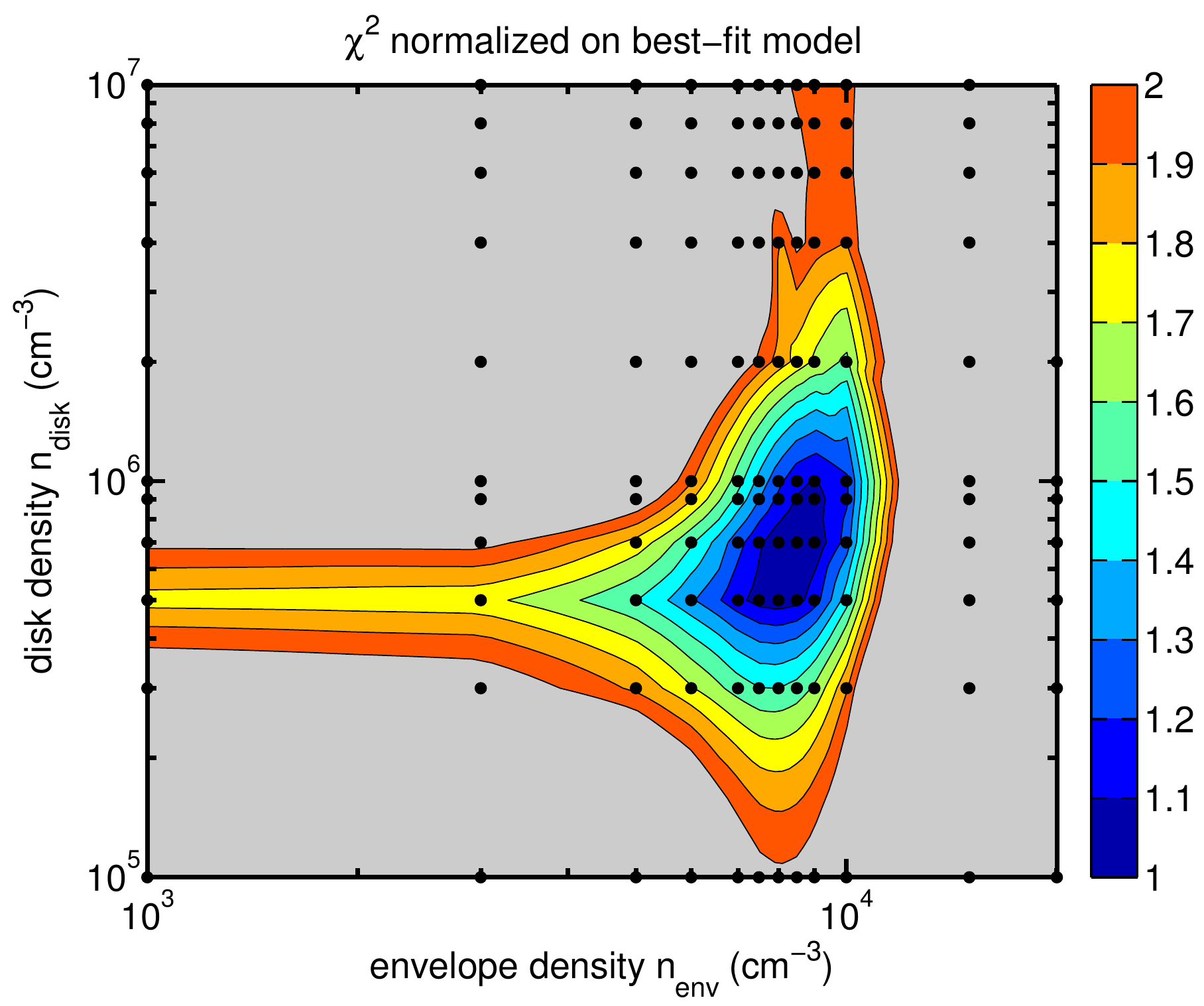}}
        }\\ %  ------- End of the first row ----------------------%
 %       \subfigure[Estimated equivalent radius]{%
  %          \label{fig_obsVsMod_contour_radius}
    %        \includegraphics[width=0.42\textwidth]{./Figures/Out/obsVsMod_contour_radius.pdf}
%        }%
 %       \subfigure[Lower limit of the disk mass]{%
%            \label{fig_obsVsMod_contour_mass}
 %           \includegraphics[width=0.42\textwidth]{./Figures/Out/obsVsMod_contour_mass.pdf}
 %       }\\ %  ------- End of the second row ----------------------%
        \subfigure[Gas temperature at the ionisation front.]{%
            \label{fig_obsVsMod_contour_gasTempIF}
            \includegraphics[width=0.42\textwidth]{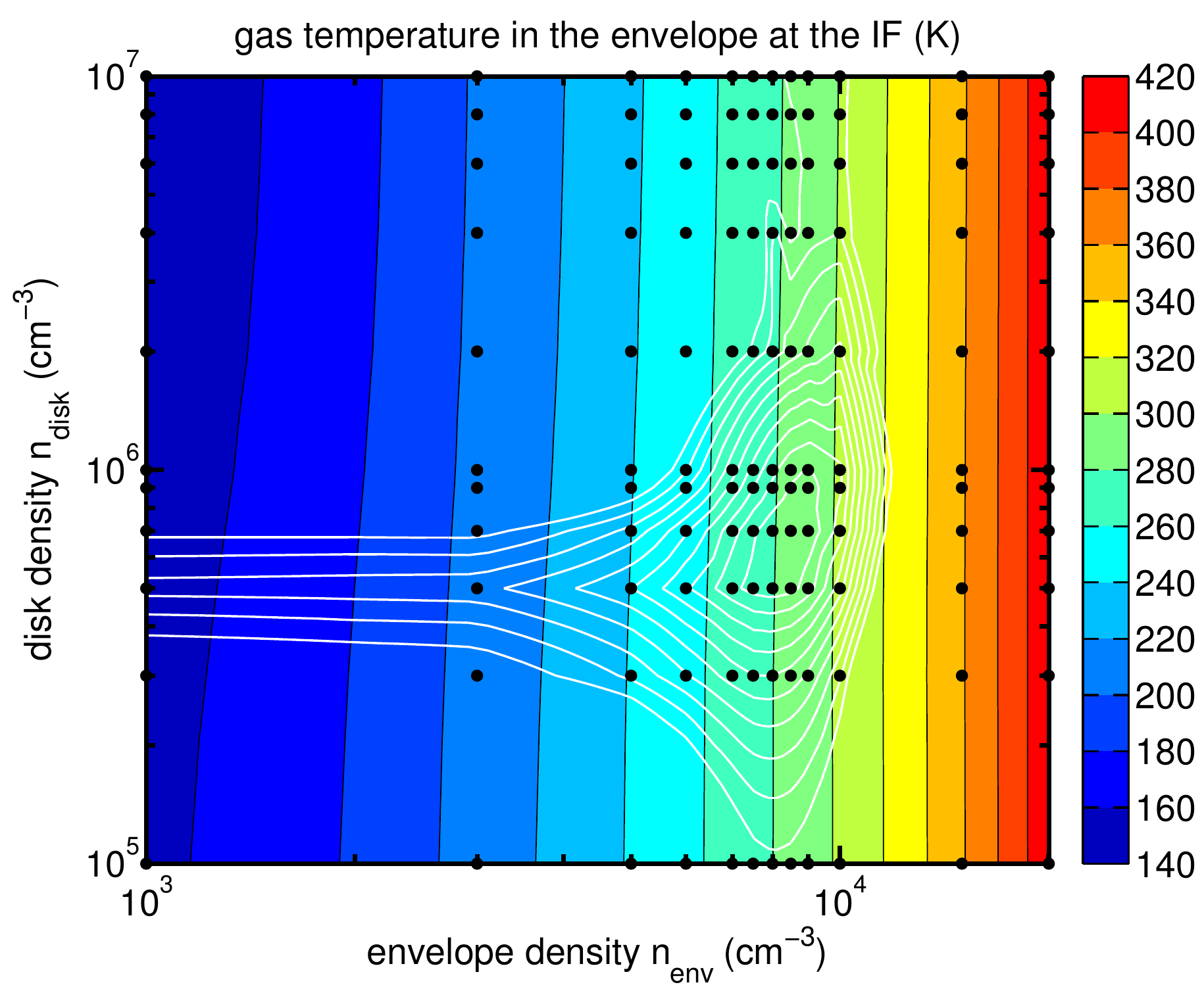}
        }%
        \subfigure[Gas temperature in the surface layer of the molecular disk.]{%
            \label{fig_obsVsMod_contour_gasTempAtSurf}
            \includegraphics[width=0.42\textwidth]{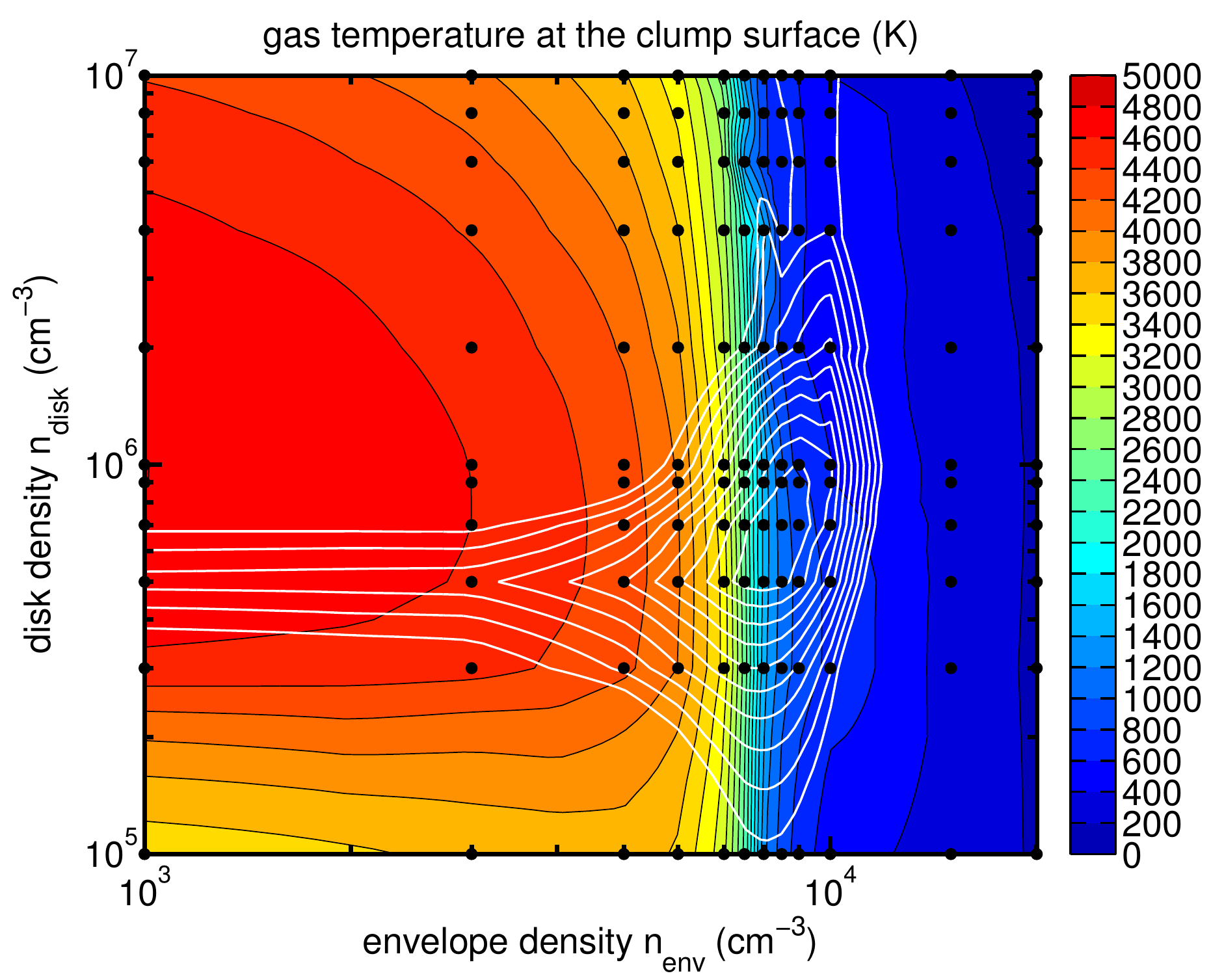}
        }%
    \end{center}
    \caption{(a) Modelled lines fluxes (105-600) for the considered density ranges (colored points) compared to observations (black diamonds). Uncertainties of observations are not plotted because they are smaller than the marker. (b) \textchi{}$^2$ comparing observed line integrated intensities of 105-600 and modeled values for the considered envelope and disk densities. (c) temperature at the ionisation front, overlaid are the 
   \textchi{}$^2$ contours.  (d) temperature at the disk surface. In panels (b,c,d), points are the positions where the models are calculated and the colors 
   and contours result from interpolating the model results over these points.}
   \label{fig_obsVsMod_contours}
\end{figure*}

\begin{figure}[]
  \begin{center}
    \includegraphics[width=\linewidth]{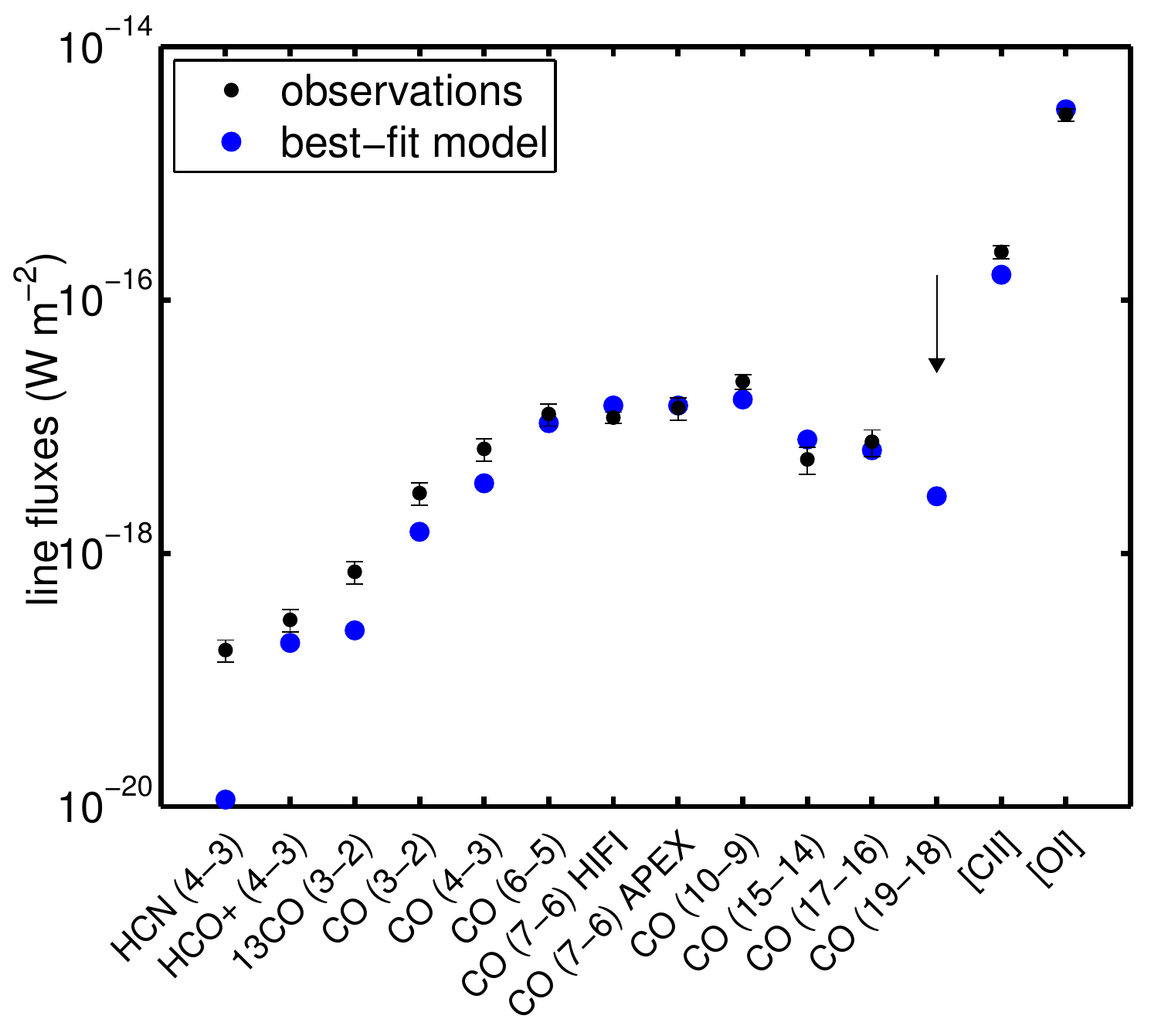}
  \end{center}
\caption{Comparison between best-fit model line integrated intensities and observed line integrated intensities for 105-600.}
\label{fig_obsVsMod_bestfit}
\end{figure}

To compare our models with observations, we have ran a grid of models with density ranges $n_\mathrm{env} = 1\times10^3 - 2 \times 10^4 ~\mathrm{cm^{-3}}$ and $n_\mathrm{disk} = 1\times10^5 - 1 \times 10^7 ~\mathrm{cm^{-3}}$, with constant grain properties as defined previously in Table \ref{table_pdrModel_fixedParam} ($a_\mathrm{min} = 3$ nm, $\delta{} = 0.01$, see App. \ref{app_grainParam} for the impact of those parameters). Fig. \ref{fig_obsVsMod_contours} presents the comparison between all the models and the observations. From the \textchi{}$^2$ analysis (Fig. \ref{fig_obsVsMod_contour_chi2}), a minimum is clearly found where the best-fit model is the one with $n_\mathrm{env} = 8.5\times10^3 ~\mathrm{cm^{-3}}$, $n_\mathrm{disk} = 7\times10^5 ~\mathrm{cm^{-3}}$ and a value $S_\mathrm{disk}=2 \times 10^7$ AU$^2$. Fig. \ref{fig_obsVsMod_bestfit} presents the comparison between the observed line integrated intensities and the best-fit model. All lines are reproduced to a factor of a few or better, with the notable exception of HCN (4-3) which is under-predicted by more than an order of magnitude. HCN emission is a tracer of dense gas, whereas our model with a constant density is adopted only for the surface layers of the disk. To go further and improve the modelling of the low-$J$ CO and HCN lines would require a detailed disk model not limited to the PDR as here. %Thermo-chemical disk models are likely more adapted for such purposes.
%We have not been able to explain this discrepancy so far. Possible reasons could be that the HCN emission is not dominated by the PDR of the disk or that we need to take into account time-dependent chemistry effects. We postpone this to further studies.

The gas temperature at the edge of the envelope (ionisation front) increases with the envelope density but in a relative small range ($\sim$ 150 to 400 K, see Fig. \ref{fig_obsVsMod_contour_gasTempIF}) compared to the variations associated to the grains properties (Fig. \ref{fig_obsVsMod_grainsRmin} and \ref{fig_obsVsMod_gtd}). The gas temperature at the surface of the disk mainly depends on the envelope density (see Sect. \ref{sect_surfaceHeating} for more details). The best-fit model corresponds to a value of about 1100 K. Models with slightly lower envelope densities, ans thus slightly higher disk surface temperatures, are still close to the observations while the decrease in disk surface temperature caused by an increase in the envelope density makes models diverge significantly from observations. This is strong evidence that the disk surface is hot with  temperature around 1000 K or more.

%Assuming a slightly lower envelope density, e.g. $n_\mathrm{env} = 7.5 - 8\times10^3 ~\mathrm{cm^{-3}}$, models predict a surface temperature between 1000 and 2000 K and can still explain the observations well \textcolor{red}{(waiting for the calculation of quantitative uncertainty, still in progress...)}. On the opposite, increasing the envelope density lowers the disk surface temperature significantly and models diverge significantly from observations.  

%			6.1 Physical structure in a proplyd: the case of 105-600		  %
%--------------------------------------------------------------------------------------------------------%

\subsubsection{Physical structure of the PDR}

\begin{figure*}[]
  \begin{center}
    \includegraphics[width=0.90\textwidth]{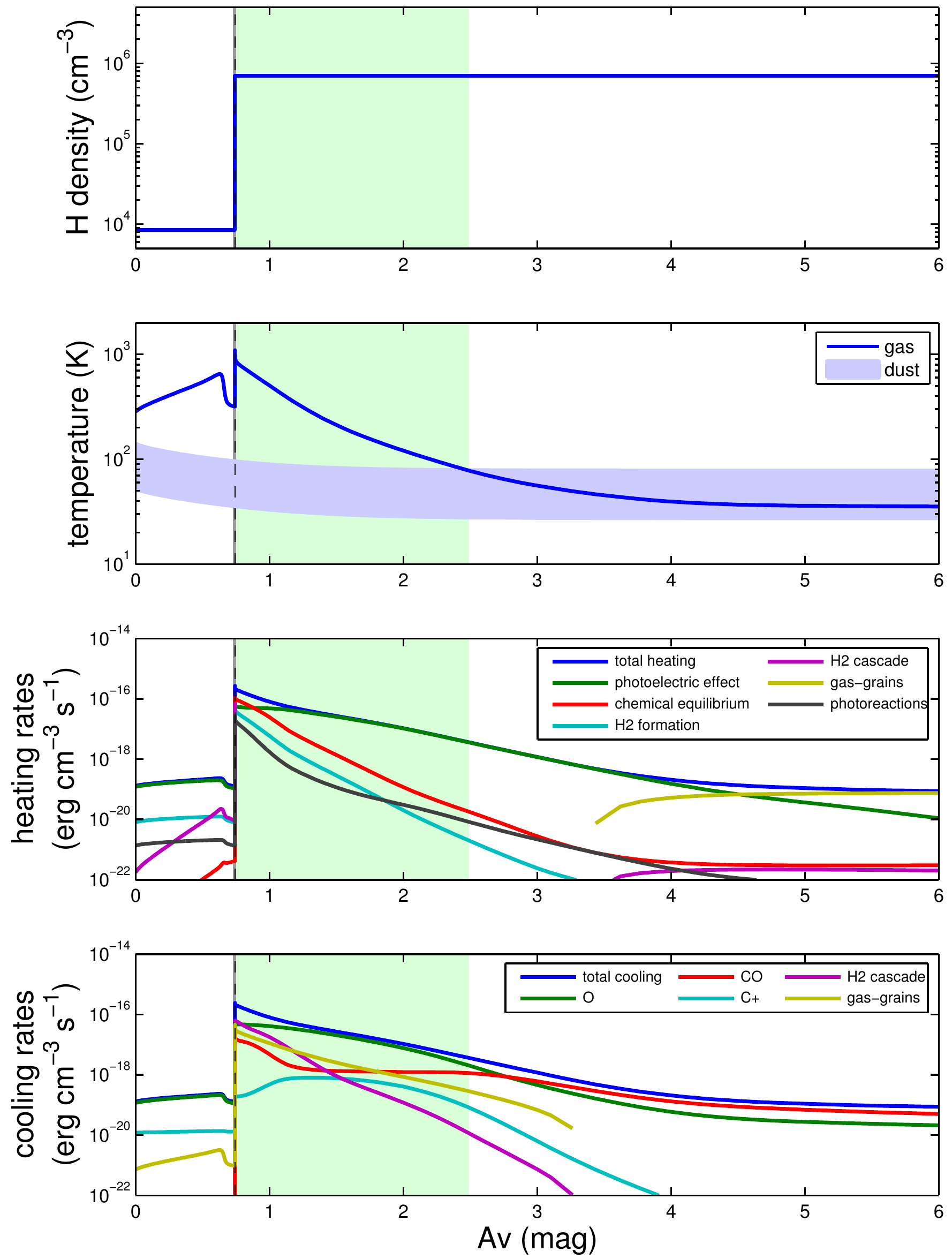}
  \end{center}
\caption{PDR structure of the best-fit model for the Carina candidate proplyd 105-600. FUV radiation field is incoming from the left where the ionisation front is located at $A_V = 0$. The disk surface, represented by a vertical dashed black line, is located at $A_V = 0.742$. The small dark grey patch close to the disk surface illustrates the H/H$_\mathrm{2}$ transition (10 to 90\% of H in H$_\mathrm{2}$) and the bright green patch illustrates the C$^+$/CO transition (10 to 90\% of C in CO). The dust temperature is given as a range corresponding to the temperature of the biggest to the lowest studied grains (radius of 3 to 300 nm). %The horizontal purple lines on the second plot indicates the escape temperature at a distance of 100, 300 and 1000 AU from a central star of 1 M$_\sun$.
Note that the x-axis is in magnitude of visual extinction. In distance, the envelope represents about 11000 AU while the green patch in the disk represents 300 AU.} 
\label{fig_pdrStructure}
\end{figure*}

As an output of the model it is possible to obtain the 1D physical structure of the PDR in the studied proplyds. Note that this structure is only limited to the PDR, i.e. the envelope and the surface layer of the disk, and so that the extracted physical properties are not valid deeper in the disk where much larger densities are expected and UV photons are strongly attenuated. %when the physics is not dominate by the FUV radiation field anymore.
%Here we chose to illustrate this with the profile obtained for 105-600 since it is the object for which we have the most constraints and because we found that it is representative of what can be derived for all sources. 
The 1D profiles for the best model of 105-600 are shown in Fig. \ref{fig_pdrStructure}. It can be seen that the H/H$_\mathrm{2}$ transition, defined here as the region where 10 to 90\% 
of H is in H$_\mathrm{2}$, encompasses the envelope - disk interface ($A_\mathrm{V}^\mathrm{jump} = 0.742$, see eq. \eqref{Eq_AvJump}) ranging from $A_\mathrm{V} = 0.729$ (inside the 
envelope, right before the interface) to $A_\mathrm{V} = 0.748$ (inside the disk). The C$^+$/CO transition starts inside the 
molecular disk, at $A_\mathrm{V} = 0.742$ and stops at $A_\mathrm{V} = 2.49$. 

The gas temperature is 285 K at the ionisation front and increases up to 650 K at $A_\mathrm{V} = 0.63$ 
(Fig. \ref{fig_pdrStructure}, second pannel). It then decreases to about 320 K before jumping to 1100 K 
at the disk surface and then slowly decreases. Note that a smoother density profile, rather than a jump, does not significantly change the temperature profile except that it is slightly smoothed too. We thus choose to keep a jump for simplicity. The initial increase inside the envelope can be attributed 
to an increase of the photoelectric-effect efficiency related to the increasing concentration of neutral 
grains (Fig. \ref{fig_pdrStructure}, third pannel). The trend is inverted just before reaching the disk (and 
the H/H$_\mathrm{2}$ transition) where the efficiency drops. The photoelectric-effect is by far the main 
heating mechanism inside the envelope. However, at the disk surface, several heating mechanisms contribute significantly (Fig. \ref{fig_pdrStructure}, third pannel): chemical reactions (37\%), photoelectric-effect (24\%), formation of H$_\mathrm{2}$ on grains (17\%,) 
collisional de-excitation of UV-pumped H$_\mathrm{2}$ (15\%) and direct ionisation/dissociation by FUV photons (7\%). Beyond $A_\mathrm{V} = 1$, the photoelectric-effect becomes 
again the main heating mechanism up to $A_\mathrm{V}= 4.5$, where the gas-grains collisions take the lead 
and keep the gas temperature close to the dust temperature. However, at this location, our simple constant density PDR model is not appropriate anymore and proper modelling will require specific disk models. %the structure is probably not the one of a PDR anymore and should be described by a disk model.

Inside the envelope, owing to the relatively high gas density, the main cooling mechanism is the fine-structure line emission of \big[OI\big] 63 \textmu{}m, which is one 
order of magnitude stronger than the \big[CII\big] 158 \textmu{}m line. At the surface of the disk, the cooling is shared between 
the cascade of H$_\mathrm{2}$, the gas-grains collisions, the emission of O and CO. Deeper in the disk, 
CO and O dominate the cooling.

%			   	   5.2 Results for the Orion proplyds	  	  	  	  	  %
%--------------------------------------------------------------------------------------------------------%

\subsection{Results for Orion proplyds} \label{subSect_resultsForOrionProplyds}

\begin{figure}[]
  \begin{center}
    \includegraphics[width=\linewidth]{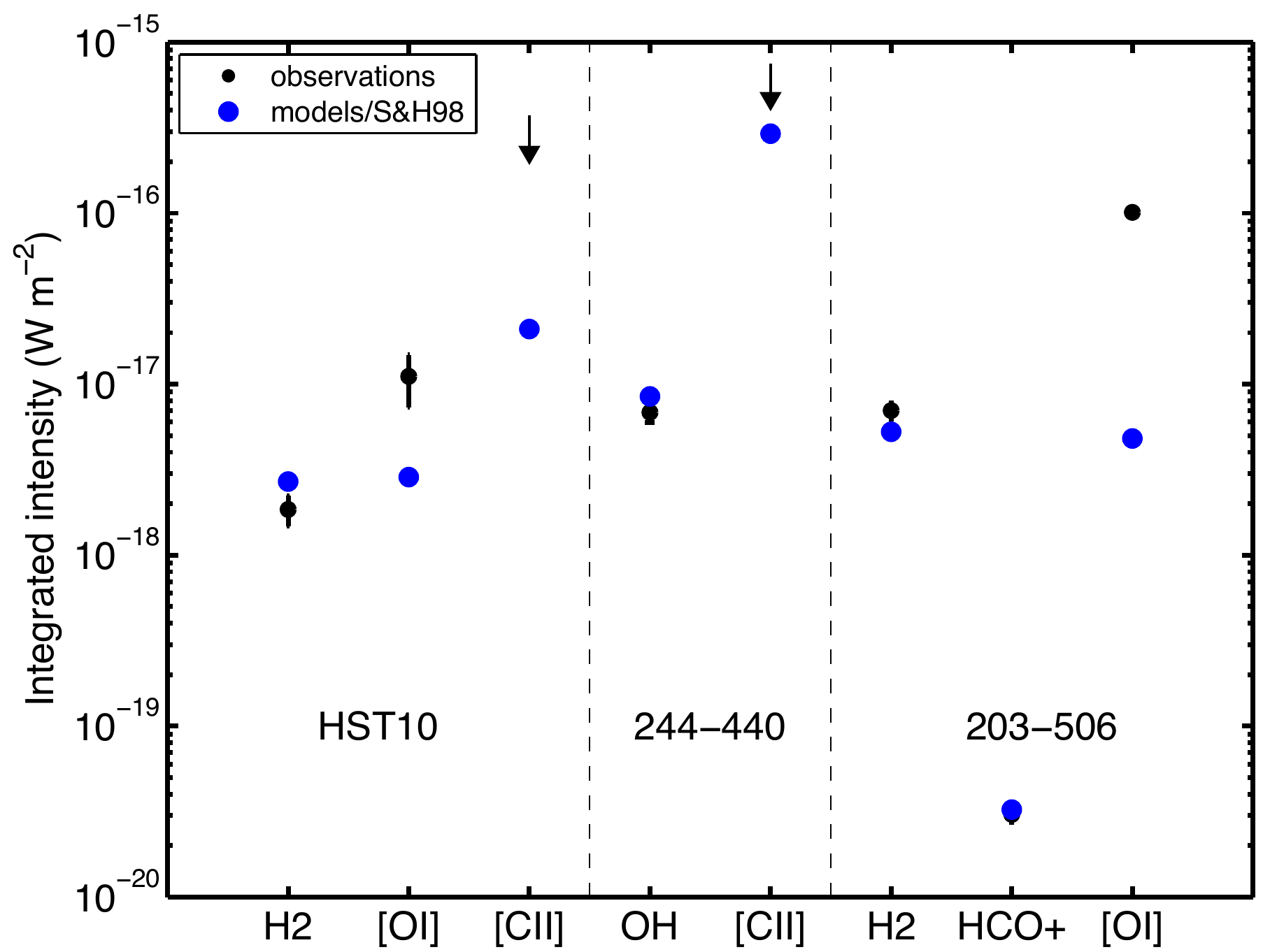}
  \end{center}
\caption{Comparison between observed line integrated intensities and best-fit models for the Orion proplyds. Observations are shown in black while estimated line integrated intensities using the outputs of the models (including the method of \citet{storzer_[o_1998} for the \big[OI\big] 6300 \AA{} line) are shown in blue. Heads of black arrows give the upper limits for \big[CII\big] emission in the case of HST10 and 244-440.}% The initial estimations of the [OI] 6300 \AA{} line integrated intensities directly extracted from the Meudon PDR code are shown in red.} 
\label{fig_obsVsMod_orionProplyds}
\end{figure}

For the three Orion proplyds, few detections of gas lines are available. For these sources the far-infrared dust emission was not detected since it is drowned in the emission from the nebula.
We therefore only attempt to reproduce the observed line emission with our model. Intense emission of the \big[OI\big] 6300 \AA{} line is observed towards the disks of HST10 and 203-506 as reported by \citet{bally_externally_1998}. This line arises from the hot neutral layers of the PDR. Its excitation is the result of 
the rapid formation of OH through the reaction of O with H$_\mathrm{2}$, followed by photodissociation leaving a large fraction of O in an electronically excited state
which radiatively decays through emission in the 6300 \AA{} line. Since this process was not included in the public version of the Meudon PDR code (it will be in the next one), we therefore calculate the intensity of the \big[OI\big] 6300 \AA{} line using the prescription of
 \citet{storzer_[o_1998}, using the column density of OH predicted by our model. In order to fit the observations, we follow the same strategy as for 105-600, i.e. 
 we vary the envelope and disk density. %except for 203-506 where no envelope is observed. In this latter case we have therefore considered the disk only in our model. 
 Comparison between observations and the best-fit model are shown in Fig. \ref{fig_obsVsMod_orionProplyds}, while parameters and outputs of the best-fit models are given in Tables \ref{table_model_param} and \ref{table_mod_orionProplyds}.
 
\begin{table*}[]
  \begin{center} 
  \caption{Physical parameters estimated with the model for the studied prolyds.}
  \label{table_mod_orionProplyds}
  \small
    \begin{tabular}{lccccccc}
 \hline\hline
 \multirow{2}{*}{Object} & \multirow{2}{*}{P$_\mathrm{trans}$}& Gas temperature at & N$_\mathrm{env}$ & r$_\mathrm{disk}$ & $\dot{M}$ supercritical & $\dot{M}$ subcritical & $\dot{M}$ observed \\
& & the disk surface (K) &(cm$^\mathrm{-2}$) & (AU) & (M$_\odot$/yr) &(M$_\odot$/yr) &(M$_\odot$/yr) \\ 
 \hline
105-600 & 1.00 &1100 & 1.4e+21 & 2500\tablefootmark{a} & 7.5e-05 & ... & $1$e-06 \\
HST10 & 1.05 & 1150 & 2.0e+21 & 85\tablefootmark{b} & 5.0e-07 & 1.7e-08 & $\left(7 \pm 5\right)$e-07 \\
244-440 & 1.01 & 1270 & 1.9e+21 & 193\tablefootmark{b} & 1.4e-06 & 8.2e-09 & $\left(15 \pm 7\right)$e-07 \\
203-506 & 1.12 & 1080 & 1.0e+21 & 150\tablefootmark{b} & 9.4e-08 & 9.9e-09 & ... \\
 \hline
\end{tabular}
  \normalsize
  \end{center}
  \tablefoot{Densities and temperature are derived form the best fit model for each object. The observed mass-loss rates derived at the ionisation front are from \citet{henney_keck_1999} for HST10 and 
  244-440, and derived by us from H$\alpha$ images for 105-600 (see text for details). A central star of 1 M$_\odot$ is used in the calculation of the mass-loss rate for 105-600 while, for the Orion proplyds, we used a 0.1 and 1 M$_\odot$ for the \textit{supercritical} and \textit{subcritical} cases respectively. \\
  \tablefoottext{a}{Disk radius roughly approximated by this study (see Sect.~\ref{subSubSect_CCP_gas}).} \\
  \tablefoottext{b}{Disk radius estimated by spatially resolved observation: HST10 from \citet{chen_2.12_1998}, 244-440 and 203-506 from \citet{vicente_size_2005}}.}
\end{table*}
% \tablefoot{1\textsigma{}-relative uncertainties of integrated intensities are given in parentheses after each value and take into account instrumental accuracy and fitting uncertainties. Other uncertainties are not indicated, they are below 5\% for the full width at half maximum (FWHM) and are estimated to be smaller than a percent for the velocity with respect to the local standard of rest ($v_\mathrm{LSR}$). \\
%  \tablefoottext{a}{Upper limit determined by the noise (peak of the line supposed to be smaller than three times the noise level).}  \\
%  \tablefoottext{b}{Upper limit determined by the observed background emission.} \\
%  \tablefoottext{c}{Values converted using the apparent size given in Table \ref{table_targets_properties} and assuming that all the lines are coming from the disk except for [CII] 158 \textmu{}m and [OI] 63 \textmu{}m lines that are supposed to originate in the envelope.} }

For HST10, models cannot produce more than $10^{-6}$ W\,m$^\mathrm{-2}$\,sr$^\mathrm{-1}$ of \big[CII\big] emission -- including changing 
the envelope density, the small grains abundance or dust extinction properties -- which is well below the upper limit extracted from observation. This is consistent with the idea that the observed emission should be dominated by the \big[$^\mathrm{13}$CII\big] emission from the HII region expected at the same frequency. Overall this suggests that the envelope properties are not well constrained here. %The C$^+$ emission is sensitive to the detailed density profile \citep{joblin_gas_2010}, hence it is also possible that our simplified constant density model is not appropriate. %however the use of line emission from the disk can also contrain the envelope column density and density. 
%However, the [OI] 6300 \AA emission is observed at the surface of a disk. 
The \big[OI\big] 6300 \AA{} line intensity is initially strongly underestimated but is better reproduced if the mechanism of formation for the excited oxygen described by \citet{storzer_[o_1998} is included in the model and if an $n_\mathrm{env} \le 4 \times 10^5$ cm$^\mathrm{-3}$ is used. Above this value, the envelope column density is large enough that
OH starts to form inside the envelope, which is incompatible with observations showing that the emission occurs close to the disk surface \citep{bally_externally_1998}. An envelope density of $n_\mathrm{env} \le 4 \times 10^5$ cm$^\mathrm{-3}$ corresponds to an envelope 
column density of about $N_\mathrm{env}=3 \times 10^{21}$ cm$^\mathrm{-2}$ which is slightly below the estimated value of 
$5.5 \times 10^{21}$ cm$^\mathrm{-2}$ from \citet{storzer_[o_1998}. The H$_\mathrm{2}$ and \big[OI\big] 6300 \AA{} lines are best explained 
with a disk density of the order of $10^6$ cm$^\mathrm{-3}$. The best-fit model gives
 $n_\mathrm{env} = 2.5 \times 10^5$ cm$^\mathrm{-3}$ and $n_\mathrm{disk} = 4 \times 10^6$ cm$^\mathrm{-3}$. 
 The disk density is consistent with the estimate of \citet{storzer_[o_1998} who found the same value. %$n_\mathrm{disk} = 2 \times 10^6$ cm$^\mathrm{-3}$. 
 The temperature at the disk surface predicted by the best fit model is 1150 K.

For 244-440, \big[CII\big] arising mainly from the envelope is probably detected but, as HST10, drowned in the nebula emission. The OH line is well detected and should arise from the the disk (under the assumptions that this line is being emitted by the proplyd and no other object of the Orion Bar in the line of sight). With two lines, whose one is an upper limit, the two densities will be poorly constrained, but we nevertheless found a clear best-fit for a model with $n_\mathrm{env} = 1 \times 10^5$ cm$^\mathrm{-3}$ and $n_\mathrm{disk} = 2 \times 10^6$ cm$^\mathrm{-3}$. The temperature at the disk surface predicted by the best fit model is 1270 K. These values are consistent with those found for the other Orion proplyds.

For 203-506, the disk density is well constrained by the observation of the H$_\mathrm{2}$ and HCO$^+$ lines and the best-fit value is $n_\mathrm{disk} = 1 \times 10^6$ cm$^\mathrm{-3}$. The "envelope" density that corresponds to the ambiant PDR resulting from the photoevaporation of the Orion bar \citep{goicoechea_compression_2016} is estimated to be $n_\mathrm{env} = 4.5 \times 10^4$ cm$^\mathrm{-3}$. The temperature at the disk surface (which we consider being at the H/H$_\mathrm{2}$ transition) predicted by the best fit model is 1080 K. Here again, we need to include the mechanism of \citet{storzer_[o_1998} in the model in order to get closer to the observed \big[OI\big] 6300 \AA{} line intensity but is still highly underestimated. This discrepancy could be explained by the emission from shocked regions or that another mechanism to form OH is missing, e.g. by reactions with vibrationally excited H$_2$.

The physical structures extracted for the Orion proplyds are not shown here because they are less well constrained than 105-600. They are found similar to the one given for 105-600 in Fig. \ref{fig_pdrStructure}, with a rapid increase of the gas temperature at the disk surface and  the H/H$_\mathrm{2}$ transition layer located close to it.

%A possible explanation to the underestimation of the [OI] line for the other Orion proplyd is that some mechanisms to form excited OH are missing in the code, especially we do not consider the formation of excited OH by reaction of O with excited H$_\mathrm{2}$ \textcolor{red}{(e.g. Marcelino's paper?)}. If this is true, the OH line should be also underestimated for this proplyd. As a consequence in that case, it means that our model overestimates the disk density to fit the observation and we should thus consider $n_\mathrm{disk} \le 2 \times 10^6$ cm$^\mathrm{-3}$. The best-fit envelope density is also an upper limit since the model predicts that increasing the density will move the H/H$_\mathrm{2}$ transition from the disk to the envelope. The emission from OH will then mainly comes from the envelope where density and temperature are lower than in the disk and so the integrated intensity will significantly drop.

%				6.3 Comparison to other models		  			  %
%--------------------------------------------------------------------------------------------------------%

\subsection{Comparison to other models}

It is interesting to compare the results obtained here with those of earlier studies of proplyds using
PDR models. When doing this comparison it is important to keep in mind that these studies usually
do not compare their predictions to actual observations of PDR tracers towards these objects.  

\citet{storzer_photodissociation_1999} and \citet{adams_photoevaporation_2004} use
a PDR model based on the classical \citet{tielens_photodissociation_1985} model. Contrary to what is presented
 here, they consider a single constant density structure. \citet{storzer_photodissociation_1999} compare
their models to the emission of vibrationally excited H$_\mathrm{2}$ and \big[OI\big] 6300 \AA{} emission observed
in Orion proplyds (including HST10) and find good agreement.They also predicted the far infrared line intensities of the \big[CII\big] line, for which they give values ranging between 2.4 and $3.1\times10^{-6}$ W\,m$^\mathrm{-2}$\,sr$^\mathrm{-1}$, which assuming an emitting size equal to the size of the envelope projected on the plane of the sky ($2.6\arcsec \times 1\arcsec$, see Table \ref{table_targets_properties}) corresponds 
to a flux of 1.15 - 1.50 $\times10^{-16}$ W.m$^\mathrm{-2}$, just below the high upper limit of $2.3 \times10^{-16}$ W.m$^\mathrm{-2}$ that we obtained. While not excluded, their value is probably overestimated since the observation is supposed to be highly contaminated by the nebula emission. They have also predicted strong \big[OI\big] 63 \textmu{}m emission close to our high upper limit of detection leading to the same possible conclusion. Detailed observation are needed to conclude clearly. The study of 
\citet{adams_photoevaporation_2004} is essentially theoretical. While focusing on different FUV field ranges, the temperature profile
is characterized in both studies by a region of constant and high temperature (> 4000 K) at low column densities
followed by a drop of temperature at the location of the H/H$_\mathrm{2}$ transition, which is situated, in their 
models, just above the disk. The temperature profile which we find here (Fig.~\ref{fig_pdrStructure})
differs greatly from that of   \citet{storzer_photodissociation_1999} and \citet{adams_photoevaporation_2004}
in that the temperature we find at low column densities is much lower (typically a few 100 K). This is mainly due
to two fundamental differences which reduce the heating in the envelope in our models: 1) the lower density
does not allow collisional de-excitation of H$_\mathrm{2}$ to be important, 2) the low abundance of small grains that
we have considered (in agreement with the observation that PAHs are underabundant in HST10 
and in 105-600 \citep[][and Sect. \ref{subSubSect_CCP_dust}]{vicente_polycyclic_2013}, reduces photoelectric heating drastically.
%It is important to note that our models with $N_{env}/N_{H-H_2}<<1$ i.e. with low envelope column densities 
%are essentially equivalent to their models with constant density. In this case, we recover temperatures at 
%that are of the order of several 1000 K Fig.~\ref{fig_gasTempAtSurf_slice_NH})
Nevertheless, an important aspect in the frame of photoevaporation is that the temperatures at the disk surface derived
by \citet{storzer_photodissociation_1999} and \citet{adams_photoevaporation_2004} are typically of the order
of a few 100 K, i.e. lower than ours. Our prediction for the H$_\mathrm{2}$ (1-0) S(1) line for HST10 is in good agreement with theirs and with our observations.

\section{Effect of the H/H$_{2}$ transition layer location on the gas thermal balance at the disk surface} \label{sect_surfaceHeating}

In this section, we explore the effect of the location of the H/H$_2$ transition layer on the gas thermal balance, and the gas temperature, at the disk surface, i.e. the density jump in our model. As it is described in the book of \citet{tielens_physics_2005}, the location of this layer, i.e. the column density of hydrogen nucleus, is proportional to the FUV flux $\mathrm{G}_0$ and inversely proportional to the density. Since the column density between the ionisation front and this layer is mainly set by the envelope density in the case of proplyds, we run our model on a grid of values for $n_\mathrm{env}$ and $\mathrm{G}_0$ from $2 \times 10^3$ to $2 \times 10^5$. We also considered different 
envelope sizes: a large one corresponding to the size of the biggest studied proplyd, 105-600, and a small one corresponding to the size of HST10.

%			6.2 The effect of the H/H2 transition on temperature			  %
%--------------------------------------------------------------------------------------------------------%

\subsection{Temperature} \label{subSect_hh2TransitionLocation}

\begin{figure}[]
  \begin{center}
    \subfigure{
      \includegraphics[width=\linewidth]{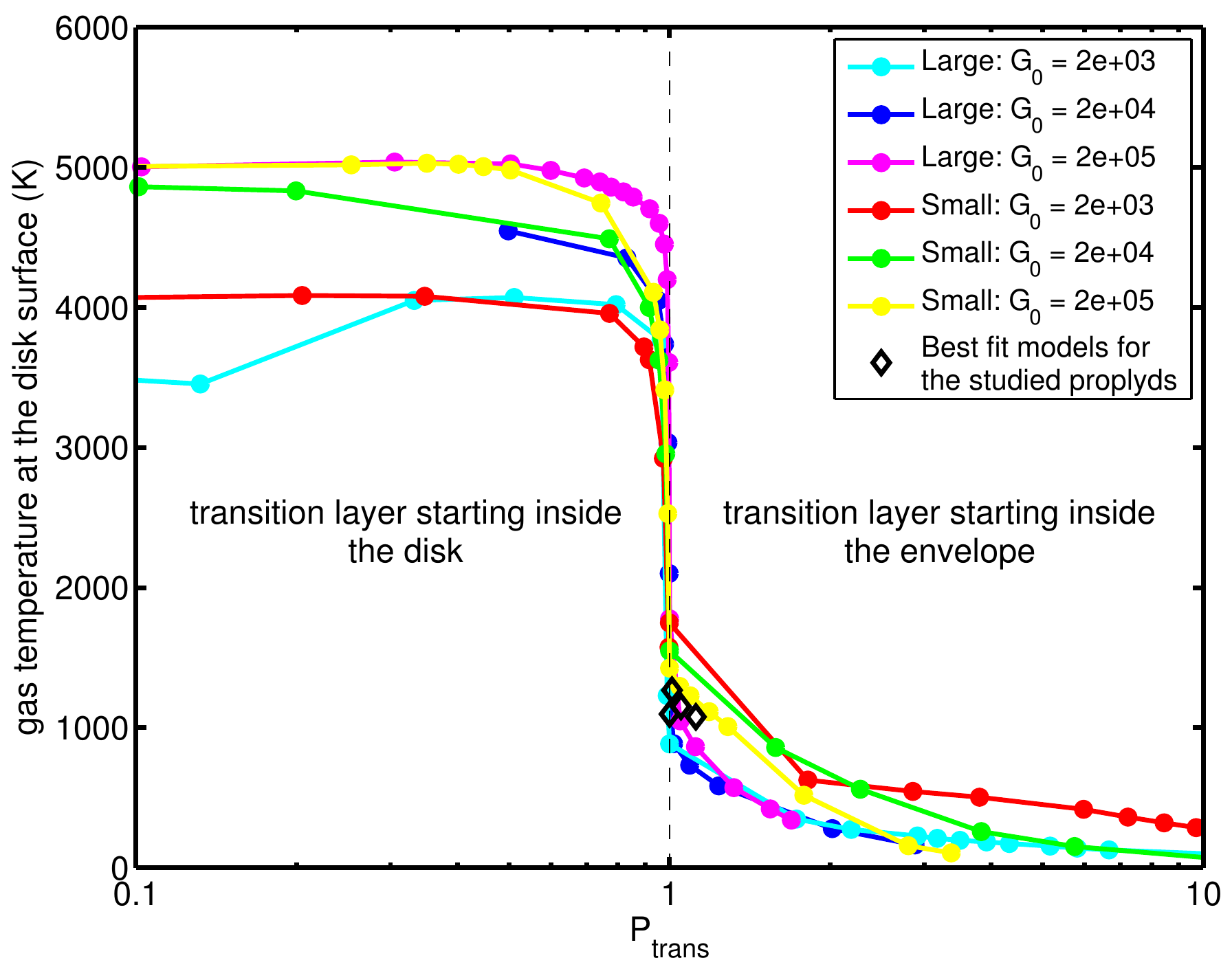}
    }\\ %  ------- End of the row ----------------------%
  \end{center}
  \caption{Modelled temperature at the disk surface as a function of the position of the H/H$_\mathrm{2}$ transition for various radiation fields and
 adopting the geometries of 105-600 as a large proplyd and HST10 as a small proplyd.  The points are the positions where the models are calculated and 
 corresponding solid lines are cubic interpolations. 
 %The error bars for the case of G$_\mathrm{0,at} = 2.2 \times 10^4$ represent the variations with the disk density in the range $10^5 - 10^7$ cm$^\mathrm{-3}$. For the two other radiation fields, models are calculated for the median value of $10^6$ cm$^\mathrm{-3}$. 
 The dotted vertical line corresponds to $P_\mathrm{trans}=1$. On the left side of this line, the H/H$_\mathrm{2}$ transition 
 is inside the disk, while on the right side of this line the H/H$_\mathrm{2}$ transition is inside the envelope.}
  \label{fig_gasTempAtSurf_slice_NH}
\end{figure}

%We have seen that the gas temperature at the surface of the disk is very sensitive to the 
%envelope density $n_\mathrm{env}$ (Fig. \ref{fig_obsVsMod_contour_gasTempAtSurf}). This is because $n_\mathrm{env}$ is connected to the 
%visual extinction of the envelope $A_\mathrm{V}^\mathrm{jump}$ through Eq.~\ref{Eq_AvJump}. The value of $n_\mathrm{env}$ is therefore particularly important because it determines
%the calculated position of the H/H$_\mathrm{2}$ transition layer (within or external to the disk). This position is critical
%to the dominant heating and cooling mechanisms at play and hence on the gas temperature. Since the FUV radiation field also moves the location of the transition layer, we make it varies from $2\times10^3$ to $2\times10^5$.
First of all, we extracted the gas temperature at the disk surface for each model of the grid to see its evolution with the position of the H/H$_2$ transition layer.
This is illustrated in Fig.~\ref{fig_gasTempAtSurf_slice_NH} which presents the gas temperature at the disk surface as a function of the position of the H/H$_\mathrm{2}$ transition which is captured by the ratio $P_\mathrm{trans} = A_\mathrm{V}^\mathrm{jump}/A_\mathrm{V}^\mathrm{H-H_2} = N_\mathrm{env}/N_\mathrm{H-H_2}$,
where $A_\mathrm{V}^\mathrm{H-H_2}$ is the visual extinction, or position, at which the transition layer starts, i.e. where 10\% of the hydrogen is in molecular form, and $A_\mathrm{V}^\mathrm{jump}$ is the visual extinction setting the position of the density jump corresponding to the disk surface in our model.
The H/H$_\mathrm{2}$ transition is thus totally located inside the disk when $P_\mathrm{trans} < 1$, and is located or starts in the envelope when $P_\mathrm{trans} > 1$. A similar behaviour is observed for all the studied cases: a \textit{hot} ($T = 1500$ - $5000$ K) regime prevails when 
the transition is located inside the disk ($P_\mathrm{trans} < 1$) while a \textit{warm} regime ($T<500$ K) is observed 
when the transition is located inside the envelope ($P_\mathrm{trans} > 1$). A \textit{transitional} regime ($T \sim  500$ - $1500$ K) lies between the \textit{hot} and \textit{warm} regimes at $P_\mathrm{trans} \sim 1$.

\subsection{Heating and cooling mechanisms}

\begin{figure*}[!p]
     \begin{center}
        \subfigure[Large proplyd (size of 105-600) - $G_\mathrm{0} = 2 \times 10^3$]{%
            \label{fig_heatingContribution_CCP_2e3}
            \includegraphics[width=0.45\textwidth]{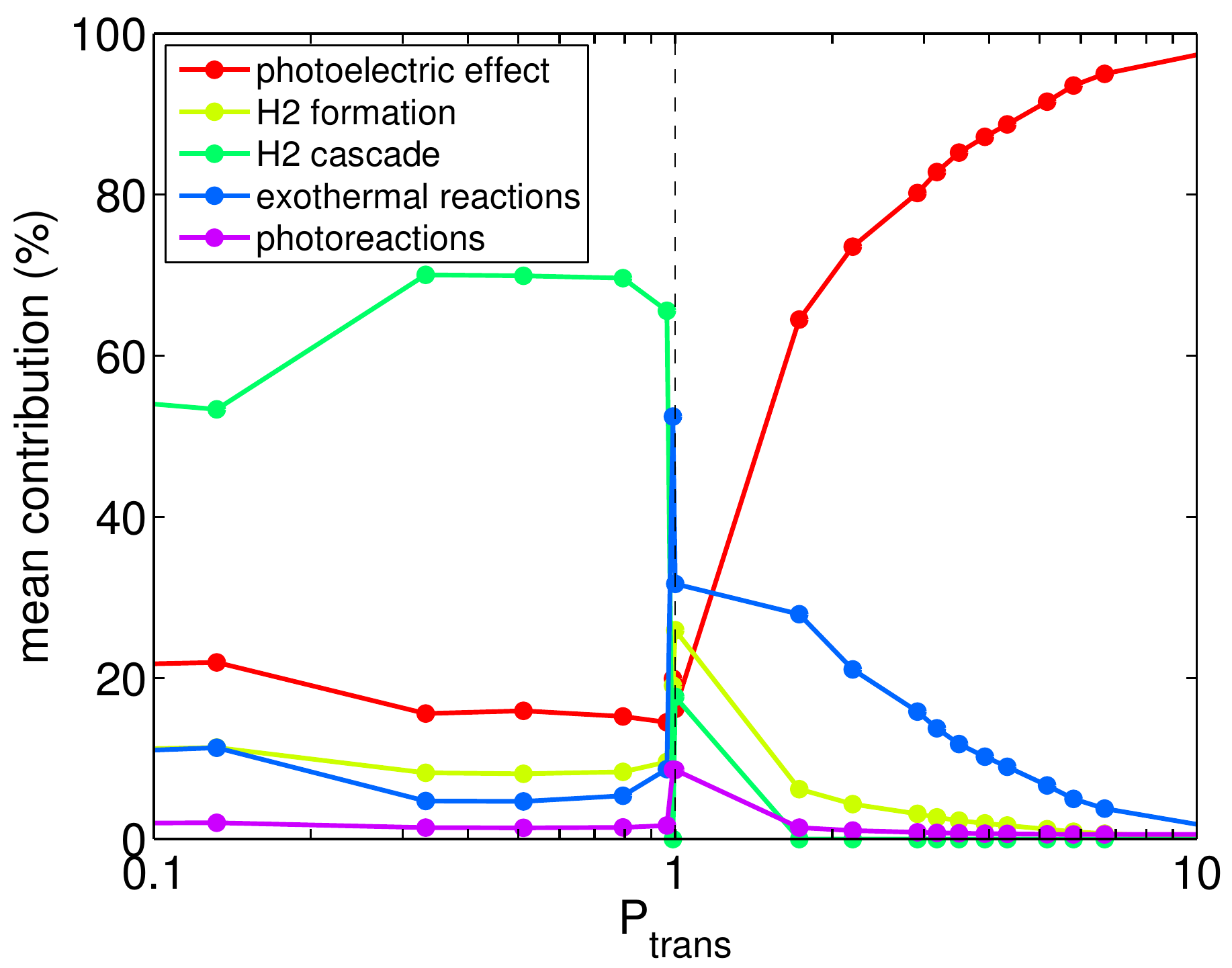}
        }%
        \subfigure[Small proplyd (size of HST10) - $G_\mathrm{0} = 2 \times 10^3$]{%
           \label{fig_heatingContribution_HST10_2e3}
           \raisebox{+0.0\height}{\includegraphics[width=0.45\textwidth]{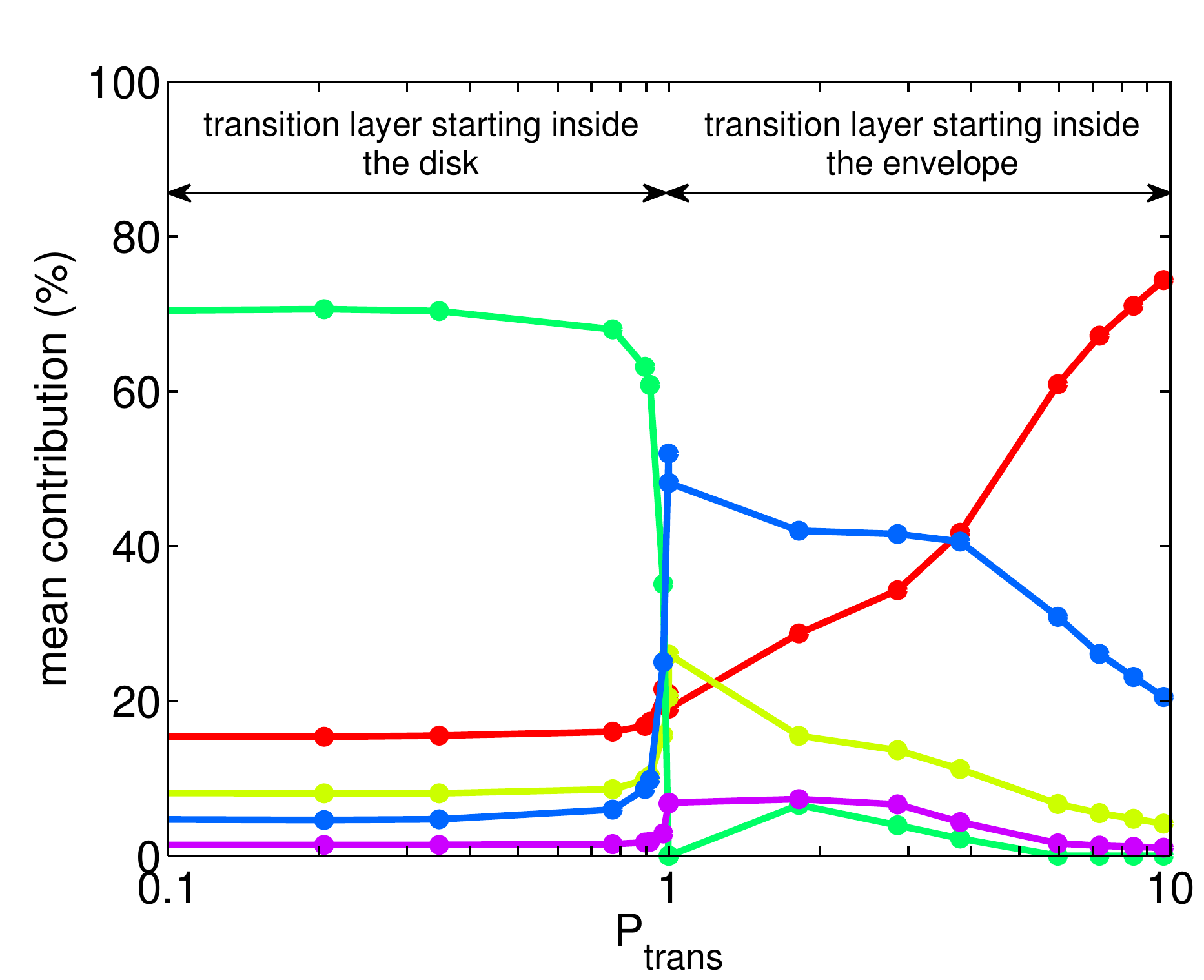}}
        }\\ %  ------- End of the first row ----------------------%
        \subfigure[Large proplyd (size of 105-600) - $G_\mathrm{0} = 2 \times 10^4$]{%
            \label{fig_heatingContribution_CCP_2e4}
            \includegraphics[width=0.45\textwidth]{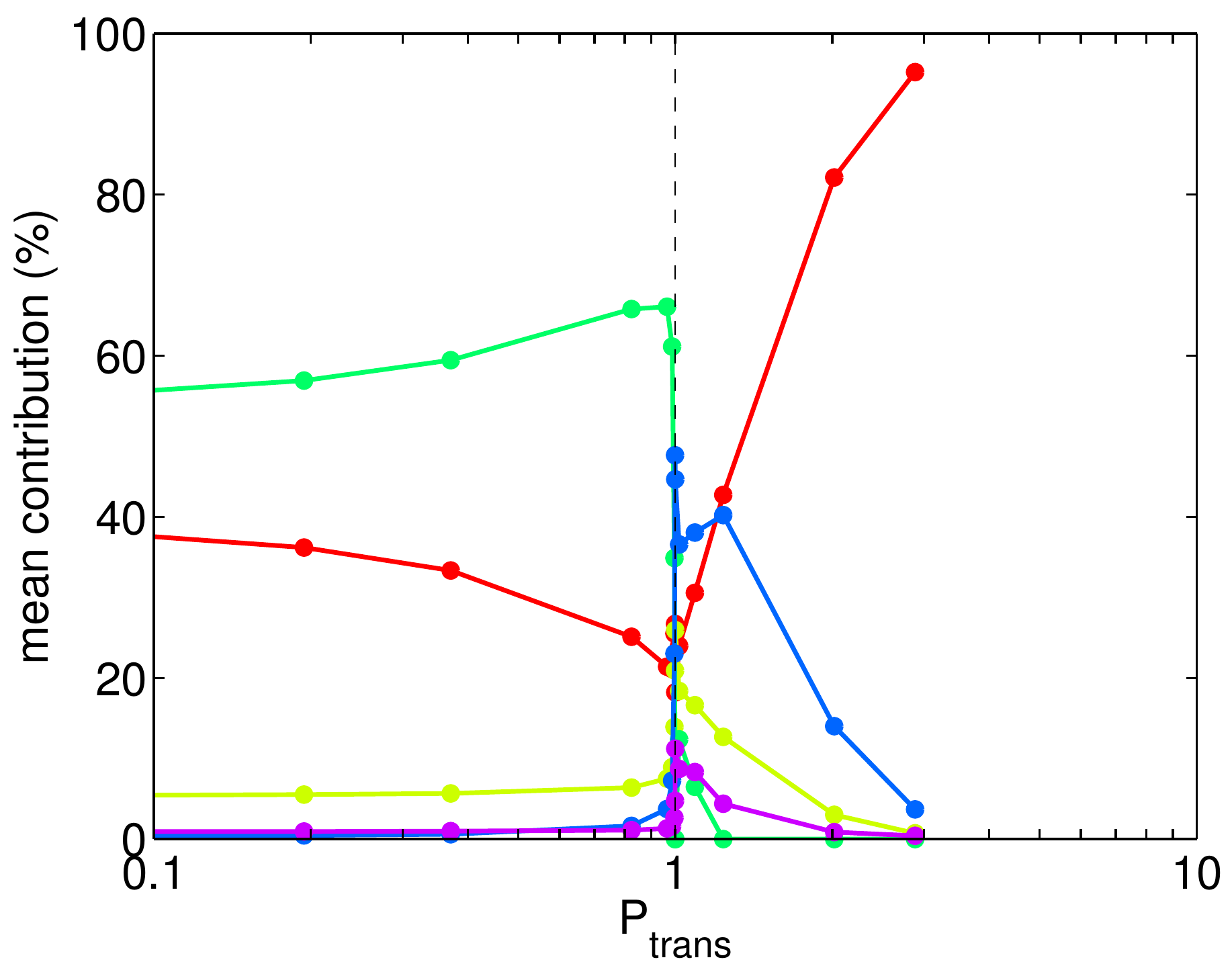}
        }%
        \subfigure[Small proplyd (size of HST10) - $G_\mathrm{0} = 2 \times 10^4$]{%
            \label{fig_heatingContribution_HST10_2e4}
            \includegraphics[width=0.45\textwidth]{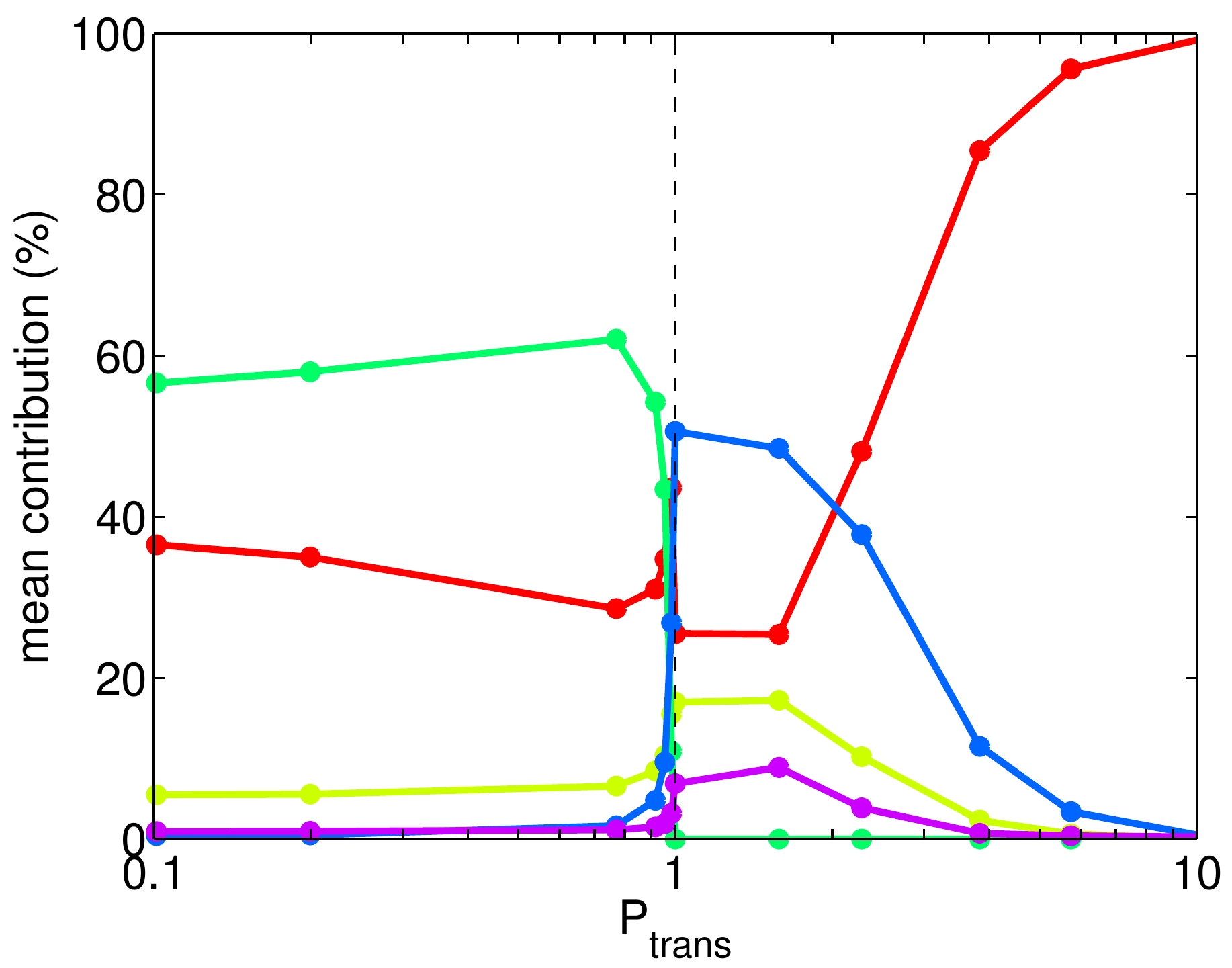}
        }\\ %  ------- End of the second row ----------------------%
        \subfigure[Large proplyd (size of 105-600) - $G_\mathrm{0} = 2 \times 10^5$]{%
            \label{fig_heatingContribution_CCP_2e5}
            \includegraphics[width=0.45\textwidth]{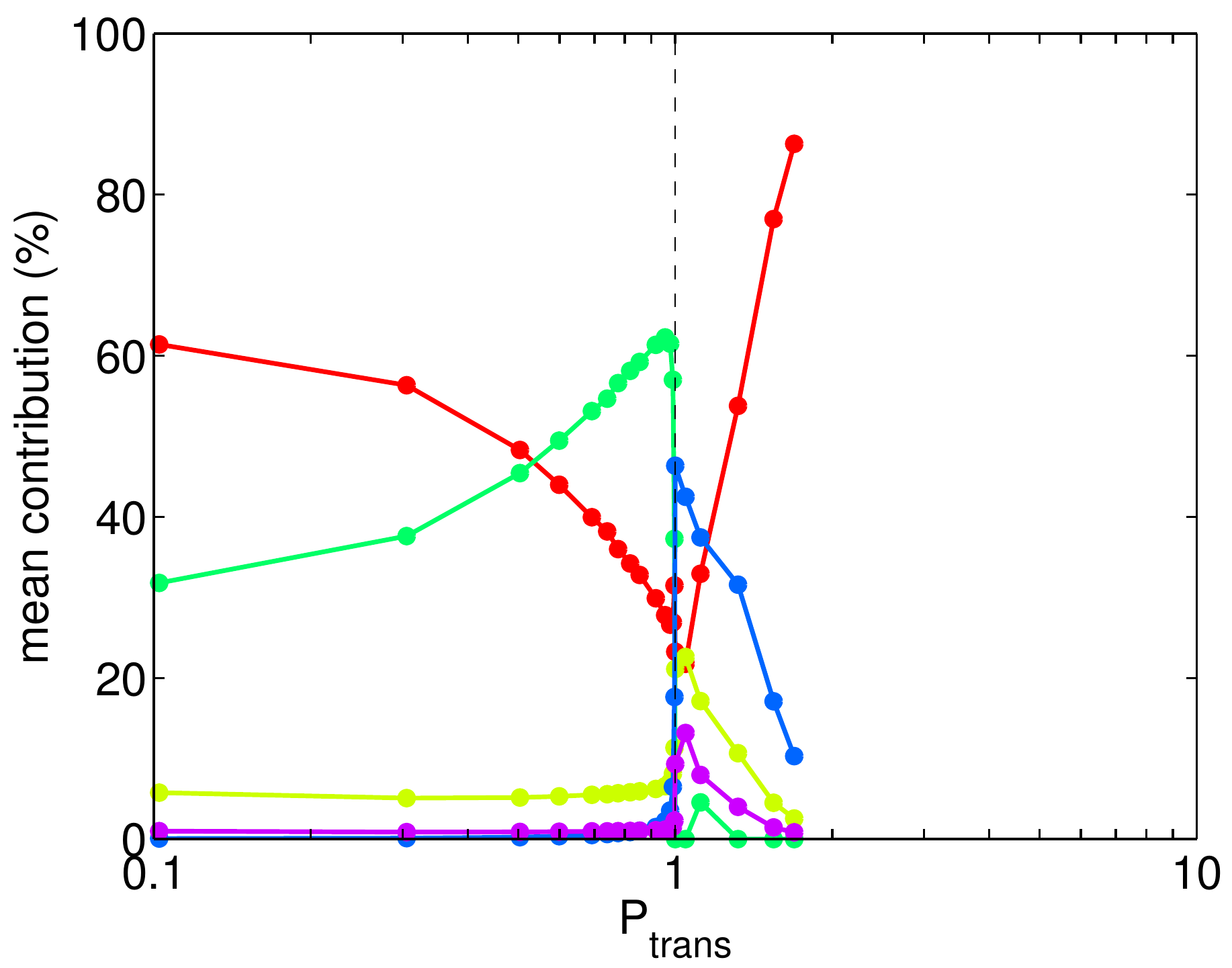}
        }%
        \subfigure[Small proplyd (size of HST10) - $G_\mathrm{0} = 2 \times 10^5$]{%
            \label{fig_heatingContribution_HST10_2e5}
            \includegraphics[width=0.45\textwidth]{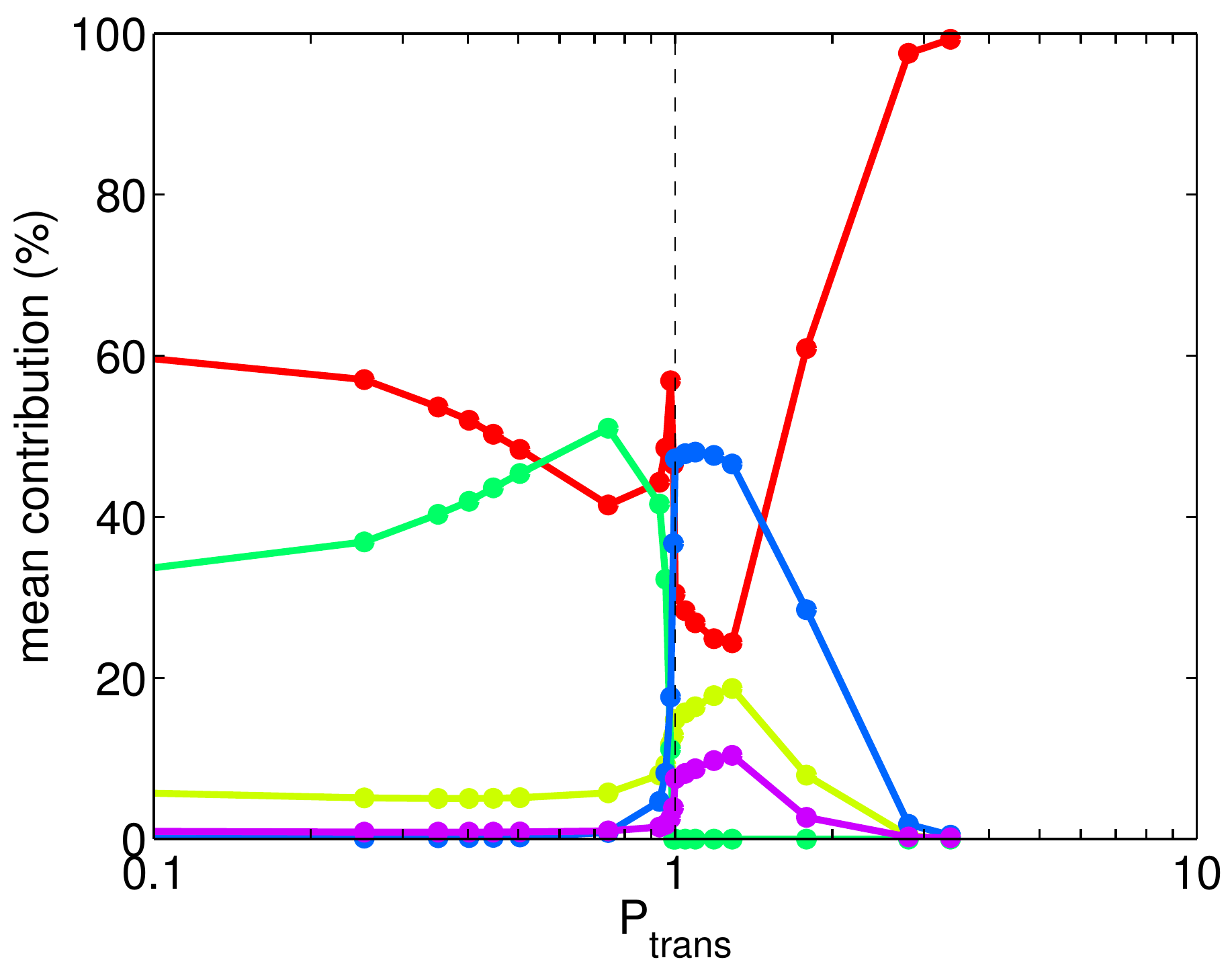}
        }%
    \end{center}
    \caption{Relative contribution of the main heating processes at the surface of the disk for different sizes of proplyd (size of 105-600 or HST10) and FUV field ($2 \times 10^3$, $2 \times 10^4$ and $2 \times 10^5$) as a function of the location of the H/H$_\mathrm{2}$ transition.}
   \label{fig_heatingContribution}
\end{figure*}

\begin{figure*}[!p]
     \begin{center}
        \subfigure[Large proplyd (size of 105-600) - $G_\mathrm{0} = 2 \times 10^3$]{%
            \label{fig_coolingContribution_CCP_2e3}
            \includegraphics[width=0.45\textwidth]{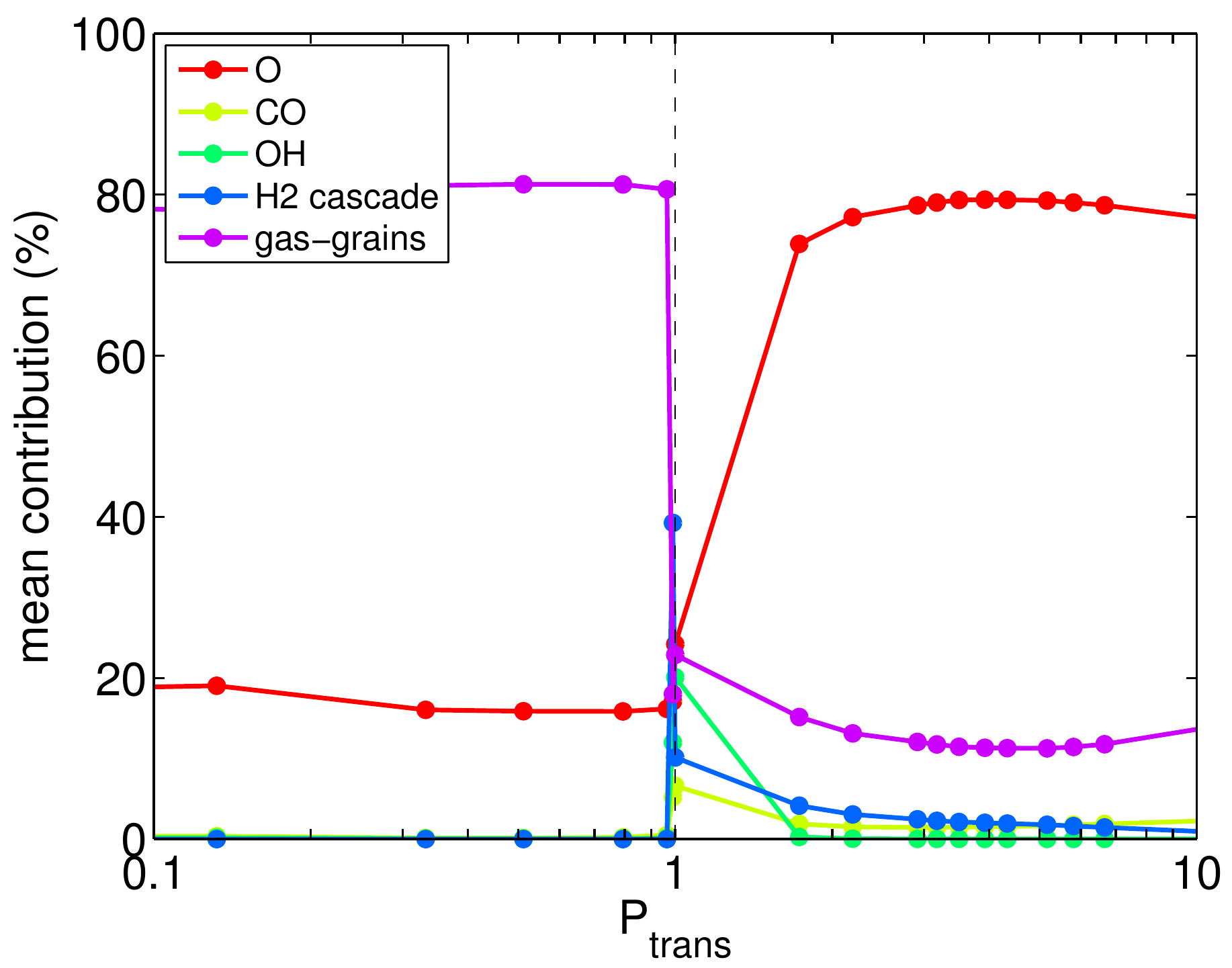}
        }%
        \subfigure[Small proplyd (size of HST10) - $G_\mathrm{0} = 2 \times 10^3$]{%
           \label{fig_coolingContribution_HST10_2e3}
           \raisebox{+0.0\height}{\includegraphics[width=0.45\textwidth]{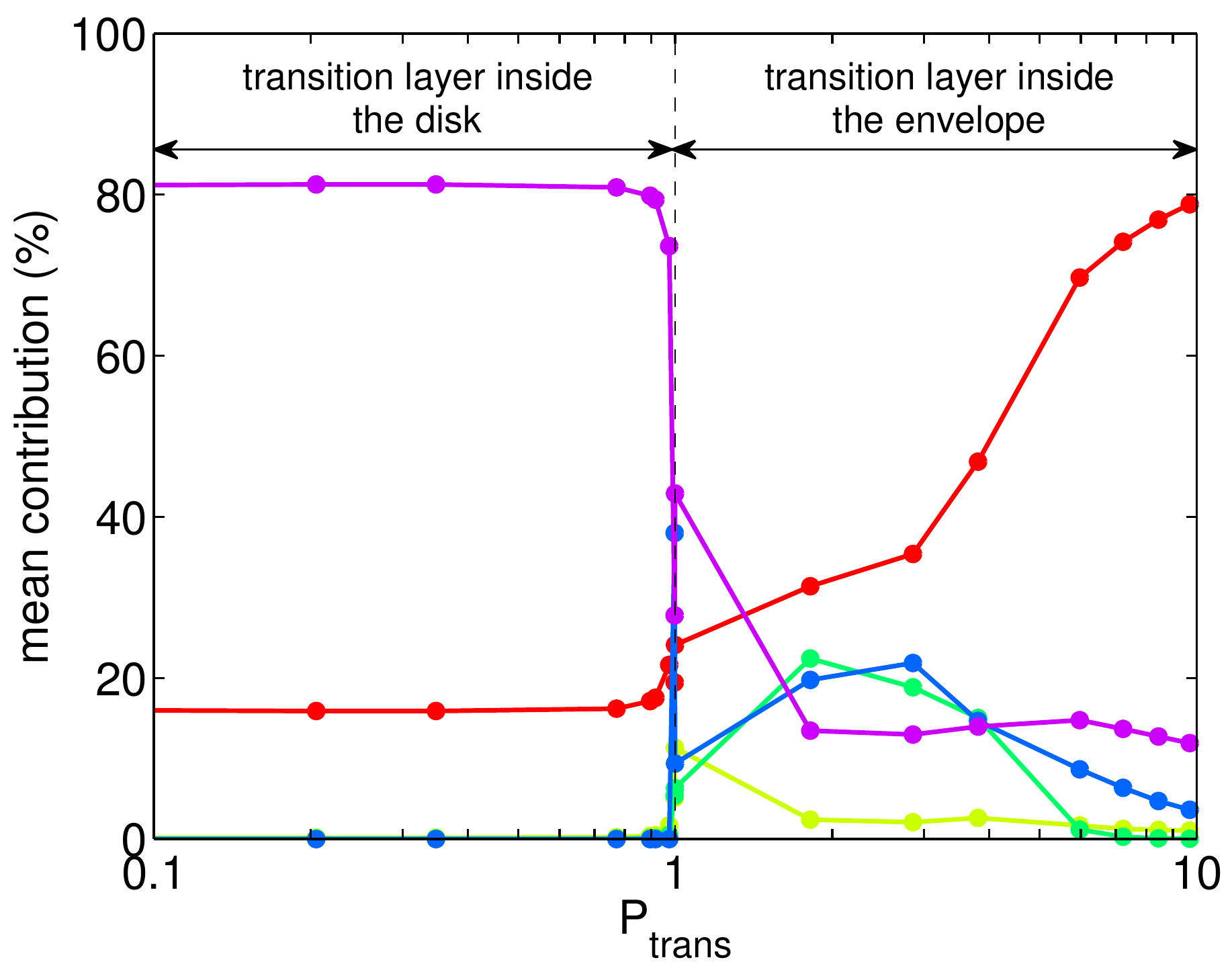}}
        }\\ %  ------- End of the first row ----------------------%
        \subfigure[Large proplyd (size of 105-600) - $G_\mathrm{0} = 2 \times 10^4$]{%
            \label{fig_coolingContribution_CCP_2e4}
            \includegraphics[width=0.45\textwidth]{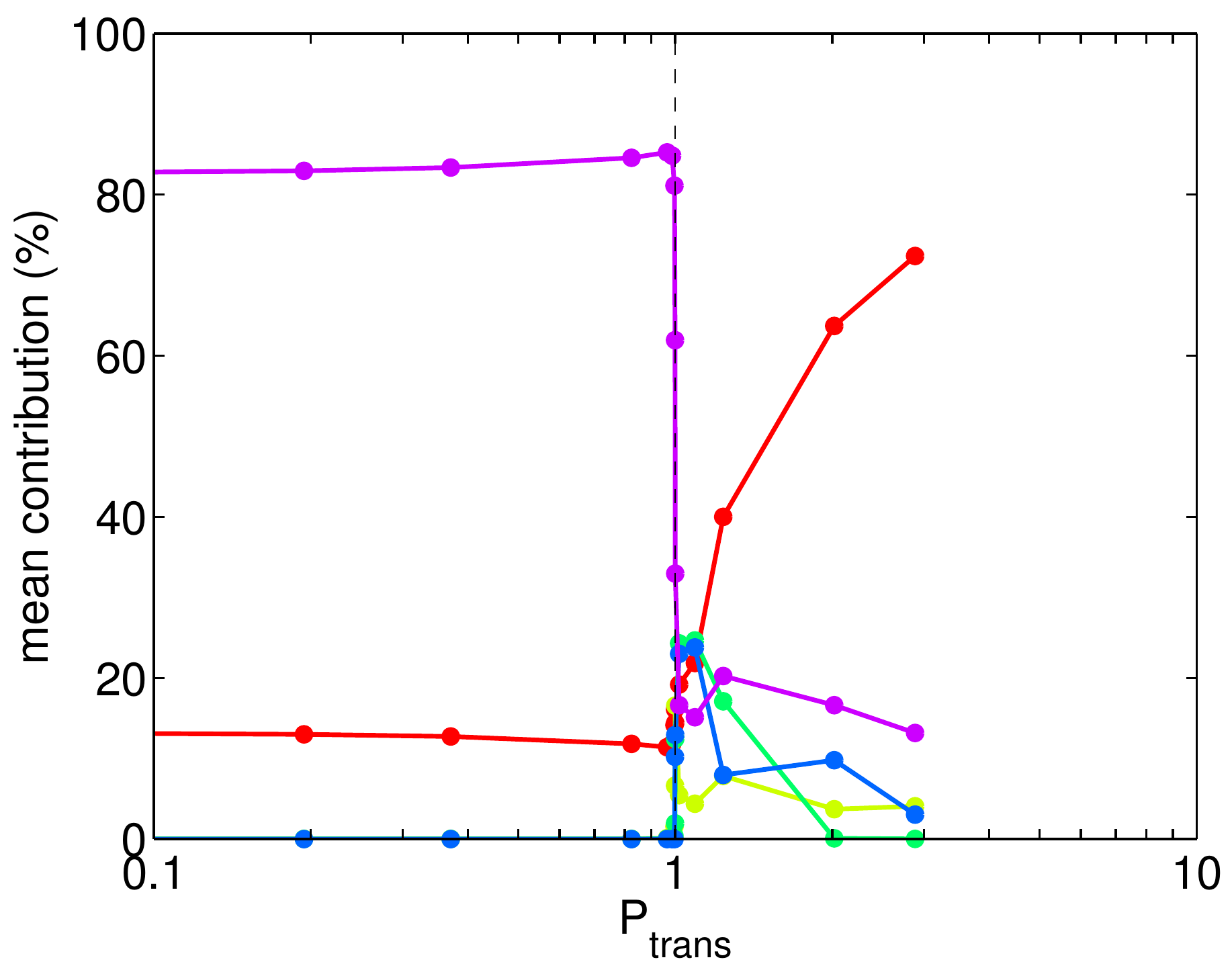}
        }%
        \subfigure[Small proplyd (size of HST10) - $G_\mathrm{0} = 2 \times 10^4$]{%
            \label{fig_coolingContribution_HST10_2e4}
            \includegraphics[width=0.45\textwidth]{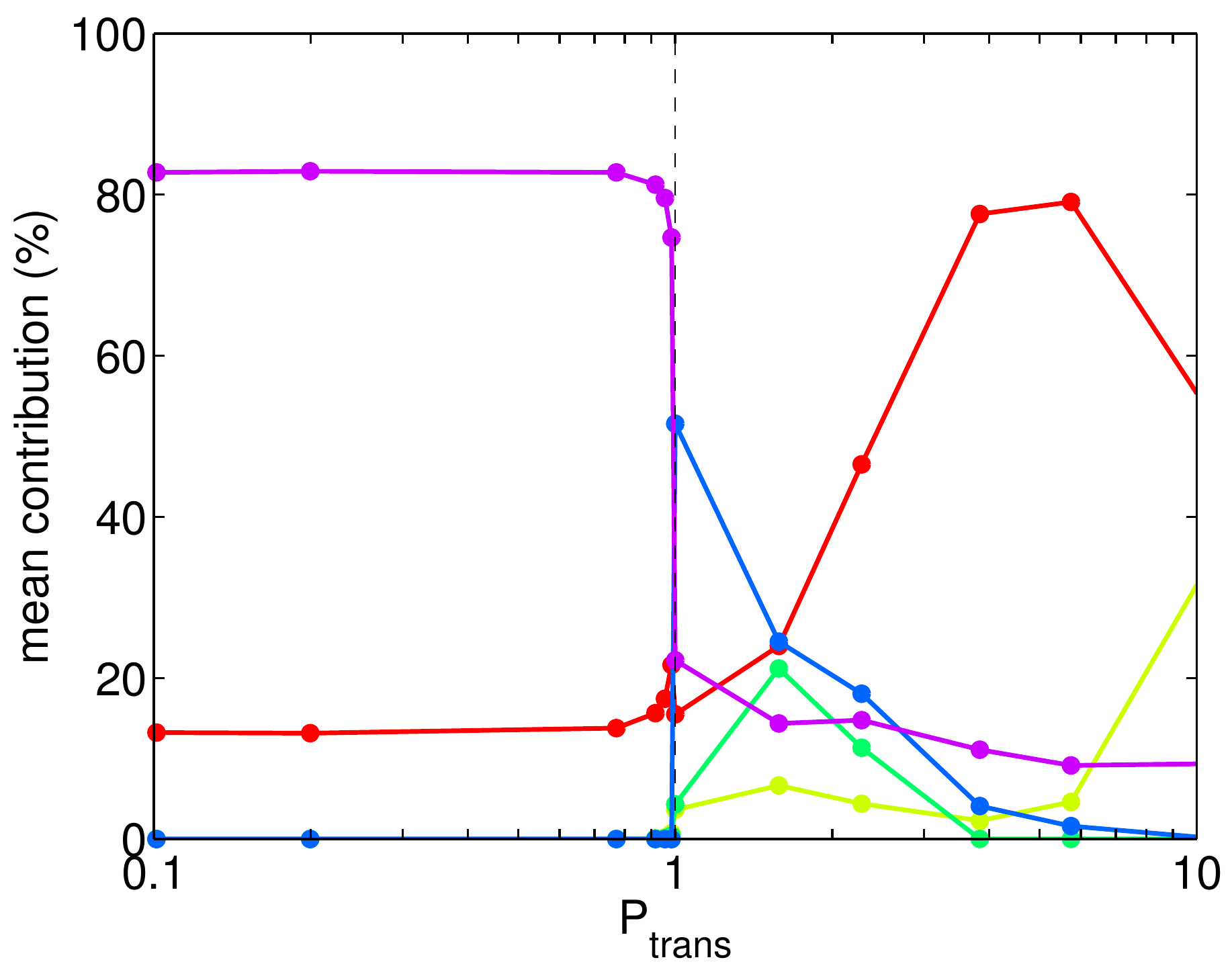}
        }\\ %  ------- End of the second row ----------------------%
        \subfigure[Large proplyd (size of 105-600) - $G_\mathrm{0} = 2 \times 10^5$]{%
            \label{fig_coolingContribution_CCP_2e5}
            \includegraphics[width=0.45\textwidth]{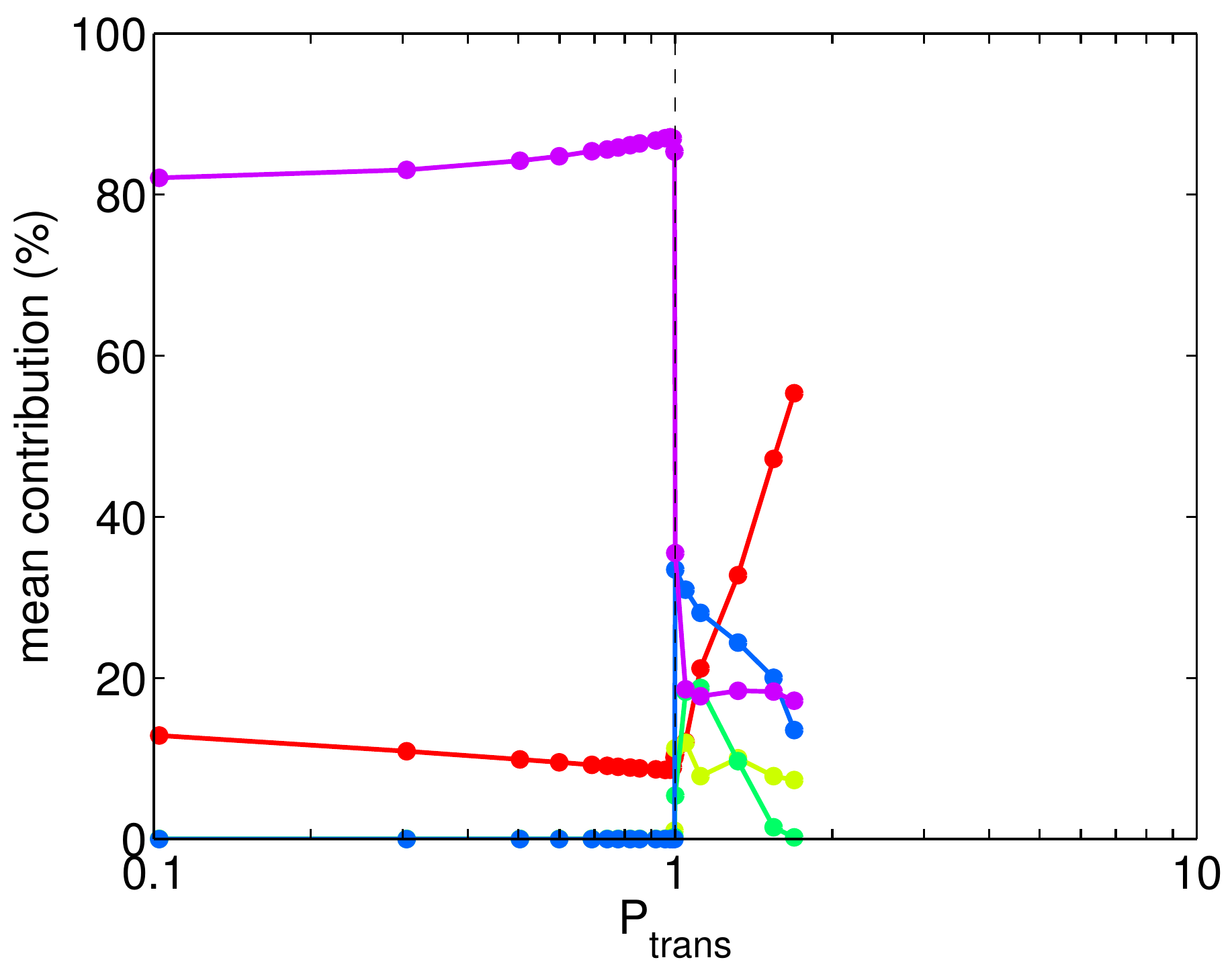}
        }%
        \subfigure[Small proplyd (size of HST10) - $G_\mathrm{0} = 2 \times 10^5$]{%
            \label{fig_coolingContribution_HST10_2e5}
            \includegraphics[width=0.45\textwidth]{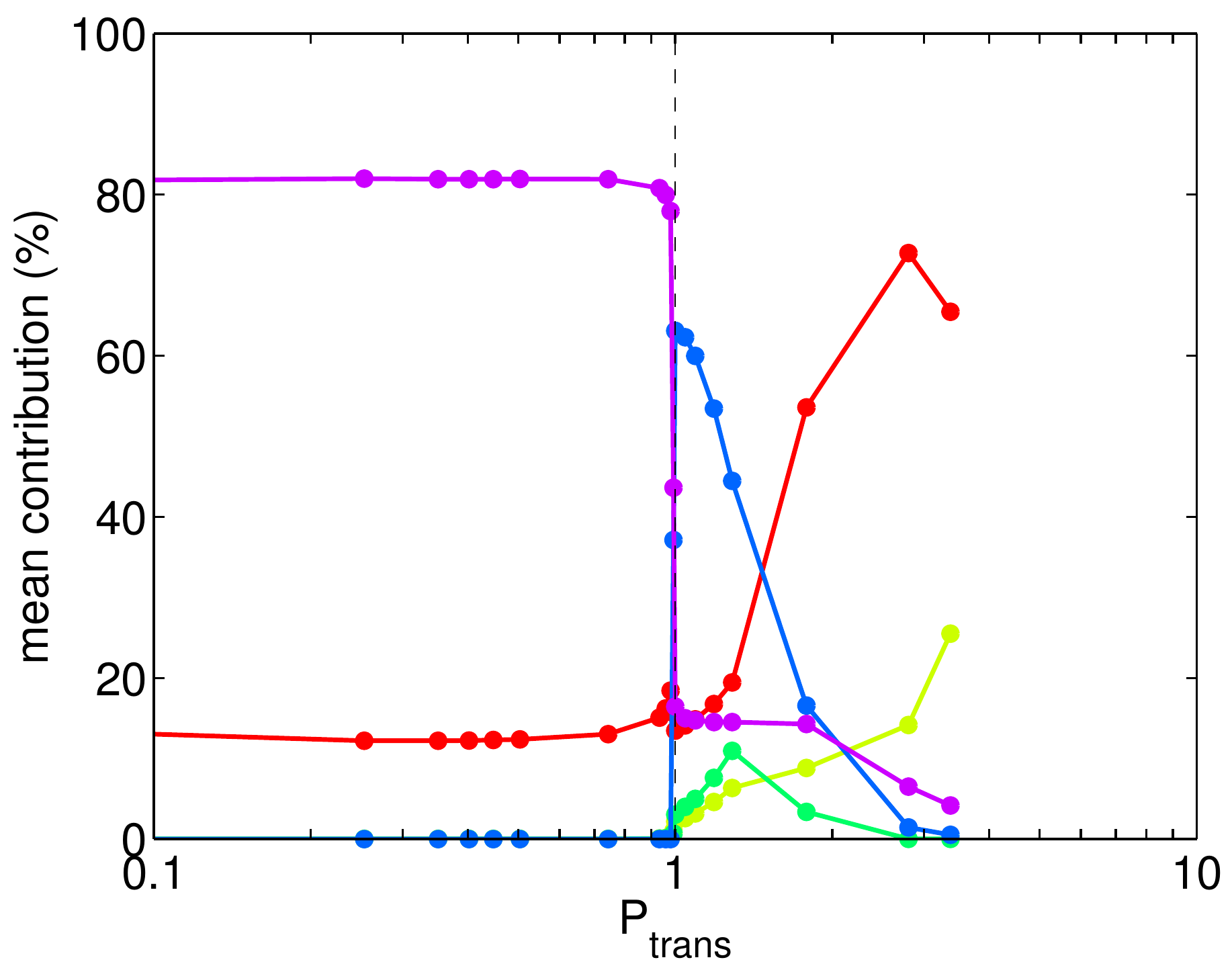}
        }%
    \end{center}
    \caption{Relative contribution of the main cooling processes at the surface of the disk for different sizes of proplyd (size of 105-600 or HST10) and FUV field ($2 \times 10^3$, $2 \times 10^4$ and $2 \times 10^5$) as a function of the location of the H/H$_\mathrm{2}$ transition.}
   \label{fig_coolingContribution}
\end{figure*}

To gain insight into the mechanisms driving this temperature behaviour, we also studied the evolution of the heating and cooling mechanisms as a function of the H/H$_2$ transition layer position. Results are given in Fig.~\ref{fig_heatingContribution} and Fig. \ref{fig_coolingContribution} respectively. The following observations can be made: when the envelope density (and visual extinction) is 
small and the transition is inside the disk ($P_\mathrm{trans} < 1$, $T \sim 1500$ - $5000$ K), the collisional de-excitation of UV-pumped H$_\mathrm{2}$ 
is very efficient and dominates the heating, except for very intense FUV fields ($G_\mathrm{0} > 10^5$) where the heating by the photo-electric effect can be even higher, while cooling is dominated by gas-grain collisions. This corresponds to the \textit{hot} regime. 
When the transition is located inside the envelope ($P_\mathrm{trans} > 1$, $T < 500$ K), the photoelectric effect on grains dominates the heating, and cooling is dominated 
by emission in fine structure lines of oxygen. This corresponds to the \textit{warm} regime which is the classical solution for thermal balance usually observed in PDRs. 
Finally, in the \textit{transitional} regime, i.e. when the transition is close to the surface but starts inside the envelope ($P_\mathrm{trans} \sim 1$, $T = 500$ - $1500$ K), 
the contribution of a collection of heating processes (exothermal reactions, H$_\mathrm{2}$ formation and direct absorption of ionizing/dissociation 
photons, exothermal chemical reactions) control the heating and cooling also occurs through multiple processes (e.g. collisional excitation of H$_\mathrm{2}$, OH and CO emission lines)

\subsection{Evidence for a specific energetic regime}

At this point it is interesting to note that for each object, the H/H$_\mathrm{2}$ transition in the best-fit model is located close to the disk surface ($P_\mathrm{trans} \gtrsim 1$, cf. Table \ref{table_mod_orionProplyds} and Fig. \ref{fig_gasTempAtSurf_slice_NH}). This is in agreement with spatially resolved observations 
which for instance show that the emission of vibrationally excited H$_\mathrm{2}$ (for HST10) and \big[OI\big] 6300 \AA{} (for HST10 and 203-506) arises 
from the disk surface or just above \citep{chen_2.12_1998, bally_externally_1998}. 
%As shown from Fig.~\ref{fig_gasTempAtSurf_slice_NH}, this lowers the disk surface temperature from a few thousand to a few hundred K. 
This implies that proplyds appear to be in the \textit{\textit{transitional} regime} with temperatures at the disk surface close to 1000 K.
Guided by Fig.~\ref{fig_gasTempAtSurf_slice_NH}, we build the following reasoning: when the isolated disk (before the formation
of the envelope) is irradiated by the external FUV photons, the  H/H$_\mathrm{2}$ transition forms inside the disk, and the temperature rises 
to some several 1000 K, driving an intense photoevaporation flow. This increases the column density of the envelope $N_\mathrm{env}$, 
dragging the H/H$_\mathrm{2}$ transition out of the disk ($P_\mathrm{trans} = N_\mathrm{env}/N_\mathrm{H-H_2} >1$) which lowers the temperature at the disk surface.
Consequently, the evaporating mass flux will decrease and so will the envelope column density due to spherical divergence.
At one point  the H/H$_\mathrm{2}$ transition will go back inside the disk ($P_\mathrm{trans}  < 1$), temperature will rise, mass-loss increase
as well as envelope column density, etc. 

Overall, this suggests that proplyds are in an equilibrium where the temperature of the disk surface is set so that mass loss 
is sufficient to keep the envelope column density to a value that maintains the H/H$_\mathrm{2}$ transition close to the disk surface. 
In this specific case, the disk surface temperature is set by a combination of numerous heating and cooling processes and 
reaches values of the order of 1000 K (for radiation fields between $G_\mathrm{0}$=2000 and $G_\mathrm{0}$=$2\times10^5$).
 
%=========== End of heating and effect of the H/H2 transition ============% 

%===========================================================%
%			   7. Photoevaporation and mass-loss rates				 %
%===========================================================%

\section{Photoevaporation and mass-loss rates} \label{sect_photoevapation_and_massloss}

At this point it is interesting to discuss photoevaporation and mass loss of proplyds in the context of our results. Namely, knowing the size of the disks and a possible range of masses for the central stars from independent studies, we can use the surface temperature and densities from our models (Tables \ref{table_model_param} and \ref{table_mod_orionProplyds}) to estimate photoevaporation rates.

%						7.1 Definitions		  					  %
%--------------------------------------------------------------------------------------------------------%

\subsection{Definitions} \label{subSect_pdrTempAndPhotoevap}

The critical radius where the kinetic energy of gas particles is equal to the gravitational potential is defined as \citep{hollenbach_photoevaporation_1994} 
\begin{equation}
r_\mathrm{g} = \frac{G M_*}{c_\mathrm{S}^2},
\label{eq_gravRadius}
\end{equation}
where $c_\mathrm{S}$ is the sound speed, i.e.
\begin{equation}
c_\mathrm{S} = \sqrt{\frac{\gamma ~\mathrm{k_B} ~T}{\mu ~m_\mathrm{H}}}.
\end{equation}
The term $\mu ~m_\mathrm{H}$ is the mean particle weight, $\gamma$ is the adiabatic index taken as 7/5 here - the value for a perfect diatomic gas - and $\mathrm{k_B}$ the Boltzmann constant. 
The mass loss rate is defined for two cases, which depend on the value of the disk radius $r_\mathrm{disk}$:
\begin{description}
\item \textbf{Supercritical case} when $r_\mathrm{disk} > r_\mathrm{g}$. \\
The mass loss rate occurs mostly from the disk surface (vertical mass-loss)
and is given by \citep{johnstone_photoevaporation_1998},
\begin{equation}
\dot{M}= \mu m_H n_\mathrm{disk} c_\mathrm{S} \times 2\pi (r_\mathrm{disk} - r_\mathrm{g})^2.
\end{equation}
\item \textbf{Subcritical case} when $r_\mathrm{disk} < r_\mathrm{g}$. \\
The mass loss rate occurs from the disk edge (radial mass-loss) and is given by \citep{adams_photoevaporation_2004},
\begin{equation}
{\dot M} = C_0~N_{C}~\mu~m_H~c_\mathrm{S}~r_g \Bigl( {r_g \over r_d} \Bigr) 
\exp\bigl[ -r_g/2r_d \bigr] \, ,
\label{eq:analytic} 
\end{equation}
where $C_0$ is a dimensionless constant of order unity and $N_{C}$ is defined as the atomic region of the PDR in their models which must be of the order of $N_\mathrm{env}$ in ours.
\end{description}

%For the rather large temperatures found at the disk surface (Table \ref{table_mod_orionProplyds})
%in this study and a 1 solar mass star, S_\mathrm{evap} roughly equal to the total disk surface. 

%					7.3 Carina candidate proplyd		  			  %
%--------------------------------------------------------------------------------------------------------% 
 
 \subsection{Carina candidate proplyd 105-600} \label{subSect_pdrTempAndPhotoevap_CCP}

For 105-600, the disk (or candidate disk) size is unknown. The value of $S_\mathrm{disk}$ derived from the model 
can be converted in an equivalent radius considering that the disk is in the plane of the sky. 
This yields a value of about 1.1\arcsec or 2500 AU. The temperature at the surface of the disk is estimated to be about 1100 K.
For such a large radius,  $r_\mathrm{disk} > r_\mathrm{g}$ for masses of central star up to 30 M$_\odot$,
and the \textit{supercritical} regime prevails everywhere. We derive for this case a mass-loss rate  $ \dot{M}= 7.5 \times 10^\mathrm{-5}$ M$_\sun/$year. 
The mass-loss at the ionisation front can be derived from the H\textalpha{} emission,
and we find a value of the order of $2 \times 10^\mathrm{-6}$ M$_\sun/$year, about 40 times smaller. %\textbf{The observed value is an upper limit since the H\textalpha{} emission is not isolated in the HST observation while it is expected to be dominant and so the real value close to this estimate.}
This discrepancy could be explained if the envelope is in a phase where it is gaining mass  
(see Sect.~\ref{subSect_hh2TransitionLocation} ) while depleting the disk. 
We note that the Orion proplyds, which seem to be at equilibrium (mass-loss at the ionisation is roughly equal
to the "disk" mass loss computed here, see Sect. \ref{subSect_pdrTempAndPhotoevap_Orion}), generally exhibit a tear-drop shaped envelope as 
HST10 (Fig. \ref{fig_halpha_HST10}) as the consequence of this equilibrium. It is thus possible that
the sharper shape of 105-600 (Fig. \ref{fig_halpha_105-600}) results from an envelope that is still growing 
in mass. Since the growth phase is expected to be quick, e.g. forming a 0.05 M$_\sun$ envelope with 
such a rate would take few hundred years, 105-600 would thus be a very young object. 
Alternatively, it is possible that the disk radius we consider here is too large 
by about an order of magnitude. This would however pose a serious problem to our model
when trying to reproduce the high-$J$ CO lines. Since a protoplanetary disk with a radius of 2500 AU appears unrealistic, what we call disk in this case could most likely refer to a young forming disk from the original cloud material \citep[similar to other large "pseudo-disks" observed, e.g. as in][]{quanz_large_2010}. Further observations at (sub)millimeter wavelengths to resolve 
the structure and kinematics of the disk will be required to settle this issue.

%						7.2 Orion proplyds		  				  %
%--------------------------------------------------------------------------------------------------------%

\subsection{Orion Proplyds} \label{subSect_pdrTempAndPhotoevap_Orion}

For HST10 and 244-440, \citet{henney_keck_1999} derived the mass-loss rates at the ionisation front
from spectroscopic observations of ionized gas tracers and found values of $\left(7\pm5\right)\times10^{-7}$ 
and $\left(15\pm7\right)\times10^{-7}$ M$_\sun$\,yr$^\mathrm{-1}$ respectively. For 203-506, since the envelope ionisation front is not 
observed because the disk is embedded in neutral gas, the mass-loss cannot be derived from ionised gas tracers.

For the Orion proplyds whose disk outer radii are known \citep{vicente_size_2005}, we can derive the mass-loss rates
from our model using the analytical prescription presented above.
The masses of their central stars are not precisely known but taking into account the stellar mass function of the cluster \citep[e.g.][]{hillenbrand_constraints_2000}, they are very likely about 0.1 M$_\sun$ or less. A surface temperature of the order of 1000 K, as obtained from our study, would result in a gravitational radius of 20 AU, or less, and implies the \textit{supercritical} case ($r_\mathrm{disk} > r_\mathrm{g}$) for all disks. The \textit{subcritical} case is still possible, but unlikely because we need to assume that central stars have a mass of about 1 M$_\sun$, resulting in a gravitational radius of the order of 200 AU. We nevertheless consider both cases in our calculations.
The mass-loss rates for each case are given in Table  \ref{table_mod_orionProplyds}. 
In the \textit{subcritical} regime, we find values of the order a few $10^{-8}$ M$_{\sun}$/yr, in agreement
with the results of \citet{adams_photoevaporation_2004}, but which are too low compared 
to the observed values. On the other hand, the \textit{supercritical} values of mass-loss are in better
agreement with the mass loss derived at the ionisation front. Hence, to reconcile the photoevaporation
rates derived from our best fit models with those derived from the observations by \citet{henney_keck_1999},
one must assume that the flows are in a \textit{supercritical} regime. This is compatible with the idea that central stars have masses well below one solar mass.
If this is indeed the case, given that disk masses do not significantly exceed $0.01$ M$_{\sun}$ in mass, this implies short lifetimes for these disks
(of the order of several 10$^4$ years) which suggests that we should not see so many of them,
or that the ionizing star \textTheta$^1$ Ori C is particularly young, a conundum that is known as the "proplyd lifetime problem"  \citep{henney_keck_1999}. 
 \citet{clarke_photoevaporation_2007} studied in details the dynamical evolution of proplyds assuming they are in
 the \textit{subcritical} regime and found that their lifetime could in this case be relatively large,
 solving this issue.  In the light of the results presented here it appears that the \textit{subcritical} 
 assumption may not be correct. Instead our results are in line with the idea that the Orion Nebula 
 is young, i.e. that $\Theta^1$ Ori C is less than a few 10$^4$ years old, and that proplyds are indeed 
 rare objects which we happen to see in Orion by chance. Another possibility is that the \textit{supercritical} regime is not sustained at the same rate during the whole proplyd life and declines with time so that the lifetimes could be not so short.

%=================== End of photoevaporation ====================%

%===========================================================%
%						8. Conclusions					 		 %
%===========================================================% 

\section{Conclusions} \label{sect_conclusions}

The main results presented in this paper are:
\begin{itemize}
\item{We have presented the first far-infrared observations of dense PDR tracers emitted by proplyds, obtained with the \textit{Herschel space observatory} and ALMA.}
\item{Based on the detailed Meudon PDR code, we have developed a 1D-model for the PDR of a proplyd to predict the line emission arising from it.}
\item{This model successfully reproduces most lines, with the exception of HCN (4-3) in 105-600. The model is also somewhat lower than 
the observed intensities of \big[OI\big] 6300 \AA{} (and possibly \big[CII\big] 158 \textmu{}m) in HST10 and 244-440. This tracer is most sensitive to the densest gas (or the most diffuse for \big[CII\big] 158 \textmu{}m) in proplyds which are not well described in our simplified model.
\item{For all the sources, at the disk surface, we find densities of $0.7$ to $ 4\times 10^6$ cm$^{-3}$ and temperatures of 1000-1300 K.}}
\item{We find that the position of the H/H$_\mathrm{2}$ transition layer is the critical parameter determining the disk surface temperature.}  
\item{Our results suggest that proplyds are in a self-regulated regime where the temperature of the disk surface is set so that mass loss 
keeps the envelope column density sufficient to bring the H/H$_\mathrm{2}$ transition out of the disk, yet close to its surface.}
\item{We find that gas energetics in this specific case results from a complex combination of several heating and cooling
mechanisms, hence it can not be captured by the classical solution where photoelectric effect on grains dominates the heating
and \big[OI\big] and \big[CII\big] far infrared emission dominates the cooling.}
%\item{The mass loss rates derived from our models are large (of the order of a few 10$^{-7}$ M$_{\sun}$/yr for the Orion proplyds and a few 
%10$^{-5}$ M$_{\sun}$/yr for 105-600). The values are consistent with spectroscopic studies of the ionisation front in these objects and strengthen the existence 
%of the proplyd "lifetime problem".}
\item{We derive mass loss rates that are large (of the order of a few 10$^{-7}$ M$_{\sun}$/yr for the Orion proplyds and a few 
10$^{-5}$ M$_{\sun}$/yr for 105-600). Most of them are consistent with spectroscopic studies of the ionisation front in theses objects and suggest that proplyds undergo \textit{supercritical} photoevaporation} and indeed may have a short lifetime.
\item{To understand proplyd evolution, in a next step, it is necessary to couple hydrodynamic simulations of a \textit{supercritical} evaporation flow with a correct treatment of PDR physics and chemistry.}
\item{Concomitant with the development of models, spatially resolved observations of molecular lines close to the surface of the disks would help to constrain the photoevaporative flows and to understand the evolution of a proplyd. Especially, the high angular resolution provided by ALMA and the future JWST will enable us to resolve photoevaporative flows and study their properties and dynamics.}
\end{itemize}

 %==================== End of conclusions ======================%

%-------------------------------------------------------------------

%\begin{acknowledgements}
%      Part of this work was supported by the German
%      \emph{Deut\-sche For\-schungs\-ge\-mein\-schaft, DFG\/} project
%      number Ts~17/2--1.
%\end{acknowledgements}
\begin{acknowledgements} 
This work was supported by the CNRS program "Physique et Chimie du Milieu Interstellaire" (PCMI). The research leading to these results has received funding from the European Research Council under the European Union's Seventh Framework Programme (FP/2007-2013)  ERC-2013-SyG, Grant Agreement n. 610256 NANOCOSMOS. Javier R. Goicoechea also thanks the Spanish MINECO for funding support under grant AYA2012-32032.
This paper makes use of the following ALMA data: ADS/JAO.ALMA\#2012.1.00352.S. ALMA is a partnership of ESO (representing its member states), NSF (USA) and NINS (Japan), together with NRC (Canada), NSC and ASIAA (Taiwan), and KASI (Republic of Korea), in cooperation with the Republic of Chile. The Joint ALMA Observatory is operated by ESO, AUI/NRAO and NAOJ. 

\end{acknowledgements}

%===========================================================%
%						    References							 %
%===========================================================% 

% \newpage %%%%% REMOVE THAT AFTERWARDS
% Forum about placing references after all figures
% http://tex.stackexchange.com/questions/28898/forcing-bibliography-to-the-end
\bibliographystyle{aa} % style aa.bst
\bibliography{champion_etal_2017} % your references Yourfile.bib

\begin{thebibliography}{79}
\expandafter\ifx\csname natexlab\endcsname\relax\def\natexlab#1{#1}\fi

\bibitem[{Adams {et~al.}(2004)Adams, Hollenbach, Laughlin, \&
  Gorti}]{adams_photoevaporation_2004}
Adams, F.~C., Hollenbach, D., Laughlin, G., \& Gorti, U. 2004, The
  Astrophysical Journal, 611, 360

\bibitem[{Alexander {et~al.}(2014)Alexander, Pascucci, Andrews, Armitage, \&
  Cieza}]{alexander_dispersal_2014}
Alexander, R., Pascucci, I., Andrews, S., Armitage, P., \& Cieza, L. 2014,
  Protostars and Planets VI, 475

\bibitem[{Bakes \& Tielens(1994)}]{bakes_photoelectric_1994}
Bakes, E. L.~O. \& Tielens, A. G. G.~M. 1994, The Astrophysical Journal, 427,
  822

\bibitem[{Bally {et~al.}(2000)Bally, O'Dell, \& McCaughrean}]{bally_disks_2000}
Bally, J., O'Dell, C.~R., \& McCaughrean, M.~J. 2000, The Astronomical Journal,
  119, 2919

\bibitem[{Bally {et~al.}(1998)Bally, Sutherland, Devine, \&
  Johnstone}]{bally_externally_1998}
Bally, J., Sutherland, R.~S., Devine, D., \& Johnstone, D. 1998, The
  Astronomical Journal, 116, 293

\bibitem[{Berné \& Tielens(2012)}]{berne_formation_2012}
Berné, O. \& Tielens, A. G. G.~M. 2012, Proceedings of the National Academy of
  Science, 109, 401

\bibitem[{Boss \& Durisen(2005)}]{boss_sources_2005}
Boss, A.~P. \& Durisen, R.~H. 2005, in Chondrites and the {Protoplanetary}
  {Disk}, Vol. 341, 821

\bibitem[{Bron {et~al.}(2014)Bron, Le~Bourlot, \& Le~Petit}]{bron_surface_2014}
Bron, E., Le~Bourlot, J., \& Le~Petit, F. 2014, Astronomy and Astrophysics,
  569, A100

\bibitem[{Bron {et~al.}(2016)Bron, Le~Petit, \&
  Le~Bourlot}]{bron_efficient_2016}
Bron, E., Le~Petit, F., \& Le~Bourlot, J. 2016, Astronomy and Astrophysics,
  588, A27

\bibitem[{Chen {et~al.}(1998)Chen, Bally, O'Dell, McCaughrean, Thompson, Rieke,
  Schneider, \& Young}]{chen_2.12_1998}
Chen, H., Bally, J., O'Dell, C.~R., {et~al.} 1998, The Astrophysical Journal
  Letters, 492, L173

\bibitem[{Clarke(2007)}]{clarke_photoevaporation_2007}
Clarke, C.~J. 2007, Monthly Notices of the Royal Astronomical Society, 376,
  1350

\bibitem[{Clarke {et~al.}(2001)Clarke, Gendrin, \&
  Sotomayor}]{clarke_dispersal_2001}
Clarke, C.~J., Gendrin, A., \& Sotomayor, M. 2001, Monthly Notices of the Royal
  Astronomical Society, 328, 485

\bibitem[{Compiègne {et~al.}(2011)Compiègne, Verstraete, Jones, Bernard,
  Boulanger, Flagey, Le~Bourlot, Paradis, \& Ysard}]{compiegne_global_2011}
Compiègne, M., Verstraete, L., Jones, A., {et~al.} 2011, Astronomy and
  Astrophysics, 525, A103

\bibitem[{de~Graauw {et~al.}(2010)de~Graauw, Helmich, Phillips, Stutzki, Caux,
  Whyborn, Dieleman, Roelfsema, Aarts, Assendorp, Bachiller, Baechtold, Barcia,
  Beintema, Belitsky, Benz, Bieber, Boogert, Borys, Bumble, Caïs, Caris,
  Cerulli-Irelli, Chattopadhyay, Cherednichenko, Ciechanowicz, Coeur-Joly,
  Comito, Cros, de~Jonge, de~Lange, Delforges, Delorme, den Boggende, Desbat,
  Diez-González, di~Giorgio, Dubbeldam, Edwards, Eggens, Erickson, Evers,
  Fich, Finn, Franke, Gaier, Gal, Gao, Gallego, Gauffre, Gill, Glenz, Golstein,
  Goulooze, Gunsing, Güsten, Hartogh, Hatch, Higgins, Honingh, Huisman,
  Jackson, Jacobs, Jacobs, Jarchow, Javadi, Jellema, Justen, Karpov, Kasemann,
  Kawamura, Keizer, Kester, Klapwijk, Klein, Kollberg, Kooi, Kooiman, Kopf,
  Krause, Krieg, Kramer, Kruizenga, Kuhn, Laauwen, Lai, Larsson, Leduc, Leinz,
  Lin, Liseau, Liu, Loose, López-Fernandez, Lord, Luinge, Marston,
  Martín-Pintado, Maestrini, Maiwald, McCoey, Mehdi, Megej, Melchior, Meinsma,
  Merkel, Michalska, Monstein, Moratschke, Morris, Muller, Murphy, Naber,
  Natale, Nowosielski, Nuzzolo, Olberg, Olbrich, Orfei, Orleanski, Ossenkopf,
  Peacock, Pearson, Peron, Phillip-May, Piazzo, Planesas, Rataj, Ravera,
  Risacher, Salez, Samoska, Saraceno, Schieder, Schlecht, Schlöder,
  Schmülling, Schultz, Schuster, Siebertz, Smit, Szczerba, Shipman, Steinmetz,
  Stern, Stokroos, Teipen, Teyssier, Tils, Trappe, van Baaren, van Leeuwen,
  van~de Stadt, Visser, Wildeman, Wafelbakker, Ward, Wesselius, Wild, Wulff,
  Wunsch, Tielens, Zaal, Zirath, Zmuidzinas, \&
  Zwart}]{de_graauw_herschel-heterodyne_2010}
de~Graauw, T., Helmich, F.~P., Phillips, T.~G., {et~al.} 2010, Astronomy and
  Astrophysics, 518, L6

\bibitem[{Facchini {et~al.}(2016)Facchini, Clarke, \&
  Bisbas}]{facchini_external_2016}
Facchini, S., Clarke, C.~J., \& Bisbas, T.~G. 2016, Monthly Notices of the
  Royal Astronomical Society, 457, 3593

\bibitem[{Fitzpatrick \& Massa(1988)}]{fitzpatrick_analysis_1988}
Fitzpatrick, E.~L. \& Massa, D. 1988, The Astrophysical Journal, 328, 734

\bibitem[{Flower \& Pineau Des~Forêts(2010)}]{flower_excitation_2010}
Flower, D.~R. \& Pineau Des~Forêts, G. 2010, Monthly Notices of the Royal
  Astronomical Society, 406, 1745

\bibitem[{Geers {et~al.}(2006)Geers, Augereau, Pontoppidan, Dullemond, Visser,
  Kessler-Silacci, Evans, van Dishoeck, Blake, Boogert, Brown, Lahuis, \&
  Merín}]{geers_C2D_2006}
Geers, V.~C., Augereau, J.-C., Pontoppidan, K.~M., {et~al.} 2006, Astronomy and
  Astrophysics, 459, 545

\bibitem[{Goicoechea {et~al.}(2016)Goicoechea, Pety, Cuadrado, Cernicharo,
  Chapillon, Fuente, Gerin, Joblin, Marcelino, \&
  Pilleri}]{goicoechea_compression_2016}
Goicoechea, J.~R., Pety, J., Cuadrado, S., {et~al.} 2016, Nature, 537, 207

\bibitem[{Goicoechea {et~al.}(2015)Goicoechea, Teyssier, Etxaluze, Goldsmith,
  Ossenkopf, Gerin, Bergin, Black, Cernicharo, Cuadrado, Encrenaz, Falgarone,
  Fuente, Hacar, Lis, Marcelino, Melnick, Müller, Persson, Pety, Röllig,
  Schilke, Simon, Snell, \& Stutzki}]{goicoechea_velocity-resolved_2015}
Goicoechea, J.~R., Teyssier, D., Etxaluze, M., {et~al.} 2015, The Astrophysical
  Journal, 812, 75

\bibitem[{Gorti \& Hollenbach(2009)}]{gorti_photoevaporation_2009}
Gorti, U. \& Hollenbach, D. 2009, The Astrophysical Journal, 690, 1539

\bibitem[{Gorti {et~al.}(2015)Gorti, Liseau, Sandor, \&
  Clarke}]{gorti_disk_2015}
Gorti, U., Liseau, R., Sandor, Z., \& Clarke, C. 2015, ArXiv e-prints, 1512,
  arXiv:1512.04622

\bibitem[{Habing(1968)}]{habing_interstellar_1968}
Habing, H.~J. 1968, Bulletin of the Astronomical Institutes of the Netherlands,
  19, 421

\bibitem[{Haisch {et~al.}(2001)Haisch, Lada, \& Lada}]{haisch_disk_2001}
Haisch, Jr., K.~E., Lada, E.~A., \& Lada, C.~J. 2001, The Astrophysical Journal
  Letters, 553, L153

\bibitem[{Henney \& O'Dell(1999)}]{henney_keck_1999}
Henney, W.~J. \& O'Dell, C.~R. 1999, The Astronomical Journal, 118, 2350

\bibitem[{Hillenbrand \& Carpenter(2000)}]{hillenbrand_constraints_2000}
Hillenbrand, L.~A. \& Carpenter, J.~M. 2000, The Astrophysical Journal, 540,
  236

\bibitem[{Hollenbach {et~al.}(1994)Hollenbach, Johnstone, Lizano, \&
  Shu}]{hollenbach_photoevaporation_1994}
Hollenbach, D., Johnstone, D., Lizano, S., \& Shu, F. 1994, The Astrophysical
  Journal, 428, 654

\bibitem[{Hollenbach \& McKee(1989)}]{hollenbach_molecule_1989}
Hollenbach, D. \& McKee, C.~F. 1989, The Astrophysical Journal, 342, 306

\bibitem[{Johnstone {et~al.}(1998)Johnstone, Hollenbach, \&
  Bally}]{johnstone_photoevaporation_1998}
Johnstone, D., Hollenbach, D., \& Bally, J. 1998, The Astrophysical Journal,
  499, 758

\bibitem[{Konigl \& Pudritz(2000)}]{konigl_disk_2000}
Konigl, A. \& Pudritz, R.~E. 2000, Protostars and Planets IV, 759

\bibitem[{Le~Bourlot {et~al.}(2012)Le~Bourlot, Le~Petit, Pinto, Roueff, \&
  Roy}]{le_bourlot_surface_2012}
Le~Bourlot, J., Le~Petit, F., Pinto, C., Roueff, E., \& Roy, F. 2012, Astronomy
  and Astrophysics, 541, A76

\bibitem[{Le~Petit {et~al.}(2006)Le~Petit, Nehmé, Le~Bourlot, \&
  Roueff}]{le_petit_model_2006}
Le~Petit, F., Nehmé, C., Le~Bourlot, J., \& Roueff, E. 2006, The Astrophysical
  Journal Supplement Series, 164, 506

\bibitem[{Marconi {et~al.}(1998)Marconi, Testi, Natta, \&
  Walmsley}]{marconi_near_1998}
Marconi, A., Testi, L., Natta, A., \& Walmsley, C.~M. 1998, Astronomy and
  Astrophysics, 330, 696

\bibitem[{Menten {et~al.}(2007)Menten, Reid, Forbrich, \&
  Brunthaler}]{menten_distance_2007}
Menten, K.~M., Reid, M.~J., Forbrich, J., \& Brunthaler, A. 2007, Astronomy and
  Astrophysics, 474, 515

\bibitem[{Mesa-Delgado {et~al.}(2016)Mesa-Delgado, Zapata, Henney, Puzia, \&
  Tsamis}]{mesa-delgado_protoplanetary_2016}
Mesa-Delgado, A., Zapata, L., Henney, W.~J., Puzia, T.~H., \& Tsamis, Y.~G.
  2016, ArXiv e-prints, 1605, arXiv:1605.08809

\bibitem[{Noel(2003)}]{noel_dynamique_2003}
Noel, B. 2003, Thèse de doctorat, Université Pierre et Marie Curie, Paris,
  France

\bibitem[{Noel {et~al.}(2005)Noel, Joblin, Maillard, \&
  Paumard}]{noel_new_2005}
Noel, B., Joblin, C., Maillard, J.~P., \& Paumard, T. 2005, Astronomy and
  Astrophysics, 436, 569

\bibitem[{O'dell {et~al.}(1993)O'dell, Wen, \& Hu}]{odell_discovery_1993}
O'dell, C.~R., Wen, Z., \& Hu, X. 1993, The Astrophysical Journal, 410, 696

\bibitem[{O'dell \& Wong(1996)}]{odell_hubble_1996}
O'dell, C.~R. \& Wong, K. 1996, The Astronomical Journal, 111, 846

\bibitem[{Oliveira {et~al.}(2010)Oliveira, Pontoppidan, Merín, van Dishoeck,
  Lahuis, Geers, Jørgensen, Olofsson, Augereau, \&
  Brown}]{oliveira_spitzer_2010}
Oliveira, I., Pontoppidan, K.~M., Merín, B., {et~al.} 2010, The Astrophysical
  Journal, 714, 778

\bibitem[{Ossenkopf {et~al.}(2013)Ossenkopf, Röllig, Neufeld, Pilleri, Lis,
  Fuente, van~der Tak, \& Bergin}]{ossenkopf_herschel/hifi_2013}
Ossenkopf, V., Röllig, M., Neufeld, D.~A., {et~al.} 2013, Astronomy and
  Astrophysics, 550, A57

\bibitem[{Ott(2010)}]{ott_herschel_2010}
Ott, S. 2010, in {ASP} {Conference} {Series}, Vol. 434, eprint:
  arXiv:1011.1209, 139

\bibitem[{Owen {et~al.}(2011)Owen, Ercolano, \& Clarke}]{owen_imprint_2011}
Owen, J.~E., Ercolano, B., \& Clarke, C.~J. 2011, Monthly Notices of the Royal
  Astronomical Society, 411, 1104

\bibitem[{Parikka {et~al.}(2016)Parikka, Habart, Bernard-Salas, Goicoechea,
  Abergel, Pilleri, Dartois, Joblin, Gerin, \& Godard}]{parikka_spatial_2016}
Parikka, A., Habart, E., Bernard-Salas, J., {et~al.} 2016, ArXiv e-prints,
  1609, arXiv:1609.04359

\bibitem[{Pilbratt {et~al.}(2010)Pilbratt, Riedinger, Passvogel, Crone, Doyle,
  Gageur, Heras, Jewell, Metcalfe, Ott, \& Schmidt}]{pilbratt_Herschel_2010}
Pilbratt, G.~L., Riedinger, J.~R., Passvogel, T., {et~al.} 2010, Astronomy and
  Astrophysics, 518, L1

\bibitem[{{Planck Collaboration} {et~al.}(2011){Planck Collaboration}, Abergel,
  Ade, Aghanim, Arnaud, Ashdown, Aumont, Baccigalupi, Balbi, Banday, Barreiro,
  Bartlett, Battaner, Benabed, Benoît, Bernard, Bersanelli, Bhatia, Bock,
  Bonaldi, Bond, Borrill, Bouchet, Boulanger, Bucher, Burigana, Cabella,
  Cardoso, Catalano, Cayón, Challinor, Chamballu, Chiang, Chiang, Christensen,
  Clements, Colombi, Couchot, Coulais, Crill, Cuttaia, Danese, Davies, Davis,
  de~Bernardis, de~Gasperis, de~Rosa, de~Zotti, Delabrouille, Delouis, Désert,
  Dickinson, Dobashi, Donzelli, Doré, Dörl, Douspis, Dupac, Efstathiou,
  Enßlin, Eriksen, Finelli, Forni, Frailis, Franceschi, Galeotta, Ganga,
  Giard, Giardino, Giraud-Héraud, González-Nuevo, Górski, Gratton, Gregorio,
  Gruppuso, Guillet, Hansen, Harrison, Henrot-Versillé, Herranz, Hildebrandt,
  Hivon, Hobson, Holmes, Hovest, Hoyland, Huffenberger, Jaffe, Jones, Jones,
  Juvela, Keihänen, Keskitalo, Kisner, Kneissl, Knox, Kurki-Suonio, Lagache,
  Lamarre, Lasenby, Laureijs, Lawrence, Leach, Leonardi, Leroy, Linden-Vørnle,
  López-Caniego, Lubin, Macías-Pérez, MacTavish, Maffei, Mandolesi, Mann,
  Maris, Marshall, Martin, Martínez-González, Masi, Matarrese, Matthai,
  Mazzotta, McGehee, Meinhold, Melchiorri, Mendes, Mennella, Mitra,
  Miville-Deschênes, Moneti, Montier, Morgante, Mortlock, Munshi, Murphy,
  Naselsky, Natoli, Netterfield, Nørgaard-Nielsen, Noviello, Novikov, Novikov,
  Osborne, Pajot, Paladini, Pasian, Patanchon, Perdereau, Perotto, Perrotta,
  Piacentini, Piat, Plaszczynski, Pointecouteau, Polenta, Ponthieu, Poutanen,
  Prézeau, Prunet, Puget, Reach, Rebolo, Reinecke, Renault, Ricciardi, Riller,
  Ristorcelli, Rocha, Rosset, Rubiño-Martín, Rusholme, Sandri, Santos,
  Savini, Scott, Seiffert, Shellard, Smoot, Starck, Stivoli, Stolyarov,
  Sudiwala, Sygnet, Tauber, Terenzi, Toffolatti, Tomasi, Torre, Tristram,
  Tuovinen, Umana, Valenziano, Verstraete, Vielva, Villa, Vittorio, Wade,
  Wandelt, Yvon, Zacchei, \& Zonca}]{planck_collaboration_planck_2011}
{Planck Collaboration}, Abergel, A., Ade, P. A.~R., {et~al.} 2011, Astronomy
  and Astrophysics, 536, A25

\bibitem[{Podio {et~al.}(2012)Podio, Kamp, Flower, Howard, Sandell, Mora,
  Aresu, Brittain, Dent, Pinte, \& White}]{podio_herschel/pacs_2012}
Podio, L., Kamp, I., Flower, D., {et~al.} 2012, Astronomy and Astrophysics,
  545, A44

\bibitem[{Poglitsch {et~al.}(2010)Poglitsch, Waelkens, Geis, Feuchtgruber,
  Vandenbussche, Rodriguez, Krause, Renotte, van Hoof, Saraceno, Cepa,
  Kerschbaum, Agnèse, Ali, Altieri, Andreani, Augueres, Balog, Barl, Bauer,
  Belbachir, Benedettini, Billot, Boulade, Bischof, Blommaert, Callut, Cara,
  Cerulli, Cesarsky, Contursi, Creten, De~Meester, Doublier, Doumayrou, Duband,
  Exter, Genzel, Gillis, Grözinger, Henning, Herreros, Huygen, Inguscio,
  Jakob, Jamar, Jean, de~Jong, Katterloher, Kiss, Klaas, Lemke, Lutz, Madden,
  Marquet, Martignac, Mazy, Merken, Montfort, Morbidelli, Müller, Nielbock,
  Okumura, Orfei, Ottensamer, Pezzuto, Popesso, Putzeys, Regibo, Reveret,
  Royer, Sauvage, Schreiber, Stegmaier, Schmitt, Schubert, Sturm, Thiel,
  Tofani, Vavrek, Wetzstein, Wieprecht, \&
  Wiezorrek}]{poglitsch_photodetector_2010}
Poglitsch, A., Waelkens, C., Geis, N., {et~al.} 2010, Astronomy and
  Astrophysics, 518, L2

\bibitem[{Press(2007)}]{press_numerical_2007}
Press, W.~H. 2007, Numerical {Recipes} 3rd {Edition}: {The} {Art} of
  {Scientific} {Computing} (Cambridge University Press)

\bibitem[{Quanz {et~al.}(2010)Quanz, Beuther, Steinacker, Linz, Birkmann,
  Krause, Henning, \& Zhang}]{quanz_large_2010}
Quanz, S.~P., Beuther, H., Steinacker, J., {et~al.} 2010, The Astrophysical
  Journal, 717, 693

\bibitem[{Ricci {et~al.}(2008)Ricci, Robberto, \&
  Soderblom}]{ricci_hubble_2008}
Ricci, L., Robberto, M., \& Soderblom, D.~R. 2008, The Astronomical Journal,
  136, 2136

\bibitem[{Richling \& Yorke(1997)}]{richling_photoevaporation_1997}
Richling, S. \& Yorke, H.~W. 1997, Astronomy and Astrophysics, 327, 317

\bibitem[{Richling \& Yorke(1998)}]{richling_photoevaporation_1998}
Richling, S. \& Yorke, H.~W. 1998, Astronomy and Astrophysics, 340, 508

\bibitem[{Richling \& Yorke(2000)}]{richling_photoevaporation_2000}
Richling, S. \& Yorke, H.~W. 2000, The Astrophysical Journal, 539, 258

\bibitem[{Sahai {et~al.}(2012)Sahai, Güsten, \& Morris}]{sahai_are_2012}
Sahai, R., Güsten, R., \& Morris, M.~R. 2012, The Astrophysical Journal
  Letters, 761, L21

\bibitem[{Shuping {et~al.}(2003)Shuping, Bally, Morris, \&
  Throop}]{shuping_evidence_2003}
Shuping, R.~Y., Bally, J., Morris, M., \& Throop, H. 2003, The Astrophysical
  Journal Letters, 587, L109

\bibitem[{Smith {et~al.}(2003)Smith, Bally, \& Morse}]{smith_numerous_2003}
Smith, N., Bally, J., \& Morse, J.~A. 2003, The Astrophysical Journal Letters,
  587, L105

\bibitem[{Smith {et~al.}(2010{\natexlab{a}})Smith, Bally, \&
  Walborn}]{smith_HST/ACS_2010}
Smith, N., Bally, J., \& Walborn, N.~R. 2010{\natexlab{a}}, Monthly Notices of
  the Royal Astronomical Society, 405, 1153

\bibitem[{Smith {et~al.}(2010{\natexlab{b}})Smith, Povich, Whitney, Churchwell,
  Babler, Meade, Bally, Gehrz, Robitaille, \& Stassun}]{smith_spitzer_2010}
Smith, N., Povich, M.~S., Whitney, B.~A., {et~al.} 2010{\natexlab{b}}, Monthly
  Notices of the Royal Astronomical Society, 406, 952

\bibitem[{Störzer \& Hollenbach(1998)}]{storzer_[o_1998}
Störzer, H. \& Hollenbach, D. 1998, The Astrophysical Journal Letters, 502,
  L71

\bibitem[{Störzer \& Hollenbach(1999)}]{storzer_photodissociation_1999}
Störzer, H. \& Hollenbach, D. 1999, The Astrophysical Journal, 515, 669

\bibitem[{Testi {et~al.}(2014)Testi, Birnstiel, Ricci, Andrews, Blum,
  Carpenter, Dominik, Isella, Natta, Williams, \& Wilner}]{testi_dust_2014}
Testi, L., Birnstiel, T., Ricci, L., {et~al.} 2014, Protostars and Planets VI,
  339

\bibitem[{Tielens(2005)}]{tielens_physics_2005}
Tielens, A. G. G.~M. 2005, The {Physics} and {Chemistry} of the {Interstellar}
  {Medium} (Cambridge University Press)

\bibitem[{Tielens \& Hollenbach(1985)}]{tielens_photodissociation_1985}
Tielens, A. G. G.~M. \& Hollenbach, D. 1985, The Astrophysical Journal, 291,
  722

\bibitem[{van~der Tak {et~al.}(2007)van~der Tak, Black, Schöier, Jansen, \&
  van Dishoeck}]{van_der_tak_computer_2007}
van~der Tak, F. F.~S., Black, J.~H., Schöier, F.~L., Jansen, D.~J., \& van
  Dishoeck, E.~F. 2007, Astronomy and Astrophysics, 468, 627

\bibitem[{van~der Werf {et~al.}(2013)van~der Werf, Goss, \&
  O'Dell}]{van_der_werf_tearing_2013}
van~der Werf, P.~P., Goss, W.~M., \& O'Dell, C.~R. 2013, The Astrophysical
  Journal, 762, 101

\bibitem[{Vicente(2009{\natexlab{a}})}]{vicente__2009}
Vicente, S. 2009{\natexlab{a}}, PhD thesis, University of Lisbon

\bibitem[{Vicente {et~al.}(2013)Vicente, Berné, Tielens, Huélamo, Pantin,
  Kamp, \& Carmona}]{vicente_polycyclic_2013}
Vicente, S., Berné, O., Tielens, A. G. G.~M., {et~al.} 2013, The Astrophysical
  Journal Letters, 765, L38

\bibitem[{Vicente \& Alves(2005)}]{vicente_size_2005}
Vicente, S.~M. \& Alves, J. 2005, Astronomy and Astrophysics, 441, 195

\bibitem[{Vicente(2009{\natexlab{b}})}]{vicente_physical_2009}
Vicente, S. M.~H. 2009{\natexlab{b}}, Ph.D. Thesis

\bibitem[{Walmsley {et~al.}(2000)Walmsley, Natta, Oliva, \&
  Testi}]{walmsley_structure_2000}
Walmsley, C.~M., Natta, A., Oliva, E., \& Testi, L. 2000, Astronomy and
  Astrophysics, 364, 301

\bibitem[{Walsh {et~al.}(2013)Walsh, Millar, \& Nomura}]{walsh_molecular_2013}
Walsh, C., Millar, T.~J., \& Nomura, H. 2013, The Astrophysical Journal
  Letters, 766, L23

\bibitem[{Weilbacher {et~al.}(2015)Weilbacher, Monreal-Ibero, Kollatschny,
  Ginsburg, McLeod, Kamann, Sandin, Palsa, Wisotzki, Bacon, Selman, Brinchmann,
  Caruana, Kelz, Martinsson, Pécontal-Rousset, Richard, \&
  Wendt}]{weilbacher_muse_2015}
Weilbacher, P.~M., Monreal-Ibero, A., Kollatschny, W., {et~al.} 2015, Astronomy
  and Astrophysics, 582, A114

\bibitem[{Williams \& Best(2014)}]{williams_parametric_2014}
Williams, J.~P. \& Best, W. M.~J. 2014, The Astrophysical Journal, 788, 59

\bibitem[{Williams \& Cieza(2011)}]{williams_protoplanetary_2011}
Williams, J.~P. \& Cieza, L.~A. 2011, Annual Review of Astronomy and
  Astrophysics, 49, 67

\bibitem[{Williams {et~al.}(2014)Williams, Mann, Di~Francesco, Andrews, Hughes,
  Ricci, Bally, Johnstone, \& Matthews}]{williams_ALMA_2014}
Williams, J.~P., Mann, R.~K., Di~Francesco, J., {et~al.} 2014, The
  Astrophysical Journal, 796, 120

\bibitem[{Yang {et~al.}(2010)Yang, Stancil, Balakrishnan, \&
  Forrey}]{yang_rotational_2010}
Yang, B., Stancil, P.~C., Balakrishnan, N., \& Forrey, R.~C. 2010, The
  Astrophysical Journal, 718, 1062

\bibitem[{Yorke \& Welz(1996)}]{yorke_photoevaporation_1996}
Yorke, H.~W. \& Welz, A. 1996, Astronomy and Astrophysics, 315, 555

\bibitem[{Zuckerman {et~al.}(1995)Zuckerman, Forveille, \&
  Kastner}]{zuckerman_inhibition_1995}
Zuckerman, B., Forveille, T., \& Kastner, J.~H. 1995, Nature, 373, 494

\end{thebibliography}
%\bibitem[Habing(1968)]{toto} Habing, H.~J.\ 1968, \bain, 19, 421
%\bibitem[goicoechea_2016]{goicoechea_2016} Goicoechea J. R., et al. 2016, Nature, submitted

%===========================================================%
%						     Appendices							 %
%===========================================================% 

\begin{appendix}

%			  		      App. C. case of the [CII]				 	  %
%--------------------------------------------------------------------------------------------------------%

\section{105-600: case of the \big[CII\big] emission}
\label{app_CII_CPP}

\begin{figure}[h]
  \begin{center}
    \includegraphics[width = 0.48\textwidth]{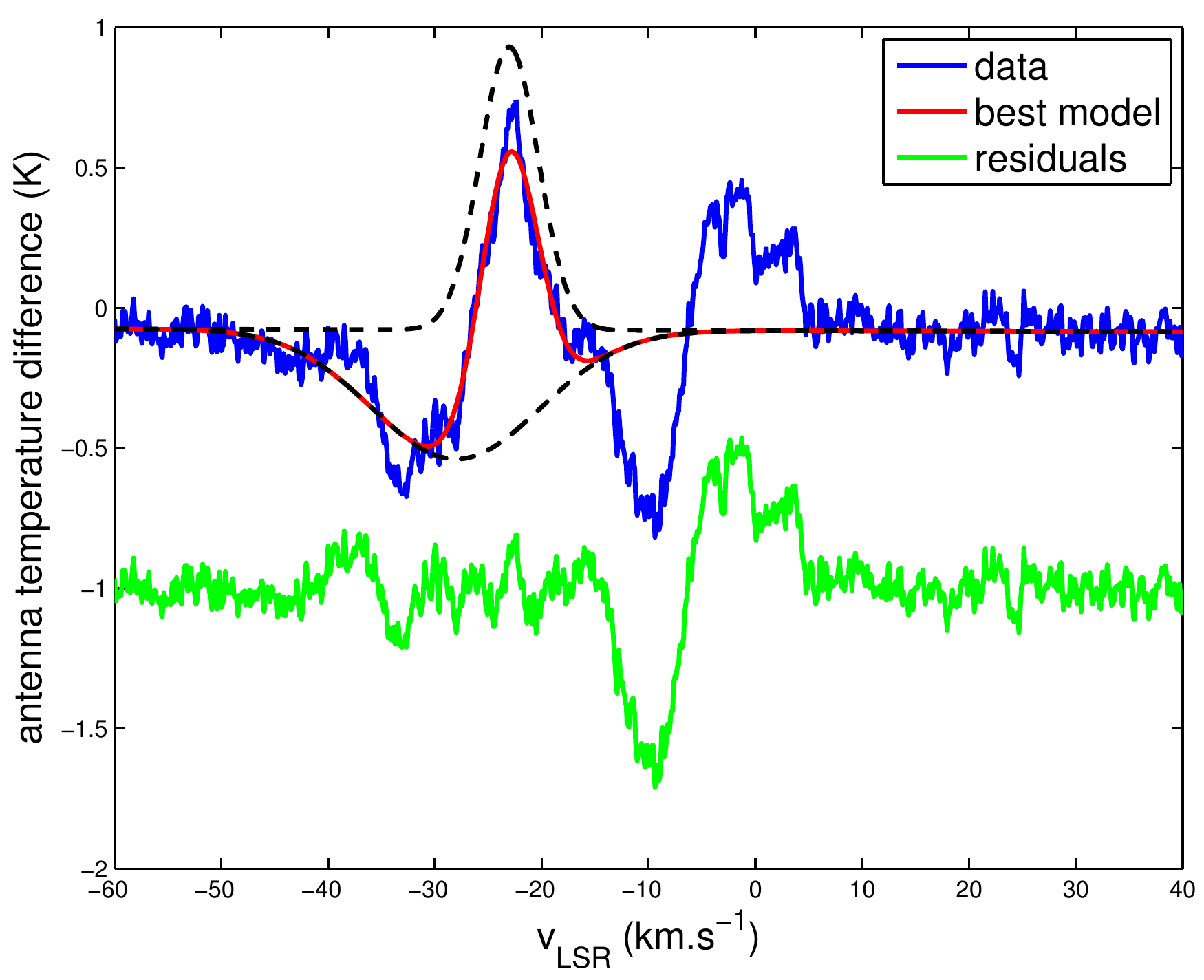}
  \end{center}
\caption{Fit of the \big[CII\big] fine-structure line. The blue curve is the level-2 calibrated spectrum which is best fitted by the red curve. The positive dashed black line is the best model without the OFF line corresponding to the observation that we could have done if there was no contamination from the nebula. The negative one is OFF contamination model. Both gaussians are fitted together and the residuals of the fit are given by the green curve.}
\label{fig_fitCII}
\end{figure}

As can be seen on Fig. \ref{fig_herschelLines_HIFI_105-600}, the fine-structure line of \big[CII\big] from the target does not seem well defined. Indeed, since the observing mode chosen was dual beam switch (see Table \ref{table_targets_observations}), the final spectrum is the result of a spectra obtained on the target (ON position) and others off the target (OFF position). Rather being flat, The off position is apparently not free of emission for this line so that we sum a positive line with a negative one. Fortunately, we have some information about the target as well as for the nebula which is seen in the OFF position. It is known from \citet{sahai_are_2012} and from our other resolved lines (see Table \ref{table_lines_features}) for this object that it has a $v_\mathrm{LSR}$ about $- ~23$ km\,s$^\mathrm{-1}$. This confirm that the positive peak seen on the resulting spectrum is the target emission. Concerning the OFF emission from the nebula, it is known that it comes from a HII region where the temperature is about 10$^\mathrm{4}$ K. At this high temperature, the width of the line is dominated by the thermal doppler broadening. With the constraints on the position of the target line and the width of the nebular line, we can fit the spectrum (Fig. \ref{fig_fitCII}).

Our model with these two gaussian lines fit quite well the part of the spectrum with $v_\mathrm{LSR} < - 10$ km\,s$^\mathrm{-1}$ while there are still some features on the remaining part which do not overlap the target emission. The constraints are sufficient to retrieved the target profile with a good uncertainty (see Table \ref{table_lines_features}).

%			  	      App. A. Effect of grain parameters			 	  %
%--------------------------------------------------------------------------------------------------------%

\section{Effect of grain parameters (test on 105-600)} \label{app_grainParam}

While grains represent only a small fraction of the total mass, a correct treatment of their role is crucial since the photoelectric effect on dust grains \citep{bakes_photoelectric_1994} is supposed to be one of the major sources of heating and because they are responsible for the UV extinction in the PDR. The grain population is generally not precisely constrained in an evolving protoplanetary disk. It should slightly differs from the one of the ISM, so we investigated the effect of changing the main grain parameters in the model: the grain size distribution given by the minimum radius ($a_\mathrm{min}$), the maximum radius ($a_\mathrm{max}$) and the power-law index, and the dust-to-gas mass ratio $\delta{}$.
%To investigate the impact and the confidence on the grains parameters we run models varying the grains minimum radius (Fig. \ref{fig_obsVsMod_grainsRmin}) and the gas-to-dust mass ratio (Fig. \ref{fig_obsVsMod_gtd}). 

\begin{figure*}
 \begin{center}
  \subfigure[]{%
  \includegraphics[width=.5\linewidth]{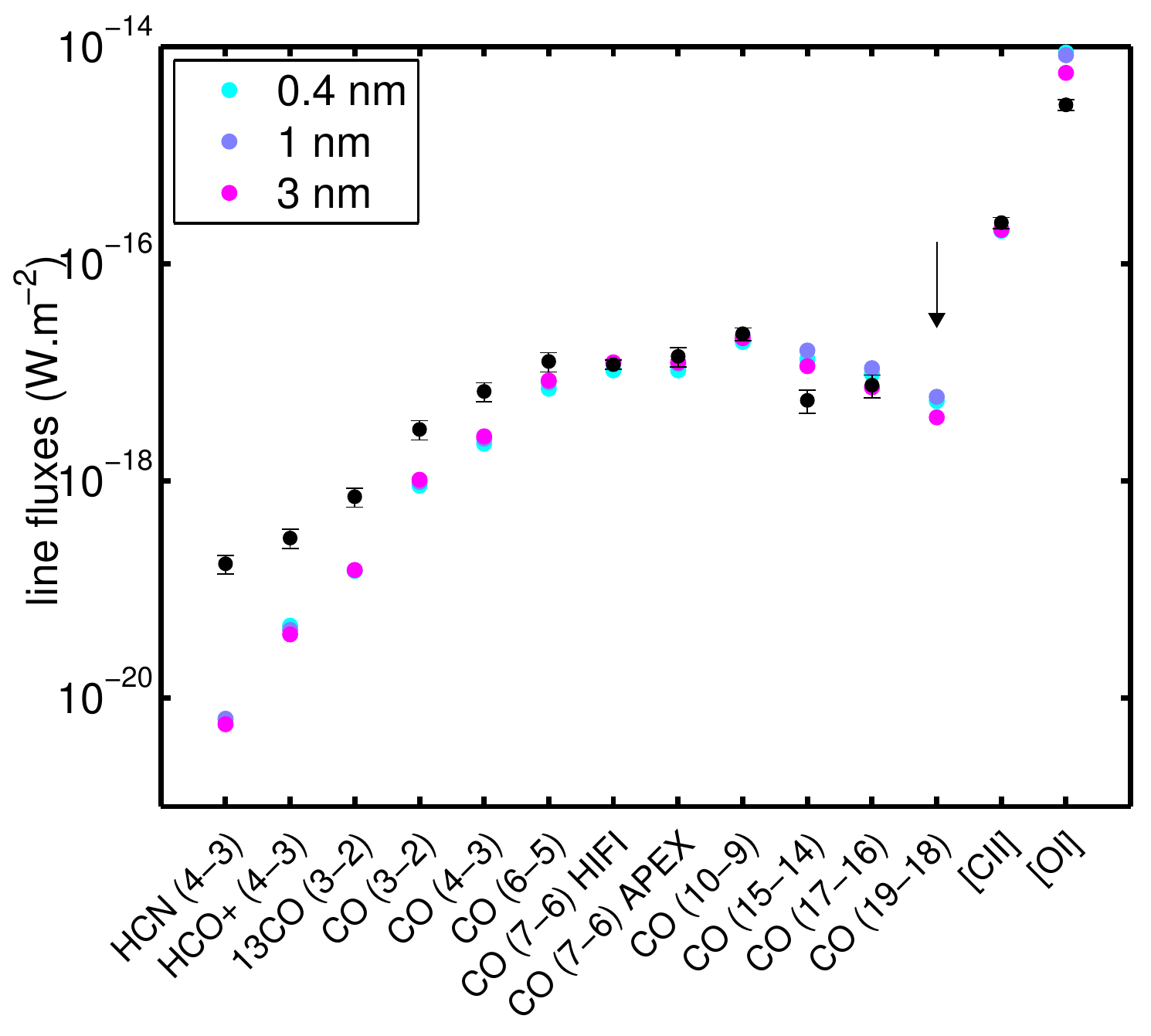}
  %\label{fig:sub1}
  }%
  \subfigure[]{%
  \raisebox{0.1\height}{\includegraphics[width=.5\linewidth]{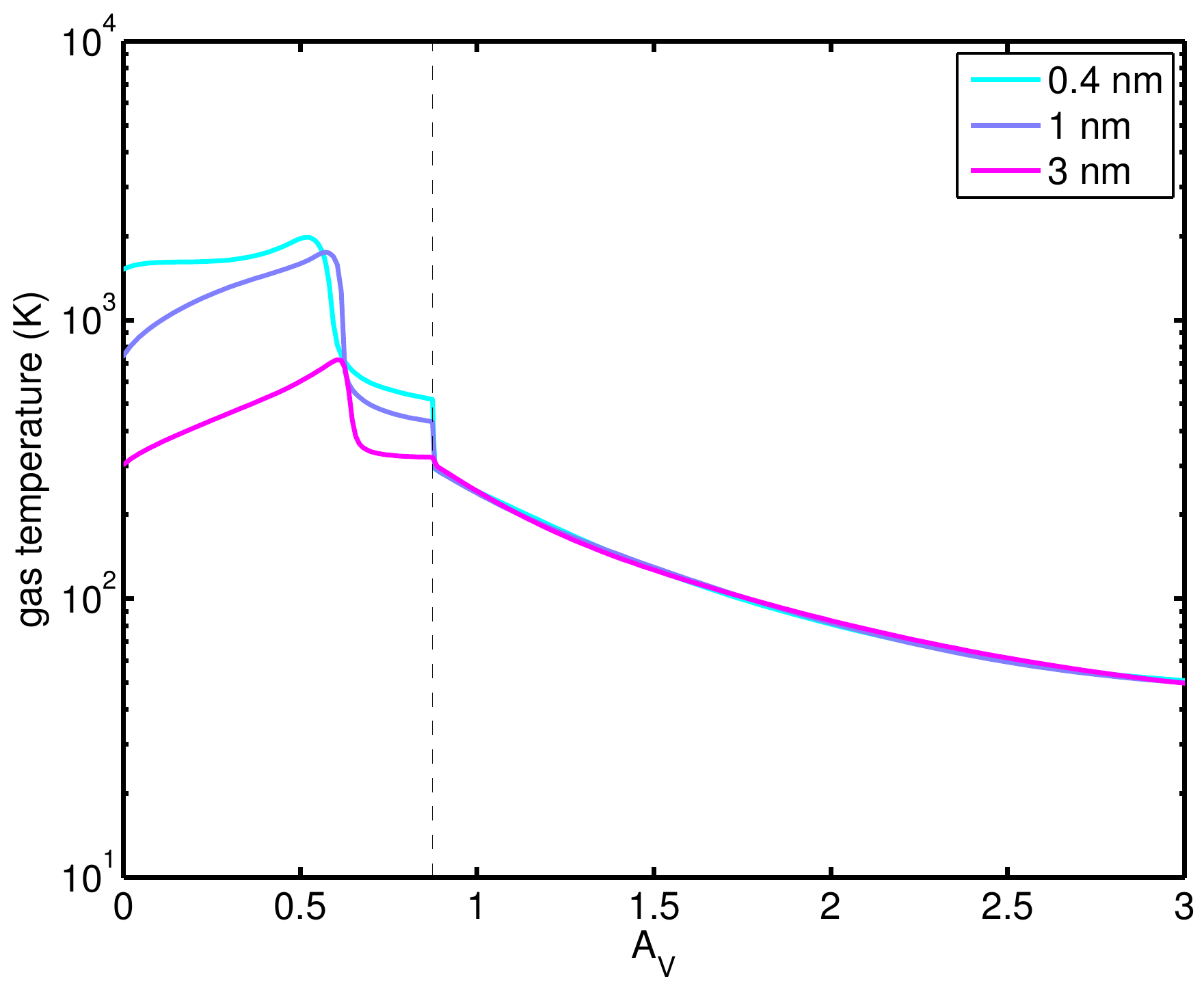}}
  %\label{fig:sub2}
  }%
 \end{center}
 \caption{Variations of the modelled line fluxes (a) and temperature profile (b) for various minimal grain size in the PDR model given in the legend. Here, densities are fixed to $n_\mathrm{env} = 1\times 10^4$ cm$^\mathrm{-3}$,  $n_\mathrm{disk} = 1\times 10^6$ cm$^\mathrm{-3}$, the FUV field is $G_\mathrm{0} = 2\times 10^4$ and a gas-to-dust mass ratio of 100 is used.}
 \label{fig_obsVsMod_grainsRmin}
\end{figure*}

If, contrary to the models presented in the main text, we include the smallest particles, i.e. we set $a_\mathrm{min}$ to 0.4 nm instead of 3 nm, we observe very minor changes in the modeled line intensities: low-$J$ CO lines are slightly lower and the \big[OI\big] 63 \textmu{}m is slightly higher (Fig. \ref{fig_obsVsMod_grainsRmin}). In the absence of strong dependency, those lines cannot significantly constrain with confidence this grain parameter. The temperature at the surface of the disk is almost constant but the thermal profile is significantly changed in the envelope with strong increase of the temperature when small grains are included because they improve heating by photoelectric effect. According to the electron density estimated at the ionisation front, $n_\mathrm{e} \leq 680 ~\mathrm{cm^{-3}}$, and assuming a pressure equilibrium (constant $P = n T$) between the neutral and ionised gas, we can derive the temperature on the side of the envelope. Note that, rigorously, the pressure in the envelope is likely somewhat larger than in the ionised gas in order to maintain the pressure gradient of the photoevaporation flow. However, the temperature on the side of the envelope should be
\begin{equation}
T_\mathrm{env,IF} \approx \frac{n_\mathrm{e}T_\mathrm{HII}}{n_\mathrm{env}},
\end{equation}
where $T_\mathrm{HII} = 10^4$ K is the temperature in the ionized region. In the case of the best-fit model for 105-600 where $n_\mathrm{env} = 8.5 \times 10^3 ~\mathrm{cm^{-3}}$, the temperature is thus estimated to be $T_\mathrm{env,IF} \leq 800$ K. Models are consistent with this result only if small grains are removed by setting the minimal radius to 3 nm or a bit less, i.e. without PAHs. This is consistent with the observed low abundance of PAHs (Sect. \ref{subSubSect_CCP_dust}). 

\begin{figure*}
 \begin{center}
  \subfigure[]{%
  \includegraphics[width=.5\linewidth]{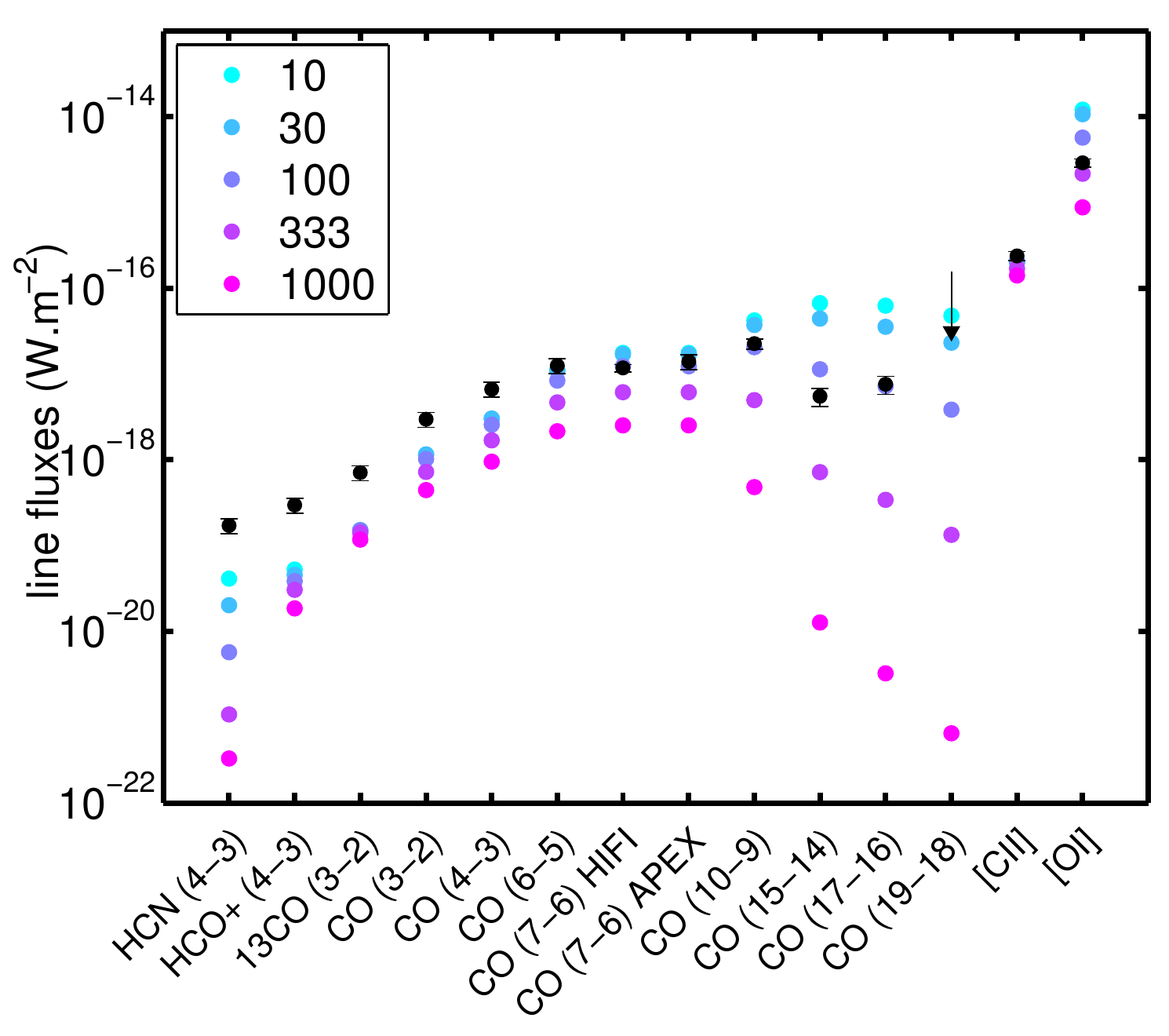}
  %\label{fig:sub1}
  }%
  \subfigure[]{%
  \raisebox{0.1\height}{\includegraphics[width=.5\linewidth]{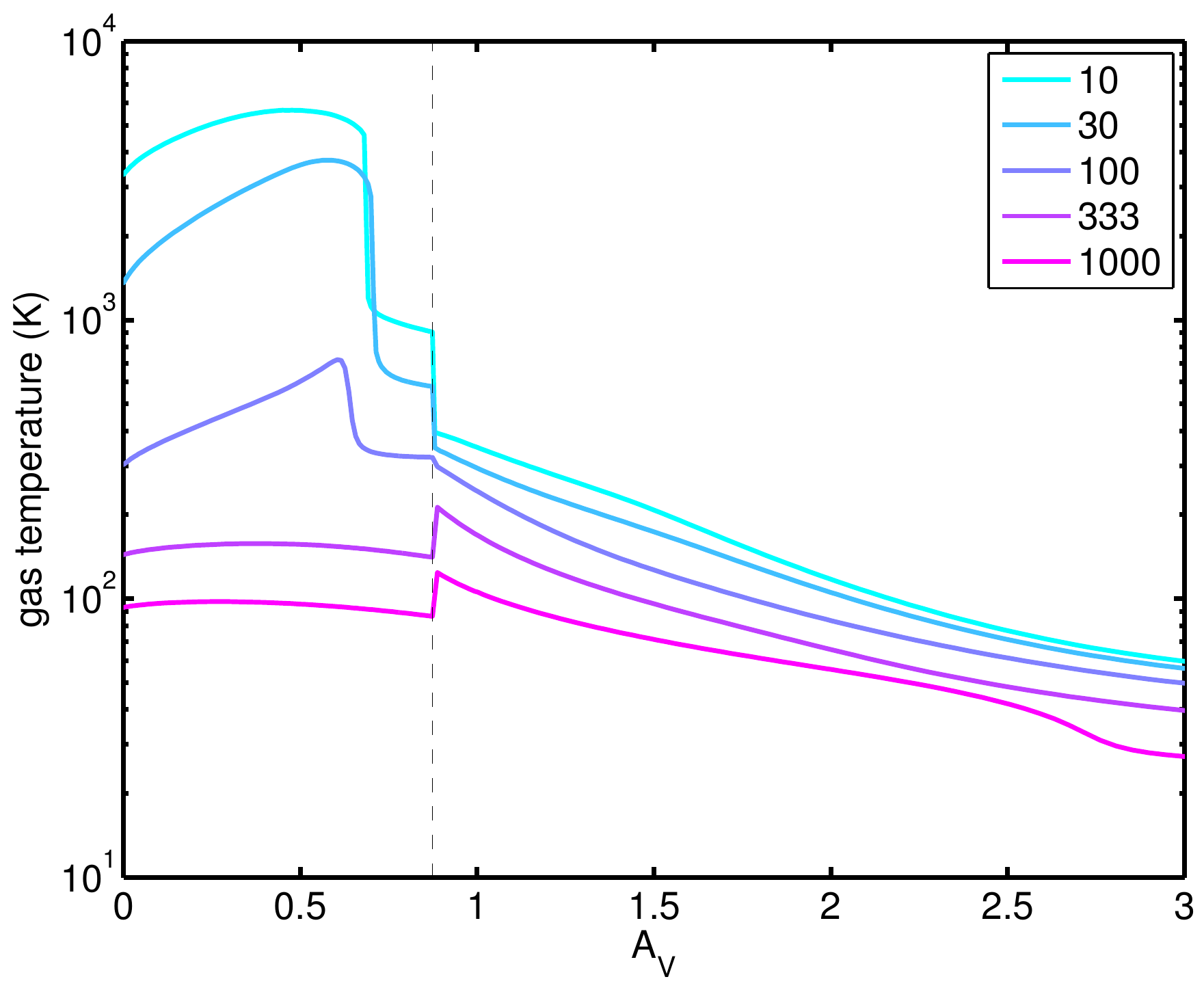}}
  \label{fig_obsVsMod_gtd_gasTemp}
  }%
 \end{center}
 \caption{Variations of the modelled line fluxes (a) and temperature profile (b) for various gas-to-dust mass ratio in the PDR model given in the legend. Here, densities are fixed to $n_\mathrm{env} = 1\times 10^4$ cm$^\mathrm{-3}$, $n_\mathrm{disk} = 1\times 10^6$ cm$^\mathrm{-3}$, , the FUV field is $G_\mathrm{0} = 2\times 10^4$ and a minimal grain radius of 3 nm is used.}
 \label{fig_obsVsMod_gtd}
\end{figure*}

We investigate the effect of changing the dust-to-gas mass ratio by varying it up to an order of magnitude (Fig. \ref{fig_obsVsMod_gtd}). In that case, line fluxes are very sensitive and the best model is the one with the standard value of 100 which is, moreover, compatible with the mass estimates (see Sect. \ref{subSect_massEst}). Assuming a slightly higher dust-to-gas mass ratio is still possible but it will increase the temperature at the ionisation front and will not be consistent with the 800 K limit for the envelope side. We therefore adopt gas-to-dust mass ratio of 100 ($\delta = 0.01$) and $a_\mathrm{min} = 3$ nm. 

We also investigated the effect of the grain growth by increasing the maximum radius and decreasing the power-law index.Adding large grains does not significantly impact the photo-electric effect but the FUV extinction curve is changed. To better describe this effect we ran tests using a standard ISM size distribution ($a_\mathrm{max} = 0.3$ \textmu{}m and a power-law index of 3.50), but with lower extinctions with the curves measured towards HD38087, HD36982, and HD37023. These span a wide range of dust FUV extinction cross section per H nucleus $\sigma_\mathrm{ext}$ including a value as low as $\simeq 8 \times 10^{-22} ~\mathrm{cm^{-2}}$ which is the value observed towards \textTheta{}$^1$ Ori C and appropriate for Orion proplyds \citep{storzer_photodissociation_1999}. The main effect of lowering the FUV extinction is to shift the H/H$_\mathrm{2}$ transition deeper in the PDR. This does not change the main heating/cooling processes involved nor the general results of our study. However, this could result in a slightly higher mean density and subsequent total column density for the envelope (as a comparison, for 105-600, the best-fitted density is $n_\mathrm{env} = 8.5 \times 10^3$ cm$^\mathrm{-3}$ for the Galaxy curve and $n_\mathrm{env} = 9.0 \times 10^3$ cm$^\mathrm{-3}$ for the one of HD37023). Since, we did not obtain better fits to our observations by changing the extinction curve, we kept a standard Milky Way extinction curve.

%			 App. B. The possibility of emissions from shocks		 	  %
%--------------------------------------------------------------------------------------------------------%

\section{105-600: the possibility of emissions from shocks} \label{app_shockPossibility}
The Carina candidate proplyd exhibits a bipolar collimated jet that is visible in HST/ACS H\textalpha{} image \citep{smith_HST/ACS_2010}. Shocks along jets around protostars at comparable evolutionary stages are known to be sources of optical and FIR emission lines \citep[see e.g.][]{podio_herschel/pacs_2012}. We should thus verify if some of the observed lines could originate in shocks rather than in the PDR.

Fast and dissociative \textit{J}-type shock models (velocity between 30 and 150 km\,s$^\mathrm{-1}$) of \citet{hollenbach_molecule_1989} could explain a fraction of the observed atomic and molecular lines but predict, for a pre-shock density of $10^4$ cm$^\mathrm{-3}$ or more, that the shocked surface should be bright in the H\textalpha{} image which is not the case (see Fig. \ref{fig_halpha_105-600}). For lower densities, the emissions  in the FIR become negligible compared to the observations. We conclude that J-shocks are unlikely to be responsible for the observed infrared lines. Moreover, except if the jet is almost perfectly in the plane of the sky, the relatively low FWHM of the \big[CII\big] line (6.4 km.$^{-1}$) suggests that this line is not coming from a high-velocity shock. 

For low-velocity shocks ($\le 30$ km\,s$^\mathrm{-1}$) associated with a jet, we looked at the estimated line fluxes from C and J-shock models of \citet{flower_excitation_2010}. C-shocks produce negligeable \big[OI\big] emission compared to the observed one while J-shocks are not able to produce the observed \big[OI\big] emission without being, at least, two orders of magnitude brighter in CO compared to the observed emission. Atomic lines are thus very unlikely produced by such shocks. Finally, we conclude that atomic lines arise from the PDR of 105-600.
%If we can modelled the intensity of some CO lines, one at a time, by going down to the lower velocities and lower densities, the relative emission between the observed transitions is not respected at all with significant differences.

The very low FWHM of CO lines observed with \textit{Herschel}/HIFI or APEX rule out the origin of this emission from relatively high-velocity shocks associated with the jet. The presence of the jet, however, suggest an accretion of matter onto a young star \citep[see e.g.][]{konigl_disk_2000}. Different low-velocity shocks can be expected in forming disk such as an accretion shock at the surface of the molecular disk, shock waves launched deeper in the disk or residual shocks from the interaction of the jet surface with the disk \citep{boss_sources_2005}. We used a C-shock code assuming a shock velocity of 4 - 6 km s$^{-1}$, a magnetic field of 1 - 2 mG and a pre-shock density of $10^6 ~\mathrm{cm^{-3}}$ to estimate the radiated energy to compare it with the UV energy absorbed. In the most favorable case, the radiated energy flux, or radiated power per unit of shocked surface, reaches $8.65 \times 10^{-5} ~\mathrm{J~m^{-2}~s^{-1}}$. The photo-electric effect injects a fraction $\epsilon{}_\mathrm{pe} \approx 1\%$ of the UV flux ($G_\mathrm{0,mol}$ times the Habing's field) into the gas. Hence, the power per unit of disk surface is $\epsilon{}_\mathrm{pe} ~ G_\mathrm{0,mol} ~1.6 \times 10^{-6} ~\mathrm{J~m^{-2}~s^{-1}}$ where $G_\mathrm{0,mol}$ is the UV radiation field incoming at the surface of the disk which is about 10\% of the initial flux incoming on the envelope for the range of studied envelope densities. With these values and $G_\mathrm{0,mol} = 2.2 \times 10^3$, the energy flux is $3.52 \times 10^{-5} ~\mathrm{J~m^{-2}~s^{-1}}$. To have an important contribution from the shock, the comparison of the two sources of energy implies that the shocked surface has to be about 45\% of the disk surface. Moreover, if it happens, some of the emitted lines can be of the order of the observations only if the shock is not FUV-illuminated. Otherwise the emission drops significantly. Hence we cannot fully exclude that shocks could have a contribution to the CO emission, however this requires that the shocked surface is large and that it is protected from UV radiation. Both constraints seem difficult to concile. Altogether, we conclude that shocks are unlikely to play a major role in the observed emission.

%			  	    App. D. Molecular mass from CO lines			 	  %
%--------------------------------------------------------------------------------------------------------%

\section{105-600: mass and mass-loss rate}

\subsection{Molecular mass from $^\mathrm{12}$CO and $^\mathrm{13}$CO (3-2) lines}
\label{app_molMass}

Column density of hydrogen in the molecular disk, and so its mass, can be estimated from the $^\mathrm{12}$CO (3-2) and $^\mathrm{13}$CO (3-2) transition lines. Under the assumption that the $^\mathrm{13}$CO line to be optically thin, the column density of the upper energy level (here 3), noted $N_\mathrm{u}$, is linked to the integrated intensity of the line,
\begin{equation}
N_\mathrm{u} = \frac{8 \pi{} k_\mathrm{B} \nu^2}{h c^3 A_\mathrm{ul}} \int{T_\mathrm{B}\left(^{13}\mathrm{CO}\right) d\nu},
\end{equation}
where $\nu$ is the frequency of the transition, $A_\mathrm{ul}$ the Einstein emission-coefficient between the upper and the lower levels of the transition. $h$ is the Planck's constant, $c$ the vacuum speed of light, $k_\mathrm{B}$ the Boltzmann's constant and  $\int{T_\mathrm{B}\left(^{13}\mathrm{CO}\right) d\nu}$ the integrated intensity of the line.
Additionally assuming that Local Thermodynamical Equilibrium (LTE) conditions are verified for $^{13}$CO, the column density of the upper level, $N_\mathrm{u}$, is linked to the total one, $N$, by
\begin{equation}
N_\mathrm{u} = \frac{N}{Z} g_\mathrm{u} \exp{\left(-\frac{E_\mathrm{u}}{k_\mathrm{B} T_\mathrm{ex}}\right)},
\end{equation}
where $g_\mathrm{u}$ and $E_\mathrm{u}$ are respectively the statistical weight and energy of the upper level. $Z$ is the partition function and  $T_\mathrm{ex}$ is the excitation temperature of the transition. This latter can be calculated thanks to the optically thick emission of $^\mathrm{12}$CO since we have
\begin{equation}
T_\mathrm{B}\left(^{12}\mathrm{CO}\right)  = \frac{h\nu}{k_\mathrm{B}} \left[\frac{1}{\exp{\left(\frac{h\nu}{k_\mathrm{B} T_\mathrm{ex}}\right)}-1} - \frac{1}{\exp{\left(\frac{h\nu}{k_\mathrm{B} T_\mathrm{bb}}\right)}-1}  \right], %\frac{1}{\exp{\left(h\nu/\left(k_\mathrm{B} T_\mathrm{bb}\right)\right)-1}}
\end{equation}
where $T_\mathrm{bb} = 2.7$ K is the background temperature.

From \citet{sahai_are_2012}, we have $T_\mathrm{B}\left(^{12}\mathrm{CO}\right) = 33.0$ K so we get $T_\mathrm{ex} = 40.8$ K. With a line integrated intensity of $\int{T_\mathrm{B}\left(^{13}\mathrm{CO}\right) d\nu} = 0.96$ K.km\,s$^\mathrm{-1}$,  we find $N\left(^{13}\mathrm{CO}\right) = 3.53 \times 10^{14}$ cm$^\mathrm{-2}$. Supposing an abundance ratio $\left[\mathrm{H_2}\right]/\left[^{13}\mathrm{CO}\right] = 7 \times 10^5$, we obtain the column density of $H\mathrm{2}$, $N_\mathrm{H_2} = 9.84 \times 10^{22}$ cm$^\mathrm{-2}$. The molecular mass, $M_\mathrm{mol}$, is finally calculated by
\begin{equation}
M_\mathrm{mol} = \mu{} ~m_\mathrm{H} ~S ~N_\mathrm{H_2},
\end{equation}
where $\mu = 2.8$ is the mean molecular weigh, $m_\mathrm{H}$ is the mass of a hydrogen atom and $S$ is the apparent surface visible is the beam of the instrument at the distance of the observed object. In our case, the molecular mass is $M_\mathrm{mol} \approx 0.187 ~\mathrm{M_\sun}$.

\subsection{Consistency of mass estimates} \label{subSect_massEst}

 We can roughly estimate of the disk mass assuming spherical geometry and making the strong hypothesis that the density is constant throughout the disk. In reality, the density deep in the disk could be much higher so that this estimate can be seen as a lower limit. From our PDR best-fit model, this mass is $M_\mathrm{disk} \gtrsim 0.19$ M$_\sun$. 
In the same way, we can evaluate the mass of the envelope assuming that the 3D shape is the one of an ellipsoid with axis of $9.5\arcsec \times 3.7\arcsec \times 3.7\arcsec$ (from Table \ref{table_targets_properties}) which gives $M_\mathrm{env} \approx 0.028$ M$_\sun$ for the best-fit model. 
%Since the estimated density is this time representative of the whole component, this calculation is a direct estimation of the mass rather than a lower limit.

We have used different methods to estimate the mass of each component in the Carina candidate proplyd. The mass of the atomic envelope is estimated to be about $0.05$ M$_\sun$ from the SED (Sect. \ref{subSubSect_CCP_dust}) and about $0.03$ M$_\sun$ from the PDR models (Sect. \ref{subSubSect_CCP_gas}). For the mass of the disk, the SED gives a value of about $0.7$ M$_\sun$ (Sect. \ref{subSubSect_CCP_dust}), consistent with the lower limit of $0.19$ M$_\sun$ from the PDR model (Sect. \ref{subSubSect_CCP_gas}). The fit on the thermal components of the dust emission using modified Planck's functions is based on empirical parameters such as the spectral index $\beta$ and the efficiency section $\sigma$. The latter is used to convert the dust opacity obtained from the fit in a total mass. Since the empirical value comes from observations in the galactic plane, it is valid only in environments relatively similar to the ISM, specially in terms of dust-to-gass mass ratio which typical value is 0.01. 
Our models with the Meudon PDR code also take into account this standard value of the dust-to-gas mass ratio and the comparison with the observations have shown that this value is most likely correct (see Appendix \ref{app_grainParam}).

Another estimation of the molecular mass, independent of the dust-to-gas mass ratio  can be obtained using emission of molecular isotopologues. \citet{sahai_are_2012} have estimated, using their measurement of $^\mathrm{12}$CO and $^\mathrm{13}$CO (3-2) line fluxes, a mass of $0.35$ M$_\sun$. With their observations, we use a slightly different method (see details in App. \ref{app_molMass}) and obtained a value of $0.187$ M$_\sun$ under the assumption that the $^\mathrm{13}$CO line is optically thin while the $^\mathrm{12}$CO is optically thick. Calculations with the online non-LTE molecular radiative transfer code RADEX \citep{van_der_tak_computer_2007} confirms that the $^\mathrm{12}$CO line is optically thick but indicates that the $^\mathrm{13}$CO starts to be thick considering the size and the density obtained from our models. Hence, this value of $0.187$ M$_\sun$ is a lower limit.

Overall, estimates of masses are consistent but it is possible that values based on the dust emission are slightly overestimated. Using a dust-to-gas mass ratio a bit higher than the typical value of 0.01, in agreement with what is often found for protoplanetary disks \citep{williams_parametric_2014}, could correct the small differences.

\subsection{Mass-loss rate at ionisation front} \label{subsubSect_masslossIF}
Recombinations of electrons and protons in the HII region give rise to the H\textalpha{} emission at the ionisation front with an intensity depending on the electron density $n_\mathrm{e}$ \citep[see][chap. 8]{tielens_physics_2005}.  % \textcolor{red}{(give the value)}
From the observed emission of $4 \times 10^{-7}$ W\,m$^\mathrm{-2}$\,sr$^\mathrm{-1}$ in the HST/ACS/F658N filter (including H\textalpha{} and \big[NII\big] 6583 \AA{} lines), 
we have derived an upper limit for the electron density of $n_\mathrm{e} \leq 680 ~\mathrm{cm^{-3}}$ when considering that all the flux is coming from the H\textalpha{} line, which is probably close to the reality. Assuming a constant electronic density at the ionisation front all around the envelope, a shape of an ellipsoid of revolution with a surface $S$, and that the gas is escaping at the sound speed $c_\mathrm{S}$ calculated here for a temperature of $10^4$ K, the mass-loss rate is 
\begin{equation}
\dot{M} = \mu ~m_\mathrm{H} ~n_\mathrm{e} ~S ~c_\mathrm{S},
\end{equation}
where $\mu m_\mathrm{H}$ is the mean particle weight. With our values, this gives a mass-loss rate of the order of $2 \times 10^\mathrm{-6}$ M$_\sun/$year 
for 105-600.

\end{appendix}

\end{document}